\documentclass[11pt, a4paper]{article}

\usepackage[height=8.85in,width=6.45in]{geometry}

\usepackage{graphicx,rotating}     
\usepackage[bookmarksopen,colorlinks=true,linkcolor=light_blue,
citecolor=light_pink,urlcolor=dark_red,linktocpage=false]{hyperref}

\usepackage{amsfonts}
\usepackage{amsmath}
\usepackage{amssymb}
\usepackage{slashed}
\usepackage{braket}
\usepackage{subfigure}
\usepackage{bm,here}
\usepackage{cite}
\usepackage{comment}
\usepackage{cases}
\usepackage{colortbl}
\usepackage{xcolor}
\usepackage[utf8]{inputenc}
\usepackage[mathscr]{eucal}
\usepackage[footnotesize]{caption}
\usepackage{lipsum}
\usepackage{subfigure}

\usepackage[utopia]{mathdesign}

\usepackage{multicol}
\definecolor{dark_blue}{rgb}{0.0, 0., 0.6}
\definecolor{dark_red}{rgb}{0.7, 0., 0.}
\definecolor{dark_green}{rgb}{0., 0.45, 0.3}
\definecolor{light_pink}{rgb}{1,0.4,0.4}
\definecolor{light_blue}{rgb}{0.284602,0.317763,0.963947}
\definecolor{red}{rgb}{1,0,0}
\definecolor{blue}{rgb}{0,0,1}
\definecolor{orange}{rgb}{1,0.5,0}

\newcommand{\vev}[1]{ \left\langle {#1} \right\rangle }

\newcommand{\GeV}{\ \text{GeV} }

\newcommand{\dd}{\mathrm{d}}
\newcommand{\Mpl}{M_{\text{Pl}}}
\newcommand{\abs}[1]{\left\vert {#1} \right\vert}

\newcommand{\cnst}{\text{const}}

\makeatletter
\@addtoreset{equation}{section}

\makeatother

\usepackage{framed}
\definecolor{shadecolor}{rgb}{0.95,0.95,0.95}

\definecolor{cred}{RGB}{180,50,40} 
\definecolor{purple}{RGB}{180,90,180} 
\definecolor{darkgreen}{RGB}{0, 100, 0}
   
\begin{document}

\hypersetup{pageanchor=false}
\begin{titlepage}

\begin{center}

\hfill DESY 18-098\\

\vskip 1.2in

{\Huge \bfseries 
Gauge Field and Fermion Production  \\
during Axion Inflation\\
}

\vskip .8in

{\Large Valerie Domcke and Kyohei Mukaida}

\vskip .3in

{\em DESY, Notkestra{\ss}e 85, D-22607 Hamburg, Germany}\\

\end{center}
\vskip .6in

\begin{abstract}
\noindent
We study the dual production of helical Abelian gauge fields and chiral fermions through the Chern-Simons (CS) coupling with a pseudo-scalar inflaton in the presence of a chiral anomaly. Through the CS term, the motion of the inflaton induces a tachyonic instability for one of the two helicities of the gauge field. We show that the resulting helical gauge field necessarily leads to the production of chiral fermions by deforming their Fermi sphere into discrete Landau levels. 
The population of the lowest Landau level leads to a chiral asymmetry as inferred from the chiral anomaly, while the higher levels are populated symmetrically through pair production. From the backreaction of the fermions on the gauge field production  we derive a conservative but stringent upper bound on the magnitude of the gauge fields. 
Consequently, we find that the scalar perturbations sourced by these helical gauge fields, responsible for enhanced structure formation on small scales, get reduced significantly.
We also discuss the fate of the primordial chiral asymmetry and of the helical gauge fields after inflation, and show that the instability in the chiral plasma tends to erase these primordial asymmetries. This result may impact scenarios where the baryon asymmetry of the Universe is connected to primordial magnetic fields.
\end{abstract}

\end{titlepage}

\tableofcontents
\thispagestyle{empty}
\renewcommand{\thepage}{\arabic{page}}
\renewcommand{\thefootnote}{$\natural$\arabic{footnote}}
\setcounter{footnote}{0}
\newpage
\hypersetup{pageanchor=true}

\section{Introduction}
\label{sec:intro}

Particle production in the early Universe is a key element in understanding the properties of the primordial plasma which sets the initial conditions for the hot Big Bang Standard Model of cosmology. The energy density stored as vacuum energy during inflation is transformed into a thermal bath of Standard Model (SM) particles, with an asymmetry between particles and anti-particles generated  along the way. A special role is played by the axial coupling of the inflaton (or of any other scalar field $\phi$) to gauge fields and fermions,
\begin{equation}
 \text{gauge fields:} \quad {\cal L} \supset g^2 \phi \, \epsilon^{\alpha \beta \mu \nu} F_{\alpha \beta}  F_{\mu \nu} \,, \qquad \text{fermions:} \quad {\cal L} \supset \phi \, \partial_\mu J_5^\mu 
 \label{eq:couplings}
\end{equation}
with $F^{\mu \nu}$ denoting the field strength tensor of the gauge fields\footnote{Throughout this paper we will for simplicity focus on Abelian gauge fields. The axial coupling to non-Abelian gauge fields has been studied under the name of chromo-natural inflation~\cite{Adshead:2012kp} and it would be interesting (and indeed necessary for a realistic connection to the SM) to extend our work to this case.}, $g$ is the gauge coupling and $J_5^\mu = \bar \psi \gamma^\mu \gamma_5 \psi$ is the axial current of the fermion $\psi$. By partial integration, these couplings can be expressed as derivative couplings of $\phi$ and hence (at the classical level) preserve any shift symmetry associated with $\phi$. This unique property renders these couplings prime candidates for couplings to the inflaton field in slow-roll inflation, as well as to any axion-like particle or more general any pseudo Nambu Goldstone Boson. On top of this, these couplings come with a range of striking phenomenological implications.

On the one hand, the spontaneous $CP$-violation induced by the rolling scalar field leads to a tachyonic instability in one of the gauge field modes~\cite{Turner:1987vd,Garretson:1992vt, Anber:2006xt}. This has a wide range of consequences, including the production of helical magnetic fields~\cite{Anber:2006xt}, an additional contribution to the scalar power spectrum and to the stochastic gravitational wave background predicted from inflation~\cite{Cook:2011hg,Barnaby:2011qe,Barnaby:2011vw,Anber:2012du,Linde:2012bt} as well as an efficient preheating mechanism~\cite{Adshead:2015pva}.  The gauge fields moreover backreact on the equations of motion for the scalar through an effective friction term. This enables inflation on fairly steep potentials~\cite{Anber:2009ua} and has also been employed to dynamically generate the electroweak scale by means of the relaxion mechanism~\cite{Hook:2016mqo,Tangarife:2017vnd,Tangarife:2017rgl,Fonseca:2018xzp}. The decay of the helical gauge fields after the end of inflation may be the source for the baryon asymmetry of the Universe~\cite{Giovannini:1997eg,Anber:2015yca,Fujita:2016igl,Kamada:2016eeb,Cado:2016kdp,Jimenez:2017cdr}.
On the other hand, implications of the axial coupling of the inflaton to fermions have been studied in Refs.~\cite{Dolgov:1994zq,Kusenko:2014uta,Adshead:2015jza,Adshead:2015kza,DeSimone:2016ofp,Anber:2016yqr}, leading to models of successful baryogenesis accompanied by chiral gravitational wave production. The impact of this coupling on the scalar power spectrum (and bispectrum) has recently been studied in~\cite{Adshead:2018oaa}.

The two couplings in Eq.~\eqref{eq:couplings} can be closely tied through the Adler-Bell-Jackiv (ABJ) anomaly which describes the anomalous nonconservation of the axial current~\cite{Adler:1969gk,Bell:1969ts},
\begin{equation}
 \partial_\mu J^\mu_5 = - \frac{g^2}{16 \pi^2} \epsilon^{\alpha \beta \mu \nu} F_{\alpha \beta} F_{\mu \nu} \,.
 \label{eq:AnomalyIntro}
\end{equation}
Consequently, background field configurations in which the right-hand side of Eq.~\eqref{eq:AnomalyIntro} is non-zero lead to a non-conservation of the number of left- and right-handed fermions (despite these numbers being conserved at the classical level). This immediately generalizes to more complex theories, as long as the group theoretical factors then appearing on the right-hand side of Eq.~\eqref{eq:AnomalyIntro} do not cancel. It is well known that in the SM this is not the case, leading to the SM chiral 
 anomaly~\cite{Peskin:1995ev}. We will here focus on theories with massless fermions. Due to the anomaly~\eqref{eq:AnomalyIntro} massless fermions charged under the gauge symmetry cannot be described by a conformally invariant theory, and the axial coupling in Eq.~\eqref{eq:couplings} cannot be eliminated by a chiral rotation.

In this paper we study the production of helical gauge fields and chiral fermions in the case where the chiral anomaly connects the two operators in Eq.~\eqref{eq:couplings}. 
In this case, the production of the gauge fields and fermions cannot be treated independently, but are linked by Eq.~\eqref{eq:AnomalyIntro}. In other words, a homogeneous scalar field with a non-vanishing velocity leads to an external chemical potential for the chiral fermions as well as to a non-vanishing Chern-Simons number on equal footing. As a result, parallel electric and magnetic fields\footnote{We will borrow the familiar terminology from electro-magnetism to describe the components of the Abelian gauge field.} created via the tachyonic instability necessarily
lead to chiral fermion production by deforming the fermion energy levels to discrete Landau levels. Populating the lowest Landau level yields the chiral asymmetry indicated by Eq.~\eqref{eq:AnomalyIntro}~\cite{Nielsen:1983rb}, while the higher levels are populated through pair-production analogue to the Schwinger effect~\cite{Heisenberg:1935qt,Schwinger:1951nm}.  The chiral fermions, accelerated in the electro-magnetic (EM) field, result in an induced current~\cite{Abramchuk:2016afc,Bavarsad:2017oyv} which backreacts on the gauge fields by inducing an EM field with destructive interference.\footnote{
	See Refs.~\cite{Kobayashi:2014zza,Hayashinaka:2016qqn} for the backreaction from the Schwinger effect in a strong electric field but without a magnetic field.
} We compute the induced current by solving the equations of motion for the fermions in the presence of a suitable gauge field background. Taking into account this backreaction {and assuming the existence of a non-trivial attractor solution for the gauge fields}, we derive upper bounds on the gauge field production which lie significantly below the results obtained in the absence of fermions. 

We briefly discuss possible implications for the wide range of applications mentioned above, focusing on  axion inflation and on baryogenesis. In the case of axion inflation, we find the effective friction arising from the gauge fields is greatly reduced and the scalar power spectrum at small scales is significantly suppressed, with implications for the possibility of primordial black hole production. Concerning baryogenesis, we note the existence of a chiral asymmetry in the fermion sector \textit{at the end of inflation}, which is a direct consequence of Eq.~\eqref{eq:AnomalyIntro}.  The survival of this asymmetry after inflation depends on the efficiency of competing erasure processes: The instability of the thermal plasma arising from the chiral magnetic effect~\cite{Fukushima:2008xe, Akamatsu:2013pjd} strives to symmetrically erase both the helical gauge field and the chiral asymmetry, whereas SM processes (\textit{i.e.}, Yukawa interactions and Sphalerons) may disrupt this balance. 

The remainder of this paper is structured as follows. In Sec.~\ref{sec:setup} we introduce the framework we will be working in and discuss the overall picture in terms of conserved Noether charges and currents. We confirm these results by an explicit computation of the gauge field and fermion production in Sec.~\ref{sec:prod}. This provides insight into the microphysical processes which ensure the equivalence of the two reference frames linked by the anomaly equation and which intimately connect the helical gauge and chiral fermion production. These results are refined in Sec.~\ref{sec:br} by taking into account the backreaction of the fermions on the gauge fields, leading to strong upper bounds on the gauge field production. In Sec.~\ref{sec:pheno} we discuss possible impacts of our results on axion inflation and leptogenesis, before concluding in Sec.~\ref{sec:conclusions}. Details on our notation and the conventions used can be found in App.~\ref{sec:nandc}.

\section{Setup}
\label{sec:setup}

\subsection{Toy model}
\label{sec:toymodel}
For simplicity, we consider the following toy model throughout this paper:
\begin{align}
	S = \int \dd^4 x\, \Bigg\{ \sqrt{-g}
	\left[
		\frac{g^{\mu\nu}}{2} \partial_\mu \phi \partial_\nu \phi - V (\phi)
		- \frac{1}{4} g^{\mu \rho} g^{\nu \sigma} \hat F_{\mu \nu} \hat F _{\rho \sigma}
		+ \hat{\overline \psi} i\hat{\slashed{\cal D}} \hat \psi
	\right] 
	+ \frac{\alpha \phi}{4 \pi f_a} \hat F_{\mu \nu} \hat{\tilde F}^{\mu \nu}
	\Bigg\},
	\label{eq:frame1}
\end{align}
where $\phi$ is a real pseudo-scalar field that could be an inflaton (but not necessarily),
$\hat A_\mu$ is a $\mathrm{U}(1)$ gauge field, and $\hat \psi$ is a massless Dirac fermion with charge $Q$ under this $\mathrm{U}(1)$ group. We will assume that, while the vector current is conserved, the axial current is anomalous.
The dual field strength is defined by $\hat{\tilde F}^{\mu\nu} \equiv \epsilon^{\mu\nu\rho\sigma} \hat F_{\rho\sigma} / 2$ with $\epsilon^{0123} =  + 1$. The $\hat F \hat{\tilde F}$ term is suppressed below the scale $\alpha/f_a$ where $\alpha = g^2/(4 \pi)$ denotes the gauge coupling of the $U(1)$ group.
Throughout this paper, we take the FLRW metric with vanishing curvature,
$\dd s^2 = \dd t^2 - a^2 (t) \dd \bm{x}^2 = a^2 (\eta) ( \dd \eta^2 - \dd \bm{x}^2 )$ with $a$ being the scale factor.
This implies the following vierbein, $e^a_\mu = a \delta^a_\mu$ and $e^\mu_a = \delta^\mu_a / a$; where $\mu$ runs over $\eta,x,y$, and $z$.
The covariant derivative acting on $\hat \psi$ involves the spin connection $\omega_\mu{}^{ab}$:
\begin{align}
	\hat{\slashed{\mathcal D}} \hat\psi 
	& = \hat \gamma^\mu 
	\left( 
		\partial_\mu + i g Q \hat A_\mu + \frac{1}{4} \omega_\mu{}^{ab}
		\gamma_{ab}
	\right) \hat \psi =
	\left[ \hat \gamma^\mu \left( \partial_\mu + i g Q \hat A_\mu \right)
	+ \frac{3}{2} a H \hat \gamma^0 \right] \hat \psi,
\end{align}
where we have inserted the FLRW metric in the second equality with
$H$ being the Hubble parameter, $H \equiv \dot a / a$.
The gamma matrices with a hat fulfill $\{ \hat \gamma^\mu, \hat \gamma^\nu \}  = g^{\mu\nu}$,
while those without a hat satisfy that in the flat spacetime $\{ \gamma^a, \gamma^b \} = \eta^{a b}$.
They are related through $\hat \gamma^\mu = e^\mu_a \gamma^a = \gamma^\mu / a$.
See Ref.~\cite{Parker:2009uva} for an introduction to QFT on curved space time as well as App.~\ref{sec:nandc} for our notations and conventions.

As is well known, massless fermions and gauge fields are conformal. That is, their dynamics does not depend on the scale factor, $a$.
To use this property explicitly, we redefine the fields as follows:
$\psi \equiv a^{3/2} \hat \psi$, $(\hat A_\mu) = (A_0, - \bm{A}) \equiv (A_\mu)$, and $(\hat A^\mu) = (A_0 / a^2, \bm{A} / a^2) \equiv (A^\mu) / a^2$,
where the index of the comoving field $A$ is raised/lowered by $\eta^{\mu\nu}$,
while for the physical field $\hat A$ this is done by $g^{\mu\nu}$.
By means of these rescaled fields, one may rewrite the action as follows:
\begin{align}
	S = \int \dd^4 x \Bigg\{ \sqrt{-g}
	\left[
		\frac{g^{\mu\nu}}{2} \partial_\mu \phi \partial_\nu \phi - V (\phi)
	\right]
		- \frac{1}{4} F_{\mu \nu} F^{\mu\nu}
		+ \overline \psi \left( i \slashed{\partial} - g Q \slashed{A} \right) \psi
	+ \frac{\alpha \phi}{4 \pi f_a} F_{\mu \nu} \tilde F^{\mu \nu}
	\Bigg\}.
	\label{eq:setup_conf}
\end{align}
Note here that the index of co-moving objects, such as $F_{\mu\nu}$ and $\gamma^\mu$, is raised/lowered by the flat metric; $F^{\mu \nu}= \eta^{\mu\rho}\eta^{\nu \sigma} F_{\rho\sigma}$ and $\gamma_\mu = \eta_{\mu\nu} \gamma^{\nu}$.
In the rest of this paper, we usually raise/lower the index by $\eta^{\mu\nu}$. If we would like to use $g^{\mu\nu}$ instead, we will explicitly write down the metric as done in the kinetic term of the scalar field.

Let us define the electric and magnetic fields here.
The \textit{physical} electric and magnetic fields, $\hat 
{\bm E}, \hat{\bm B}$, are given by
\begin{align}
	\hat{\bm{E}} &= \frac{1}{a^2 (\eta)} \left( -
		\frac{\partial}{\partial \eta} \bm{A} - \nabla A_0
		\right) = \frac{\bm E}{a^2 (\eta)}, \label{eq:E}\\
	\hat{\bm{B}} & = \frac{1}{a^2 (\eta)} \nabla \times \bm{A}
	= \frac{\bm{B}}{a^2 (\eta)}. \label{eq:B}
\end{align}
Here we have also defined the \textit{comoving} electric and magnetic fields, $\bm{E}$ and $\bm{B}$. In terms of these fields the kinetic and Chern-Simons (CS) term read\footnote{Here and in the following, we use the dot product to denote the three-dimensional scalar product over the spatial indices.}
\begin{align}
	F_{\mu\nu}F^{\mu\nu} = 2 \left(  \bm{B}^2 - \bm{E}^2 \right) 
	= 2 a^4 (\eta) \left(   \hat{\bm{B}}^2 - \hat{\bm E}^2 \right), ~~~
	F_{\mu\nu} \tilde F^{\mu\nu} = - 4 \bm{E} \cdot \bm{B} 
	= - 4 a^4 (\eta) \hat{\bm E} \cdot \hat{\bm{B}}.
\end{align}
The energy density of this system can be conveniently expressed as
\begin{align}
	\rho = \frac{1}{2} \dot \phi^2 + V(\phi) 
	+ \frac{1}{a^4}\left[ \frac{\bm{E}^2 + \bm{B}^2}{2}
	+ \overline \psi \left(- i \bm{\nabla} \cdot \bm{\gamma} - g Q \bm{A} \cdot \bm{\gamma}  \right) \psi
	\right].
	\label{eq:energy}
\end{align}
Again, one can see that the fermion and gauge fields are conformal. 
For later convenience, we divide the energy density into three contributions and take the expectation value:
\begin{align}
\label{eq:energy_scalar}
	\rho_\phi &\equiv
	\frac{1}{\text{vol}\, (\mathbb{R}^3)} \int \dd^3 x\,
	\vev{\frac{1}{2} \dot \phi^2 + V (\phi) }, \\
\label{eq:energy_gauge}
	\rho_A &\equiv 
	\frac{1}{\text{vol}\, (\mathbb{R}^3)} \int \dd^3 x\,
	\frac{1}{2a^4} \vev{ \bm{E}^2 + \bm{B}^2 },\\
\label{eq:energy_fermion}
	\rho_\psi & \equiv 
	\frac{1}{\text{vol}\, (\mathbb{R}^3)} \int \dd^3 x\,\frac{1}{a^4}
	\vev{ \overline \psi \left(
		- i \bm{\nabla} \cdot \bm{\gamma} - g Q \bm{A} \cdot \bm{\gamma}
	\right) \psi }.
\end{align}
Here we have explicitly written down the spatial average,
\textit{i.e.}, $\int \dd^3 x/\text{vol}\,(\mathbb R^3)$.
Practically, one may omit it because we are mostly interested in a state with translational invariance, \textit{i.e.}, $[\hat \rho, \hat P] = 0$, with $\hat P$ denoting the spatial translation operator.
In this case, the one point function does not depend on the spatial coordinate, $\vev{O(x)} = O(t)$, and hence the spatial average becomes trivial.
In the following, we usually drop the spatial average for this reason, but
we sometimes recover it to avoid confusions and to make the physical meaning clear.

\subsection{Conservation equations}
\label{sec:consv}

To capture the dynamics of this system intuitively, it is instructive to see it from the viewpoint of conserved quantities.
Here we summarize the current equations that the system obeys.

Suppose that $V(\phi)$ does not depend on $\phi$. 
If this is the case, the scalar field enjoys the shift symmetry $\phi \mapsto \phi + \cnst$.
This observation implies a conserved quantity associated with the shift symmetry which is broken by $V'$. This is in particular useful for the description of slow-roll inflation, which is often characterized by an approximate shift symmetry. 
We can explicitly see this structure after reorganizing the equation of motion for the scalar field:
\begin{align}
	\partial_\mu \left( \sqrt{-g} J_\phi^\mu - \frac{1}{2} K_\text{CS}^\mu	 \right) = - \sqrt{- g} f_a V', \label{eq:shift_sym}
\end{align}
where
\begin{align}
	J_\phi^\mu &\equiv f_a g^{\mu\nu} \partial_\nu \phi, \\
	K_\text{CS}^\mu & \equiv \frac{\alpha}{\pi} \epsilon^{\mu\nu\rho\sigma} A_\nu \partial_\rho A_\sigma.
\end{align}

Let us move on to the fermions. Classically, for massless fermions, we have two independent symmetries, $\mathrm{U}(1)_\text{L} \times \mathrm{U}(1)_\text{R}$. 
However, their axial summation is modified in the presence of an anomaly while the vector one is kept intact, $\mathrm{U}(1)_\text{L} \times \mathrm{U}(1)_\text{R} \to \mathrm{U}(1)_\text{V}$: $\psi \mapsto e^{i \theta_V} \psi$.
Note here that the CS coupling with the inflaton, $\phi F_{\mu\nu} \tilde F^{\mu\nu}$, never changes the symmetry structure of our setup.
Thus one may derive the ABJ anomaly equation as done in Fujikawa's method~\cite{Fujikawa:1979ay,Fujikawa:1980eg}.
The equations of motion for the vector/axial currents are given by
\begin{align}
	0 &= \partial_\mu J_\psi^\mu, \label{eq:vectorcurrent}\\
	0 &= \partial_\mu \left( J_5^\mu + Q^2 K_\text{CS}^\mu \right),
	\label{eq:anomalous}
\end{align}
where
\begin{align}
\label{eq:current}
	J_\psi^\mu 
	&\equiv \overline \psi \gamma^\mu \psi \\
	&= \sqrt{- g}\, \hat{\overline \psi} \hat \gamma^\mu \hat\psi
	\equiv \sqrt{-g} \hat J_\psi^\mu,\\
	J_5^\mu 
	&\equiv \overline \psi \gamma^\mu \gamma_5 \psi \\
	&= \sqrt{-g}\, \hat{\overline \psi} \hat \gamma^\mu \gamma_5 \hat \psi
	\equiv \sqrt{-g} \hat J_5^\mu.
\end{align}
Here we have clarified the relation of these currents in terms of the original field before the rescaling.
In our notation (see App.~\ref{sec:nandc}), the vector/axial current is given by the right-handed current plus/minus the left-handed current:
$J_{\psi/5}^\mu = J_\text{R}^\mu \pm J_\text{L}^\mu$ with
$J_H \equiv \overline \psi \gamma^\mu {\cal P}_H \psi$ for $H = \text{R}, \text{L}$.
Throughout this paper, we write down the charge densities with respect to those quantities as
\begin{align}
	q_\bullet \equiv \frac{1}{\text{vol}\, (\mathbb{R}^3)} \int \dd^3 x\, \vev{ J_\bullet^0 }.
	\label{eq:charge}
\end{align}

\begin{figure}
	\centering
	\includegraphics[width=.50\textwidth]{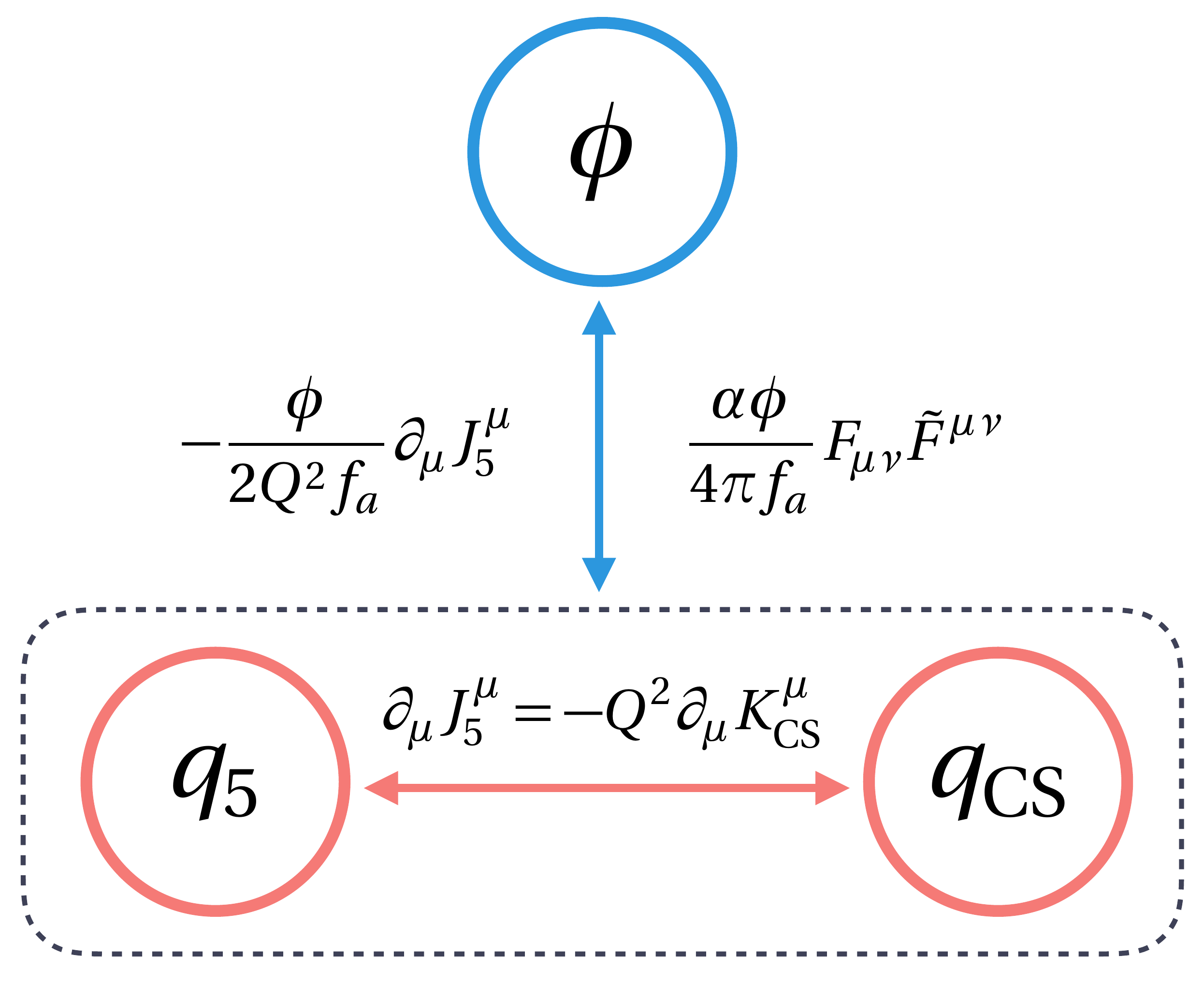}
	\caption{A schematic figure of our setup.
	The interaction with $\phi$ never breaks the symmetry of the fermion-gauge system. Rather, it generates a non-vanishing chemical potential for the system.}
	\label{fig:setup}
\end{figure}

Also, one can reorganize Eqs.~\eqref{eq:shift_sym} and \eqref{eq:anomalous} to obtain the following current equation:
\begin{align}
	\partial_\mu \left( \sqrt{-g} J_\phi^\mu + \frac{1}{2 Q^2} J_5^\mu	 \right) = - \sqrt{- g} f_a V'.
	\label{eq:shift_sym_2}
\end{align}
This fact is related to the redundancy of the description of the system.
By performing the chiral rotation, we can replace the CS term with a term proportional to $\phi \partial_\mu J_5^\mu$:
\begin{align}
		S = \int \dd^4 x\, \Bigg\{ \sqrt{-g}
	\left[
		\frac{g^{\mu\nu}}{2} \partial_\mu \phi \partial_\nu \phi - V (\phi)
	\right]
		- \frac{1}{4} F_{\mu \nu} F^{\mu\nu}
		+ \overline \psi \left( i \slashed{\partial} - g Q \slashed{A} \right) \psi
	- \frac{\phi}{2 Q^2 f_a} \partial_\mu J_5^\mu
	\Bigg\}.
	\label{eq:frame2}
\end{align}
In this frame, the shift symmetric charge is $q_\phi + q_5 / 2 Q^2$ which is consistent with Eq.~\eqref{eq:shift_sym_2}.
While the two theories [Eqs.~\eqref{eq:frame1} and \eqref{eq:frame2}] are inequivalent classically, the anomalous equation \eqref{eq:anomalous} makes them identical.
See Fig.~\ref{fig:setup} for illustration of our setup.
In a word, the interaction with $\phi$ never breaks the symmetry of the fermion-gauge system.
Rather, it pumps up the chemical potential of the system.

Finally, let us explicitly write down the energy conservation which clarifies how the energy is converted from the scalar field to the gauge/fermion field.
By using equations of motion, 
one can show that the energy densities for each component given in Eqs.~\eqref{eq:energy_scalar}, \eqref{eq:energy_gauge}, and \eqref{eq:energy_fermion} obey
\begin{align}
\label{eq:dot_rho_scalar}
	\dot \rho_\phi + 3 H \vev{\dot \phi^2} 
	&= - 
	\frac{\alpha \dot \phi}{\pi f_a} \frac{\vev{\bm{E} \cdot \bm{B}}}{a^4} , \\
\label{eq:dot_rho_gauge}
	\dot \rho_A + 4 H \rho_A
	&= +  
	\frac{\alpha \dot \phi}{\pi f_a} \frac{\vev{\bm{E} \cdot \bm{B}}}{a^4} 
	-  \frac{\vev{ \bm{E} \cdot (gQ \bm{J_\psi}) }}{a^5} 
	, \\
\label{eq:dot_rho_fermion}
	\dot \rho_\psi + 4 H \rho_\psi
	&= + 
	\frac{\vev{ \bm{E} \cdot (gQ \bm{J_\psi}) }}{a^5}.
\end{align}
One can easily see that the energy density is conserved up to the cosmic dilution by taking $a \to 1$ and $H \to 0$.
The CS term converts the energy of the scalar field into the gauge field,
while $\vev{\bm{E} \cdot \bm {J}_\psi }$ transfers this energy into the fermionic sector.
Of course, all the processes have to be consistent with the charge conservation laws depicted above.
Note that we must have $\dot \phi \vev{\bm{E} \cdot \bm{B}} > 0$ otherwise the system runs into contradiction.
This indicates the relation between the sign of $\dot \phi$ and $\dot q_\text{CS}$. See also discussion in the next subsection.

Thus, the rapid production of the helical gauge field should occur also in from the action~\eqref{eq:frame2}, though it may not be apparent at a first glance from the equation of motion for the gauge field.
In the Sec.~\ref{sec:vac}, we will see how these two descriptions give the same physics. 

\subsection{Overall picture via current equations}
\label{sec:overall}

Before going into details,
here we briefly summarize the dynamics of this system.
For intuitive understanding, the current equations derived in the previous section are useful.
In what follows, the initial state of our interest is a homogeneous background of the scalar field with a non-vanishing velocity: $\dot \phi \neq 0$, as arises \textit{e.g.}, if $\phi$ is identified with the inflaton.
Such a background spontaneously breaks $T$ or $CP$, and when communicated to the visible sector this leads to interesting phenomenology.

As an illustration, let us suppose that the scalar field respects $\phi \mapsto \phi + \cnst$, \textit{i.e.}, $V' = 0$,
and let us neglect the cosmic expansion.
Initially, the shift-symmetric charge stored in the scalar field is
$q_\phi = f_a \dot \phi$.
The question is how this charge will be distributed among the other charges.
If one changes the basis to Eq.~\eqref{eq:frame2}, the question becomes more evident.
Since such a background can be regarded as an external chemical potential $\mu_\text{eff}$ imposed on the fermion-gauge system,
\begin{equation}
 (\phi/f_a) \partial_\mu J^\mu_5 = - (\dot \phi/f_a) J^0_5 \equiv - \mu_\text{eff} J^0_5 \,,
\end{equation}
the system would like to increase/decrease $q_5$ for $\dot \phi \gtrless 0$
via $q_5$ breaking processes.
Eq.~\eqref{eq:anomalous} tells us that $q_5$ can be broken by the production of $q_\text{CS}$, but it costs finite energy because the Abelian gauge theory does not have a non-trivial vacuum structure contrary to non-Abelian gauge theories. 
Interestingly, as we will see in Sec.~\ref{sec:vac}, such a background of $\dot \phi \neq 0$ can create helical gauge fields yielding non-zero $q_\text{CS}$.
Through this process, the system tries to approach non-vanishing $q_5$ ($q_\text{CS}$) corresponding to the chemical potential from $\dot \phi \neq 0$.

Let us perform a consistency check.
The production of non-zero $q_\text{CS}$ costs finite energy,
and hence $| \dot \phi |$ must decrease according to energy conservation.
This means $\dot q_\phi \lessgtr 0$ for $\dot \phi \gtrless 0$.
Then, the current equation for the shift-symmetric charge, Eq.~\eqref{eq:shift_sym}, yields
$\dot q_\text{CS} = 2 \dot q_\phi \lessgtr 0$ for $\dot \phi \gtrless 0$.
This, in turn, implies the production of the chiral charge because of the anomalous equation \eqref{eq:anomalous}:
$\dot q_5 = - q^2\dot q_\text{CS} \gtrless 0$ for $\dot \phi \gtrless 0$.
We can see that the direction of the dynamics is consistent with the chemical potential which can be read off from Eq.~\eqref{eq:frame2}.
Putting it the other way around, 
one can understand which helicity of the gauge field is amplified from $\dot \phi \neq 0$ without explicitly solving the equation of motion for the gauge field.
See also Eqs.~\eqref{eq:wave_eq} and \eqref{eq:def_xi}; and discussion below. 
Moreover, in the background of the strong helical gauge field,
we may have another fermion production processes which create the right- and left-handed fermions in a symmetric way, \textit{i.e.}, $0 = \dot q_5 = \dot q_\text{R}- \dot q_\text{L}$.
This process must also be consistent with Eq.~\eqref{eq:vectorcurrent}, $0 =  \dot q_\psi =  \dot q_\text{R} + \dot q_\text{L}$.
Hence we expect $\dot q_\text{R} = \dot q_\text{L} = 0$ for a symmetric production, $\dot q_5 = 0$.
This observation implies that such a process, $\dot q_5 = 0$, must be a pair-production of particle and anti-particle, analogous to the Schwinger effect without a magnetic field. 
We will confirm these properties by explicit computations in the rest of this paper.

\section{Production of helical gauge fields and chiral fermions}
\label{sec:prod}

In this section, we investigate helical gauge field and chiral fermion production induced by the background of $\dot \phi \neq 0$.
In particular, we emphasize that these processes are inevitably related through the anomaly equation~\eqref{eq:anomalous}.
In addition, besides the process related to Eq.~\eqref{eq:anomalous} which yields the net $q_5$ charge in the plasma, we will find another fermion production that does not give non-zero $q_5$. This is similar to the Schwinger process in a strong electric field.
We will clarify the relations between these two channels of the fermion production.

\subsection{Helical gauge field production without backreaction}
\label{sec:vac}

Let us start with the helical gauge field production by assuming that the backreaction from the fermion production can be safely neglected. This part of the discussion closely resembles the nonperturbative helical gauge field production in axion inflation~\cite{Turner:1987bw,Garretson:1992vt,Anber:2006xt}.
The validity of this approximation is discussed later in Sec.~\ref{sec:br}.
Throughout Sec.~\ref{sec:prod}, we assume that the change of $\dot \phi$ is much slower than all the production processes we will discuss.
For instance, this is the case for a slow-rolling $\phi$ during inflation.

Taking a variation with respect to $A$ in Eq.~\eqref{eq:setup_conf},
one gets
\begin{align}
	0 = \partial_\mu \left( F^{\mu\nu} - \frac{\alpha \phi}{\pi  f_a} \tilde F^{\mu\nu} \right) - g Q J_\psi^\nu + \frac{1}{\rho} \partial^\nu \left( \partial_\mu A^\mu \right),
	\label{eq:eom_gauge}
\end{align}
where the last term is a gauge fixing term with $\rho$ denoting a gauge fixing parameter.
We take the Feynman gauge, $\rho = 1$, in the following.

Once the fermion is generated by the gauge field,
the gauge field drives the motion of fermion.
As a result, the current $J_\psi$ is induced.
This induced current in turn affects the equation of motion for the gauge field.
We can also see this from the equations for the energy densities
[Eqs.~\eqref{eq:dot_rho_gauge} and \eqref{eq:dot_rho_fermion}].
Now it is clear that this term is responsible for the backreaction from the fermion production.
We leave the discussion of its effect in Sec.~\ref{sec:br}, focusing on the weak backreaction regime in this section.
Then, one may simplify the Eq.~\eqref{eq:eom_gauge} as
\begin{align}
	0 = \left( \Box \eta^{\mu\nu} - a \frac{\alpha \dot \phi}{\pi f_a} \epsilon^{0\mu\sigma \nu} \partial_\sigma \right) A_\nu.
	\label{eq:helical_vac}
\end{align}

The quantization of $A$ can be performed in the usual way.
The mode expansion of $A$ is given by
\begin{align}
	A_\mu (x) = \int \frac{\dd^3 k}{(2 \pi)^{3/2}} 
	\sum_{\sigma = \pm, L, S} \left[
		\epsilon^{(\sigma)}_\mu (\bm{k}) a^{(\sigma)}_{\bm{k}} A_\sigma (\eta, \bm{k}) e^{i \bm{k}\cdot \bm{x}}
		+ \text{H.c.}
	\right],
	\label{eq:Adecom}
\end{align}
where the polarization tensor fulfills the following relations: $(\epsilon^\mu{}^{(\pm)})= (0, \bm{\epsilon}^{(\pm)})^t$,
$\bm{k} \cdot \bm{\epsilon}^{(\pm)} (\bm{k}) = 0$,
$\bm{\epsilon}^{(\pm)} (\bm{k})^\ast \cdot \bm{\epsilon}^{(\pm)} (\bm{k}) = 1$, 
$\bm{k} \times \bm{\epsilon}^{(\pm)} = \mp i k \bm{\epsilon}^{(\pm)}(\bm{k})$;
$\epsilon_\mu^{(L)} = - i k_\mu$, and $(\epsilon^{(S)}_\mu ) = i (k , - \bm{k})^t/2k^2$.
As can be seen from Eq.~\eqref{eq:helical_vac}, 
the background of $\dot \phi$ has no effects on modes with $\sigma =L , S$. 
Thus, we may focus on transverse modes $\sigma = \pm$ to study the gauge production from $\dot \phi \neq 0$ as expected.
The resulting wave equation is obtained from Eq.~\eqref{eq:helical_vac}:
\begin{align}
	0 = \left[ \partial_\eta^2 + k \left( k \pm 2 \lambda \xi a H \right) \right] A_\pm (\eta, \bm{k}) \,,
	\label{eq:wave_eq}
\end{align}
where $\lambda = \pm$ for $\dot \phi \gtrless 0$ encodes the sign of $\dot \phi$ and
\begin{align}
	\xi \equiv \frac{\alpha { \lambda \dot \phi}}{2 \pi f_a H} > 0 \,.
	\label{eq:def_xi}
\end{align}
We take the normalization of Wronskian as $A_\pm \partial_\eta A_\pm^\ast - ( \partial_\eta A_\pm) A_\pm^\ast = i$, which implies the following normalization for the commutators, $[a^{(\pm)}_{\bm{k}}, a^{(\pm)}_{\bm{q}}{}^\dag] = \delta (\bm{k} - \bm{q})$.

\paragraph{Flat spacetime.}
Let us first look at how the wave function behaves without the cosmic expansion.
We can recover the flat spacetime result by taking $a \to 1$ and $H \to 0$.
It is clear that one of the polarizations exhibits an exponential growth in its infrared mode as can be seen from in Eq.~\eqref{eq:wave_eq}:
$A_\pm \propto e^{ \omega_k t}$ for $\dot \phi \lessgtr 0$ with $\omega_k = \sqrt{(\alpha | \dot \phi | / \pi f_a  - k)k}$.
Plugging this solution into the CS term, we find
$\dot q_\text{CS} \lessgtr 0$ for $\dot \phi \gtrless 0$.

It is instructive to see what happens from the viewpoint of Eqs.~\eqref{eq:shift_sym}, \eqref{eq:anomalous}, and also \eqref{eq:shift_sym_2}.
Lets assume for a moment a flat potential $V' = 0$ and take the initial condition as $\dot \phi \neq 0$.
The production of the helical gauge field indicates that $|\dot \phi|$ decreases due to energy conservation; namely
$\dot q_\phi \lessgtr 0$ for $\dot \phi \gtrless 0$. Then 
Eq.~\eqref{eq:shift_sym} tells us that $\dot q_\text{CS} = 2 \dot q_\phi \lessgtr 0$ for $\dot \phi \gtrless 0$, which is consistent with the above result obtained from the direct computation.
The anomaly equation \eqref{eq:anomalous} determines the growth of the chiral charge: $\dot q_5 = - \dot q_\text{CS} \gtrless 0$ for $\dot \phi \gtrless 0$.
In Sec.~\ref{sec:fermion}, we will also confirm that this result is consistent with the direct computation of the equation of motion for the fermion.

\paragraph{De Sitter.}
Next, we move on to a de Sitter Universe.
Contrary to flat spacetime, the exponential expansion of the Universe dilutes away the gauge field and red-shifts its momentum. In addition, the scalar field now experiences Hubble friction. 
Hence, to have an almost constant $\dot \phi$, $V'$ is needed.
The background of $| \dot \phi | \simeq \cnst$ is more violent than that of  flat spacetime because the instability scale, $2 \xi a H$, grows exponentially. 
In other words, the gauge field production in de Sitter is twofold; due to
the tachyonic instability and the broken conformal invariance, induced by
the background of $|\dot \phi| \simeq \cnst $.\footnote{
	If the scalar field has a conformal non-minimal coupling $\xi_R = 1/6$ to the Ricci scalar and respects the shift symmetry, one may have $|\partial_\eta \phi | \simeq \cnst$ initially. 
	This limit is exactly parallel to the case in the flat spacetime because we can completely factor out the scale factor.
	To be more specific, one can directly observe it from $\xi_R a H \propto \partial_\eta \phi$.
}
As a result, the gauge field becomes superhorizon and approaches a constant amplitude.
Requiring that the wave function becomes a plane wave, $A_\pm \propto e^{- i \omega_k \eta}$, for deep inside the horizon, $- k \eta \gg 1$, we 
get the following analytical solution for the growing mode: 
\begin{align}
	A_{-\lambda} (\eta , \bm{k}) = \frac{e^{\pi \xi / 2 }}{\sqrt{2 k}}
	W_{- i \lambda \xi,1/2} (2 i k \eta),
	\label{eq:gauge_desitter}
\end{align}
where $W_{\kappa,\mu} (x)$ is the  Whittaker function.
One can explicitly check that this solution approaches a constant value for superhorizon modes, $ - k \eta \ll 1$:
\begin{align}
	\lim_{k \eta \to - 0} A_{-\lambda} (\eta ,\bm{k}) = 
	\frac{1}{\sqrt{2k}} \frac{e^{\pi \xi / 2 }}{\Gamma(1+ i \lambda \xi)}.
	\label{eq:gauge_sh}
\end{align}

By plugging the growing mode into $F \tilde F$, one may estimate $\partial_\eta q_\text{CS}$ in de Sitter Universe:
\begin{align}
	\partial_\eta q_\text{CS}
	&= \frac{1}{\text{vol}\,(\mathbb{R}^3)} 
	\int \dd^3 x\, \frac{\alpha}{2 \pi}\vev{F_{\mu\nu} \tilde F^{\mu\nu}} \\
	&= - a^4 (\eta) \frac{2 \alpha}{\pi} 
	\vev{\hat{\bm{E}} \cdot \hat{\bm{B}}},
	\label{eq:QCS}
\end{align}
where
\begin{align}
	\vev{\hat{\bm{E}} \cdot \hat{\bm{B}}}
	&= 
	\lambda \frac{e^{2 \pi \xi}}{\xi^4} H^4
	\times \left( - \frac{\xi^4}{8 \pi^2} e^{- \pi \xi}
	\int^{\kappa_\text{UV}}_0 \dd \kappa\, \kappa^3 \frac{\partial}{\partial \kappa} \abs{W_{-i \lambda \xi,1/2} (-2i \kappa)}^2 \right) \nonumber \\
	&\simeq 2.6 \times 10^{-4} \lambda \frac{e^{2 \pi \xi}}{\xi^4} H^4.
	\label{eq:EB0}
\end{align}
with $\kappa = - k \eta = k/(a H)$. 
Recall that $\lambda = \pm$ for $\dot \phi \gtrless 0$.
Here we have taken $\kappa_\text{UV} = 2 \xi$, which is a natural value of the UV cutoff since we do not have instabilities above this value. For $\xi \gtrsim 3$, the results are insensitive to the choice of this UV cut-off~\cite{Jimenez:2017cdr}. For these values of $\xi$, the last term in the parenthesis takes a constant, $\xi$-independent value, as given in the last line of Eq.~\eqref{eq:EB0}. 
Also, one can check that this integral is dominated by $\kappa \lesssim 1$.
The produced gauge fields have a constant amplitude with a super-horizon correlation length.
The fourth power of the scale factor arising in Eq.~\eqref{eq:QCS} means that the charge stored in a \textit{comoving} volume must grow to have such a constant \textit{physical} electromagnetic field.

We can estimate the growth of the chiral charge compared to some initial time $t_0$ by using Eq.~\eqref{eq:anomalous}:
\begin{align}
	&\partial_t (a^3 \hat q_5) = \dot q_5 = \frac{1}{a} \partial_\eta q_5 
	\simeq \lambda Q^2 a^3 \left( 2.6 \times 10^{-4} \frac{2\alpha}{\pi} \frac{e^{2 \pi \xi}}{\xi^{4}} H^4 \right),\\
	\rightarrow&\, \Delta \hat q_5(t) \sim \lambda Q^2
	\left(
		10^{-4} \frac{2\alpha}{\pi} \frac{e^{2 \pi \xi}}{\xi^4} H^3
	\right) \left( 1 - e^{- 3 H (t-t_0)} \right).
	\label{eq:Q5_dS}
\end{align}
On time-scales larger than the Hubble expansion rate, one may drop $e^{- 3 H(t - t_0)}$. 
Note that the physical charge, $\hat q$, is related to its comoving charge, $q$, via $a^3 \hat q = q$.
Hence the net chiral charge in the Universe becomes non-zero and constant over the superhorizon scale,
if a helical gauge field is generated under which the fermion is charged (and if there is a chiral anomaly).
We will reproduce this result directly from the equation of motion for the fermion in Sec.~\ref{sec:fermion}.

\begin{figure}
	\centering
	\includegraphics[width=.30\textwidth]{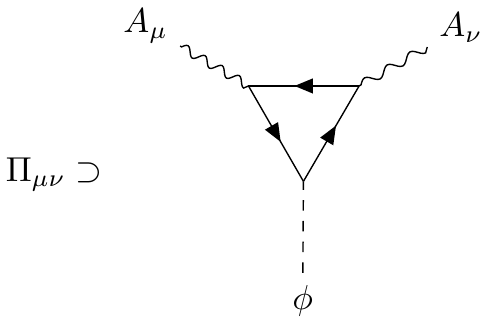}
	\caption{The diagram leading to Eq.~\eqref{eq:equivalence}.}
	\label{fig:triangle}
\end{figure}

\paragraph{~~}
 Before closing this section, we would like to clarify the equivalence of Eqs.~\eqref{eq:setup_conf} and \eqref{eq:frame2} in terms of the equation of motion for the gauge field.
The equation of motion for the gauge field in Eq.~\eqref{eq:frame2} is
\begin{align}
	0 = \partial_\mu F^{\mu\nu} - g Q J_\psi^\nu + \frac{1}{\rho} \partial^\nu (\partial_\mu A^\mu).
\end{align}
One may express the current by
\begin{align}
	gQJ_\psi^\nu (x) = i \int_z \Pi^{\nu\mu}_\text{Ret} (x,z) A_\mu (z),
\end{align}
at leading order in the coupling expansion, where $\Pi^{\nu\mu}_\text{Red}$ denotes the retarded fermion propagator.
At first glance, this seems to be independent of the background with $\dot \phi \neq 0$, but this is not the case once you include the quantum effects on $J^\nu_\psi$: namely a fermion loop in the background of $\dot \phi \neq 0$ (see Fig.~\ref{fig:triangle}).

The equation of motion for the fermion in Eq.~\eqref{eq:frame2} is given by
\begin{align}
	0 = \left( i \slashed{\partial} - g Q \slashed{A} + \frac{a \dot \phi}{2 Q^2 f_a} \gamma^0 \gamma_5 \right) \psi.
\end{align}
From this, we can obtain the fermion propagator in the presence of $\dot \phi \neq 0$.
By plugging the propagator into the self energy, one can estimate the impact of $\dot \phi$.
In the following, we will be interested in the UV part of the loop integral, and hence one may regard $a \dot \phi$ as essentially constant.
Also, we assume that the phase space density of the fermion gets suppressed for a sufficiently large momentum.
Hence, the self energy with a large loop momentum may be regarded as the vacuum state for these particles.
After some computation, we arrive at
\begin{align}
	i\Pi^{\mu\nu}_\text{Ret} (P) \supset \frac{\alpha a \dot \phi}{\pi f_a} \epsilon^{0\mu\sigma\nu} (-i P_\sigma)
	~~~\rightarrow~~~
	\partial_\nu \left( \frac{\alpha \phi}{\pi f_a} \tilde F^{\nu\mu} \right),
	\label{eq:equivalence}
\end{align}
which reproduces the equation of motion for $A_\mu$ as obtained from Eq.~\eqref{eq:frame1}, see Eq.~\eqref{eq:helical_vac} or Eq.~\eqref{eq:eom_gauge}. Here $P$ denotes the external momentum in the Fourier-transformed propagator.
The connection more evident in the language of Feynman diagrams.
The relevant diagram is nothing but the one which leads to the triangle anomaly as can be seen from Fig.~\ref{fig:triangle}.
To sum up, the two theories, Eqs.~\eqref{eq:setup_conf} and \eqref{eq:frame2}, are independent classically,
and hence we have to look at the loop contributions to see the equivalence explicitly from the equations of motion.

\subsection{Chiral fermion production from the helical gauge field}
\label{sec:fermion}

Now we are in a position to discuss the fermion production in the presence of the helical gauge field.
The aim of this section is twofold.
On the one hand, the production of fermions is expected from the anomalous current equation given in Eq.~\eqref{eq:anomalous}.
On the other hand, we expect another production channel in the presence of a strong electric field, namely the Schwinger effect.
We would like to clarify the relation among them 
and also reproduce the result inferred by Eq.~\eqref{eq:anomalous} directly from the equation of motion for the fermions.

The rigorous way to study this fermion production may be to track the real time evolution of all the correlators, such as gauge bosons and fermions, simultaneously in the presence of the slowly rolling $\phi$, starting for instance from first principles like the closed-time-path formalism~\cite{Kasper:2014uaa,Fukushima:2015tza,Mueller:2016aao}. This treatment is however beyond the scope of this paper.
Instead, we would like to approximate the situation. This allows us to investigate the process intuitively and analytically.

In flat spacetime, the approximation we will employ is easy to understand:
stop the gauge field production at a given time $t$, 
take one patch within a correlation length of the generated gauge field,
and study the fermion production inside each patch.
If the fermion production is much faster than the growth rate of the gauge field, this approximation is justified a posteriori.

In de Sitter spacetime, the situation is more involved due to the cosmic expansion.
As indicated by the discussion in the previous section,
the electric and magnetic fields are correlated 
over superhorizon scales.
The averaged values of $\hat E^2$, $\hat B^2$, and $\hat E \hat B$ can be expressed as~\cite{Jimenez:2017cdr}
\begin{align}
	\vev{\hat{\bm{E}}^2}
	&=
	\frac{e^{2 \pi \xi}}{\abs{\xi}^3} H^4 \times 
		\left( \frac{\abs{\xi}^3}{4 \pi^2} e^{- \pi \xi}
			\int^{\kappa_\text{UV}}_0 \dd \kappa\, \kappa^3
			\abs{\frac{\partial}{\partial \kappa}W_{- i \lambda \xi,1/2} (-2i \kappa)}^2 \right) \nonumber \\
	& \simeq 2.6 \times 10^{-4} \,
	\frac{e^{2 \pi \xi}}{\abs{\xi}^3} H^4,
	\label{eq:E2}
\end{align}
\begin{align}
	\vev{\hat{\bm{B}}^2}
	&=
	\frac{e^{2 \pi \xi}}{\abs{\xi}^5} H^4 \times 
	\left(
		\frac{\abs{\xi}^5}{4 \pi^2} e^{- \pi \xi}
			\int^{\kappa_\text{UV}}_0 \dd \kappa\, \kappa^3 \abs{W_{- i \lambda \xi,1/2} (-2i \kappa)}^2
	\right) \nonumber \\
	& \simeq
	3.0 \times 10^{-4} \,
	\frac{e^{2 \pi \xi}}{\abs{\xi}^5} H^4,
	\label{eq:B2}
\end{align}
\begin{align}
	\vev{\hat{\bm{E}}\cdot \hat{\bm{B}}}
	&=
	\lambda \frac{e^{2 \pi \xi}}{\xi^4} H^4 \times
	\left( -
		\frac{\xi^4}{8 \pi^2}	e^{- \pi \xi}	\int^{\kappa_\text{UV}}_0 \dd \kappa\, \kappa^3 \frac{\partial}{\partial \kappa} \abs{W_{- \lambda i \xi,1/2} (-2i \kappa)}^2
	\right) \nonumber \\
	& \simeq 
	2.6 \times 10^{-4} \, \lambda
	\frac{e^{2 \pi \xi}}{\xi^4} H^4,
	\label{eq:EB}
\end{align}
where $\lambda = \pm$ for $\dot \phi \gtrless 0$. Moreover, inserting the helical structure of Eq.~\eqref{eq:Adecom} into Eqs.~\eqref{eq:E} and~\eqref{eq:B}, we see that the electric and magnetic fields are parallel.
All the integrals above are dominated by $\kappa \equiv k / (aH) \lesssim 1$,
namely by superhorizon modes.
Let us take one Hubble patch at a time $t$.
We approximate the electric/magnetic field as a uniform field in this patch;
$E \sim 10^{-2} H^2 e^{\pi \xi} / \xi^{3/2}$ and $B \sim 10^{-2} \lambda H^2 e^{\pi \xi} / \xi^{5/2}$.\footnote{Note that this is only possible locally, but not globally since averaged over all space $\langle E \rangle = 0$. For this reason it will not be possible to construct a vector potential for constant, parallel $E$ and $B$-fields in de Sitter spacetime. This problem is circumvented by considering a sufficiently small patch and fast processes so that expansion of the Universe can be neglected.
To clarify the situation, suppose that the production stops at $t$.
Then, the gauge field decays proportional to $\hat E,\hat B \propto a^{-2}$.
This observation indicates that the background of $\dot \phi \neq 0$ compensates the decay by the exponential production.
Putting it the other way around,
we may neglect the cosmic expansion (and the gauge field production through the tachyonic instability) if the process of our interest is much faster than it.} 
In the rest of this section, we assume that the production of fermions is much faster than the cosmic expansion and discuss the validity of this assumption a posteriori. This enables us to perform a first estimate of the effects of the fermion production. We hope our results will motivate more refined studies in the future.

With this, we now turn to the production of fermions in the presence of constant, parallel electric and magnetic fields. Parts of this discussion closely follow Refs.~\cite{Nielsen:1983rb,Bavarsad:2017oyv}. 
The equation of motion for the fermion can be obtained from a variation with respect to $\overline \psi$ in Eq.~\eqref{eq:setup_conf}.
For later convenience, we consider the left-/right-handed fermions separately:
\begin{align}
	0 = \left( i \partial_\eta  \pm i \bm{\nabla} \cdot \bm{\sigma}- g Q A_0 \pm gQ \bm{A} \cdot \bm{\sigma}\right) \psi_\text{R/L} \, .
\end{align}
Let us define an auxiliary field
\begin{align}
	\psi_\text{R/L} \equiv 
	\left( i \partial_\eta  \mp i \bm{\nabla} \cdot \bm{\sigma}- g Q A_0 \mp gQ \bm{A} \cdot \bm{\sigma}\right) \Phi_\text{R/L} \, .
	\label{eq:aux}
\end{align}
Then, one can reorganize the equation of motion as
\begin{align}
	0 = \left[
		\left( - \, \Box - 2 i g Q A \cdot \partial + g^2 Q^2 A^2 \right)
		+ gQ a^2 \left( \hat{\bm{B}} \pm i \hat{\bm{E}} \right) \cdot \bm{\sigma}
	\right] \Phi_\text{R/L} \,.
\end{align}

Under the aforementioned approximation,
we may consider a uniform background of electric and magnetic fields
pointing the same (opposite) direction for $\dot \phi > 0$ ($\dot \phi<0$).
We also neglect the cosmic expansion in the following, \textit{i.e.}, $a \to 1$. Note that this also implies $\eta \mapsto t$. We adopt the following vector potential, without loss of generality pointing along the $z$-axis, 
$(A_\mu) = (0, 0, - \lambda Bx, E t)$ with $\lambda = \pm$ for $\dot \phi \gtrless 0$.
For later convenience, we perform the Fourier transform with respect to the spatial coordinates $y$ and $z$:
\begin{align}
	\Phi_\text{R/L} (t,\bm{x})
	= \int \frac{\dd p_y \dd p_z}{2 \pi}\,
		e^{i ( p_y y + p_z z )} \Phi_\text{R/L} (t,x; p_y, p_z) \,.
\end{align}
Then, one may rewrite the equation of motion as follows:
\begin{align}
	0 = \left[
		- \frac{\partial^2}{\partial t^2} + \frac{\partial^2}{\partial x^2}
		- \left( g Q \lambda B x - p_y \right)^2
		 - \left( g Q E t + p_z \right)^2 
		 + g Q \left( \lambda B \pm i E \right) \sigma_z
	\right] \Phi_\text{R/L} \, .
\end{align}
Noticing that the $x$-dependent part of this equation is nothing but the Harmonic oscillator,
one can solve it by the separation of variables.
Let the auxiliary function be $\Phi_\text{R/L} = h_n (x_{-s}) \,\, g_\text{R/L} (t;n,p_z) \,\, \chi_s$ with $x_{-s} \equiv \sqrt{g |Q| B} x - s p_y / \sqrt{g |Q| B}$, $\sigma_z \chi_\pm = \pm \chi_\pm$, and $s = \pm$ for $Q\lambda \gtrless 0$.
One can separate the equation into two parts:
\begin{align}
	\left[ \frac{\partial^2}{\partial x^2} - g^2 Q^2 B^2 \left( x - \frac{p_y}{g Q \lambda B} \right)^2 \right] h_n &= - \left( 2 n + 1 \right) g \abs{Q} B h_n \,, 	\label{eq:eom_l} \\
	\left[
		\frac{\partial^2}{\partial t^2} + \left( p_z + g Q E t \right)^2
		- g | Q | (B \pm i \lambda E) 
	\right] g_\text{R/L} &= - \left( 2 n + 1 \right) g \abs{Q} B g_\text{R/L} \,,
	\label{eq:eom_s}
\end{align}
where $h_n$ is expressed by the Hermite function $H_n$:
\begin{align}
	h_n (x_{-s}) \equiv \left( \frac{1}{2^n n ! } \right)^{1/2} \left( \frac{g |Q| B}{\pi} \right)^{1/4} e^{- x_{-s}^2 / 2} H_n (x_{-s}) \,.
\end{align}
It is straightforward to show that $h_n$ spans a complete and orthogonal set, namely 
$\int \dd x \, h_n (x_{-s}) h_{\bar n} (x_{-s}) = \delta_{n, \bar n}$
and $\sum_n h_n (x_{-s}) h_n (\bar x_{-s}) = \delta (x - \bar x)$.

Let us count degrees of freedom before proceeding further.
A Weyl fermion has two degrees of freedom.
Nevertheless, the solution of this form, $\Phi_\text{R/L} = h_n g_\text{R/L}\chi_\pm$, apparently has $4$ degrees of freedom: positive/negative energy solution of $g_\text{R/L}$ and two spin wave functions $\chi_\pm$.
This observation implies its redundancy.
In fact, one may take one of two spin wave functions, $\sigma_z \chi = \mp \chi$ for $Q \lambda \lessgtr 0$, without loss of generality.
We have already taken this into account in Eq.~\eqref{eq:eom_s}.

\paragraph{Landau levels.}

\begin{figure}
	\centering
	\subfigure{
	\includegraphics[width=.48\textwidth]{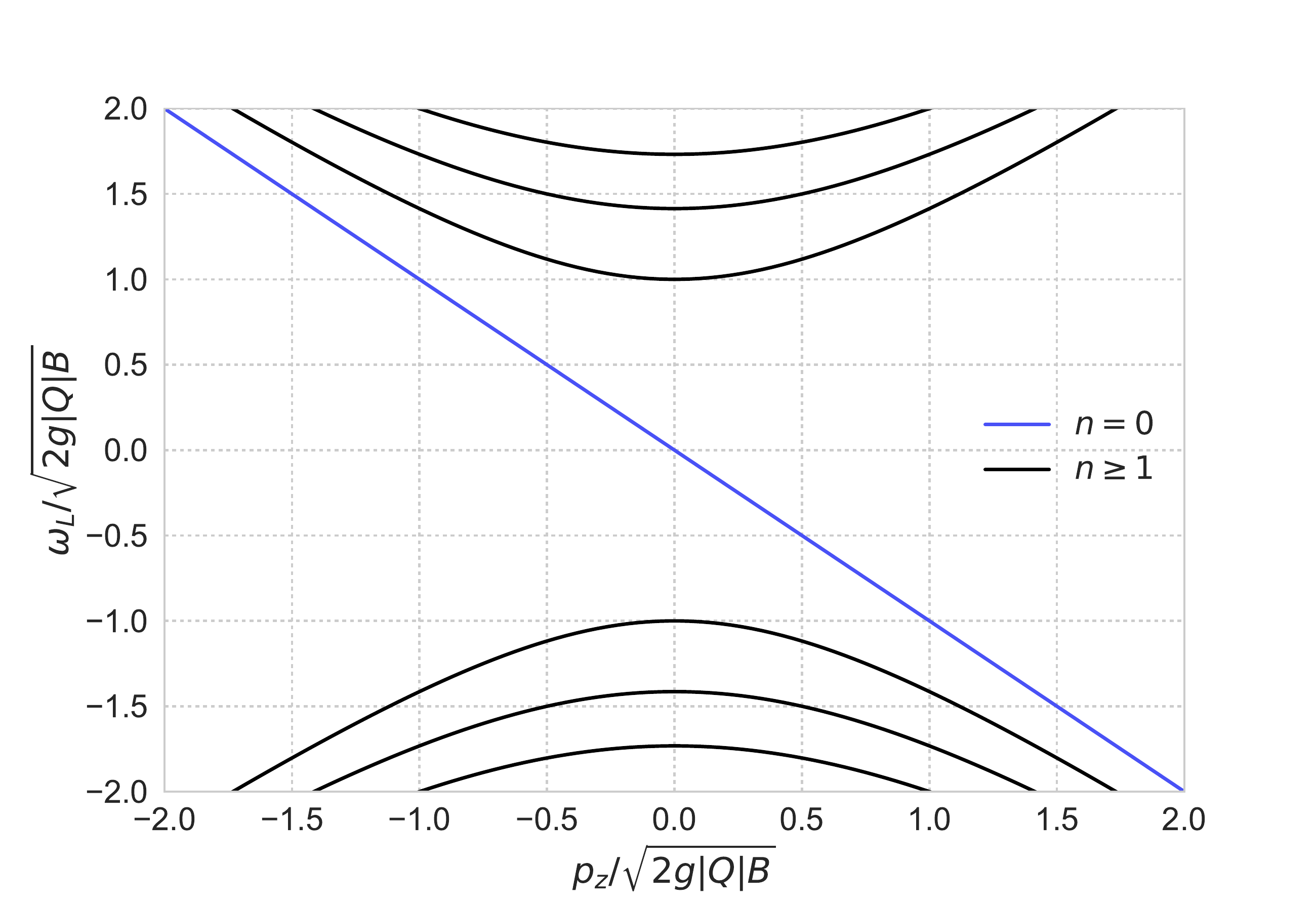}} \hfill
	\subfigure{
	\includegraphics[width=.48\textwidth]{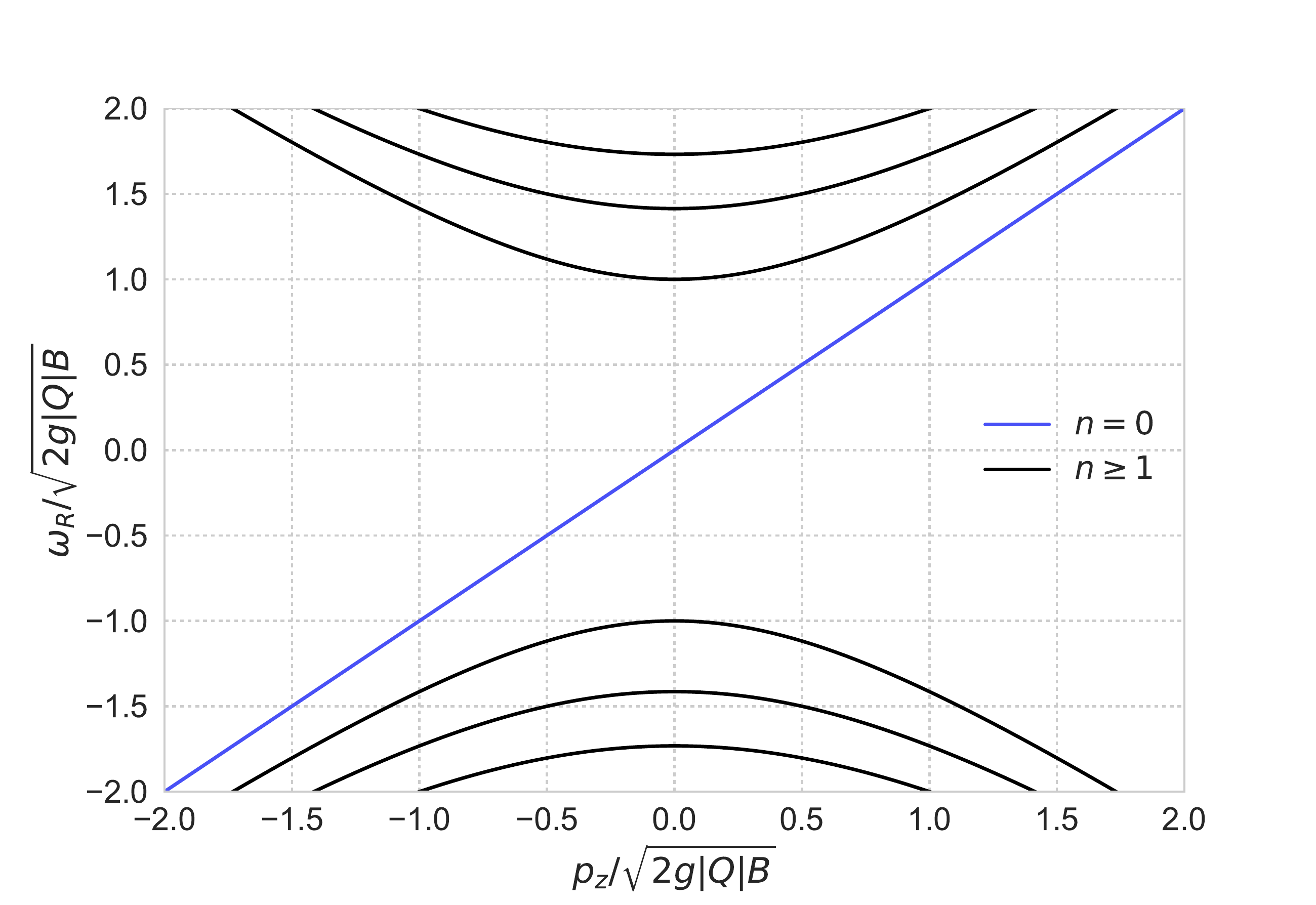}}
	\caption{The Landau levels for left-/right-handed fermions for $s = +$, \textit{i.e.,} $\dot \phi Q > 0$, are shown in the left/right figure. The lowest Landau level is depicted by the blue line while the higher ones are drawn as black lines. One can readily see that the higher ones are symmetric but the lowest one is asymmetric with respect to the interchange of left- and right-handed fermions.}
	\label{fig:LL}
\end{figure}

First, we study the spectrum of Eq.~\eqref{eq:eom_s} when we turn off the electric field.
This consideration is useful for understanding the relation of two fermion production channels via the anomalous equation and via the Schwinger-like effect.
Eq.~\eqref{eq:eom_s} then becomes
\begin{align}
	0 = \left( 
		\frac{\partial^2}{\partial t^2} + p_z^2 + 2 n g \abs{Q} B 
	\right) g_\text{R/L} \,.
	\label{eq:eom_s_E0}
\end{align}

Let us focus on the right-handed fermion.
Its dispersion relation is obtained from Eq.~\eqref{eq:eom_s_E0}:
\begin{align}
	\omega_\text{R} = 
	\begin{cases}
		\pm \sqrt{p_z^2 + 2n g \abs{Q} B} &\text{for}~~n = 1,2,\dots \,, \\
		s p_z &\text{for}~~n=0 \,,
	\end{cases}
	\label{eq:ll_R}
\end{align}
where $s = \pm$ for $Q\lambda \gtrless 0$.
One can see that the energy spectrum is discretized, which is known as \textit{Landau levels}.
Intuitively, this is because the uniform magnetic field restricts the transverse motion of a charged particle by the Lorentz force.
Note here that, for given $Q$ and $\lambda$, the lowest Landau level (LLL) with $n = 0$ has a unique dispersion relation while the higher Landau levels (HLLs) with $n \geq 1$ have positive/negative frequencies.
To understand the reason, let us move back to the definition of the auxiliary field, Eq.~\eqref{eq:aux}.
Evidently, Eq.~\eqref{eq:eom_s_E0} allows two independent solutions $g_\text{R} = N e^{\mp i p_z t}$ for $n=0$ with $N$ being a normalization factor.
However, one of them yields $\psi_\text{R} = 0$ if we insert the solution into Eq.~\eqref{eq:aux}:
\begin{align}
	\psi_\text{R} 
	&= \left( \pm p_z
	- i \frac{\partial}{\partial x} \sigma_x + p_y \sigma_y + p_z \sigma_z 
	- s g \abs{Q \lambda} B x \sigma_y
	\right) N h_0 (x_{-s}) e^{ \mp i p_z t} \chi_s \nonumber \\
	&= \left( \pm p_z + s p_z \right) N h_0 (x_{-s}) e^{ \mp i p_z t} \chi_s \,,
\end{align}
where $s = \pm$ for $Q\lambda \gtrless 0$.
Now it is clear that we need $\omega_\text{R} = \pm p_z$ for $s = \pm$ to have non-vanishing $\psi_\text{R}$.
More intuitively, since the LLL can be regarded as a fermion moving along with the magnetic field ($z$-direction),
the right-handed fermion must have a spin, $\chi_s$, parallel to its motion.

Similarly, we get the dispersion relation for the left-handed fermion as follows:
\begin{align}
	\omega_\text{L} = 
	\begin{cases}
		\pm \sqrt{p_z^2 + 2n g \abs{Q} B} &\text{for}~~n = 1,2,\dots \,, \\
		- s p_z &\text{for}~~n=0 \,,
		\label{eq:ll_L}
	\end{cases}
\end{align}
where $s = \pm$ for $Q\lambda \gtrless 0$.
Note that the HLLs have exactly the same structure for left- and right-handed fermions,  while the LLL has opposite sign.
Intuitively, this is because the left-handed fermion has a spin anti-parallel to its motion contrary to the right-handed fermion.
We can also show the sign of the LLL explicitly by plugging the solution into Eq.~\eqref{eq:aux} as done for the right-handed fermion:
\begin{align}
	\psi_\text{L} 
	&= \left( \pm p_z
	+ i \frac{\partial}{\partial x} \sigma_x - p_y \sigma_y - p_z \sigma_z 
	+ s g \abs{Q \lambda} B x \sigma_y
	\right) N h_0 (x_{-s}) e^{ \mp i p_z t} \chi_s \nonumber \\
	&= \left( \pm p_z - s p_z \right) N h_0 (x_{-s}) e^{ \mp i p_z t} \chi_s,
\end{align}
which implies $\omega_\text{L} = \mp p_z$ for $s = \pm$.

To sum up, in the presence of a uniform magnetic field,
the dispersion relation is discretized to form the Landau levels.
For the HLLs, the right- and left-handed fermions have exactly the same spectrum.
Meanwhile, the LLL is asymmetric between the right- and left-handed fermions.
See Fig.~\ref{fig:LL} for an illustration.
This implies that the LLL is related to the production of the chiral charge, while the HLLs contribute to the symmetric production of the right- and left-handed fermions.
We will confirm this understanding explicitly in the following.

\paragraph{Lowest Landau level and anomaly equation.}

Here, we focus on the particle production in the LLL
and discuss its relation to the anomalous current equation \eqref{eq:anomalous}.
As we will see soon, the electric field (anti-)parallel to the magnetic field drives the particle production from the vacuum.

As indicated previously, the LLL can be regarded as a particle moving along with the magnetic field, namely $z$-direction.
To be concrete, let us take $Q > 0$ and $\lambda = +$ for which the dispersion relation then becomes $\omega_\text{R/L} = \pm p_z$.
Its dependence on $Q$ and $\lambda$ will be discussed later.
If we add an electric field parallel to the magnetic field,
the charged particles get accelerated as $\dot p_z = gQE$.
As a result, even if we start from the vacuum state where all the negative energy modes are filled while the positive energy modes are empty,
the positive energy modes are continuously generated/destroyed for the right-/left-handed fermions because $p_z$ increases with time.
See Fig.~\ref{fig:LL} for an illustration.
Thus, we have the particle/anti-particle production for the right-/left-handed fermions, which contributes to the asymmetry in the chirality, \textit{i.e.,} to $q_5$.
In this case, $q_5$ increases through this particle production
since $q_5$ measures $q_\text{R} - q_\text{L}$ with $q_H$ ($H = \text{R}, \text{L}$) being the number of $H$-handed particles minus anti-particles.

Now we discuss how the particle production depends on the sign of $Q$ and $\dot \phi$.
Let us first flip the sign of $Q$. The dispersion relation gets flipped $\omega_\text{R/L} = \mp p_z$ [See Eqs.~\eqref{eq:ll_R} and \eqref{eq:ll_L}].
At the same time, a particle with a negative charge  is decelerated, $\dot p_z = - g |Q| E$.
As a result, the dynamics of the chiral charge remains the same, namely $q_5$ increases.
Next, we move on to the dependence on the sign $\lambda$.
If one takes $\lambda = -$ but $Q > 0$, the dispersion relation becomes $\omega_\text{R/L} = \mp p_z$ while a particle is accelerated.
In this case, the chiral charge $q_5$ decreases.
In a word, we find $\dot q_5 \gtrless 0$ for $\dot \phi \gtrless 0$
which is consistent with the result indicated from the current equation
[see \textit{e.g.}, Eq.~\eqref{eq:Q5_dS}].
Recall here that the sign of $\lambda$ encodes the sign of the velocity of the scalar field:
$\lambda = \pm$ for $\dot \phi \gtrless 0$.

Finally, let us quantitatively estimate the particle production rate
and reproduce the anomalous current equation \eqref{eq:anomalous}.
As an illustrative example, we assume the following evolution of the $z$-component vector potential:
\begin{align}
	A_z (t) =
	\begin{cases}
		0 & \text{for} ~~ t \leq 0\\
		-Et &\text{for} ~~ 0 < t < \tau \\
		-E\tau  &\text{for} ~~ \tau \leq t
	\end{cases}
	~~~\rightarrow~~~
	E_z (t) =
	\begin{cases}
		0 & \text{for} ~~ t \leq 0\\
		E &\text{for} ~~ 0 < t < \tau\\
		0 &\text{for} ~~ \tau \leq t
	\end{cases} \,.
	\label{eq:electric_field}
\end{align}
For $t \leq 0$ and $\tau  \leq t$, we can unambiguously define the positive/negative frequency modes, and thus count how many particles are generated due to the electric field imposed during $0 < t < \tau$.
For $t \leq 0$, the right-handed fermions of the LLL can be expressed as
\begin{align}
	\psi_\text{R} = \int \frac{\dd p_y \dd p_z}{2 \pi} e^{i p_y y + i p_z z} h_0 (x_{-s}) \chi_s e^{-isp_z t}
	\left[
		b_{0,p_y,p_z} \theta (s p_z) 
		+ d^\dag_{0, - p_y, - p_z} \theta (- sp_z)
	\right] \,,
	\nonumber
\end{align}
where $b_{n,p_y,p_z}^{(\dag)}$, $d_{n,p_y,p_z}^{(\dag)}$ are the annihilation (creation) operators and the vacuum is defined by $b\ket{0} = d \ket{0} = 0$.
Here the Heaviside $\theta$ function encodes the positive/negative energy condition.
After $t \geq \tau$, this becomes 
\begin{align}
\label{eq:sol_LLL}
	\psi_\text{R} = \int & \frac{\dd p_y \dd p_z}{2 \pi} e^{i p_y y +  i p_z z}
	h_0 (x_{-s}) \chi_s e^{-i s (p_z + gQ E \tau)t} \nonumber \\
	&\Bigg[
		\underbrace{
		\left( b_{0,p_y,p_z} \theta(s p_z)
		+ d^\dag_{0,-p_y,-p_z} \theta (- s p_z ) \right)}_{B_{0,p_y,p_z}}
		\theta (s (p_z + g Q E \tau))
		\nonumber \\
		&+ 
		\underbrace{
		\left( b_{0,p_y,p_z} \theta(s p_z)
		+ d^\dag_{0,-p_y,-p_z} \theta (- s p_z ) \right)
		}_{D_{0, - p_y, -p_z}^\dag}
		\theta (- s(p_z + g Q E \tau) )
	\Bigg] \,,
\end{align}
where $B^{(\dag)}$ and $D^{(\dag)}$ represent the annihilation (creation) operators associated with the positive and negative frequency modes for $t \geq \tau$.
To avoid unnecessary complication, let us take $Q > 0$ and $\lambda > 0$, \textit{i.e.}, $s = +$.
Now one can compute the expectation value of $\overline \psi \gamma_0 P_\text{R} \psi$ for the LLL explicitly:
\begin{align}
	\left. q_\text{R} \right|_{n=0} 
	&= \frac{1}{\text{vol}\, (\mathbb{R}^3)} \int \dd^3 x
	\left. \vev{:\psi_\text{R}^\dag \psi_\text{R}:} \right|_{n=0}	 \nonumber \\
	&= \frac{1}{\text{vol}\, (\mathbb{R}^3)} \int \dd^3 x
	\int \frac{\dd p_y \dd p_z}{(2\pi)^2}
	h_0 (x_-) h_0 (x_-) \theta (-p_z) \theta (p_z + g Q E \tau) \nonumber \\
	&=
	\frac{L_y L_z \tau}{\text{vol}\, (\mathbb{R}^3)}
	\left( \frac{g Q E}{2 \pi} \right) \left( \frac{gQB}{\pi} \right)^{1/2}
	\int \frac{\dd x \dd p_y}{2 \pi} e^{- \left( \sqrt{gQB} x - \frac{p_y}
	{\sqrt{gQB}} \right)^2} \nonumber \\
	&= \tau \times \frac{\alpha Q^2}{\pi} EB \,,
\end{align}
where $:O:$ represents the normal ordering of $O$.
It is straightforward to show that the result with $\lambda = -$ has the opposite sign.

One can perform a similar computation for the left-handed fermions.
Recall that $q_H$ ($H = \text{R}, \text{L}$) counts the number of particles minus that of anti-particles: $B^\dag B - D^\dag D$.
For the left-handed fermions, a non-vanishing contribution comes from the anti-particles, \textit{i.e.}, $\vev{D^\dag D}$.
As a result, we get the opposite sign for the left-handed fermions:
\begin{align}
	\left. q_L \right|_{n=0} = 
	\tau \times \left( - \frac{\alpha Q^2}{\pi} EB \right) \,.
\end{align}
for $\lambda = +$. Again, the overall sign is flipped for $\lambda = -$.

Collecting the obtained results so far, we finally arrive at
\begin{align}
	\dot q_5 &= \left. \dot q_\text{R} \right|_{n = 0} 
	-  \left. \dot q_\text{L} \right|_{n=0}
	= \frac{2 \alpha Q^2}{\pi} E (\lambda B) \label{eq:dnLLL} \\
	&=
	- \frac{\alpha Q^2}{2 \pi} F_{\mu\nu} \tilde F^{\mu\nu} \,,
\end{align}
where $\lambda = \pm$ for $\dot \phi \gtrless 0$.
Following the procedure of Ref.~\cite{Nielsen:1983rb}, we have here reproduced the anomaly equation \eqref{eq:anomalous}.
Also, we directly obtain $\dot q_5 \gtrless 0$ for $\dot \phi \gtrless 0$
which is indicated from the current equations: Eqs.~\eqref{eq:shift_sym}, \eqref{eq:anomalous}, and \eqref{eq:shift_sym_2}.
As we will see in the section,
the particle production from 
the HLLs does not contribute to the chiral charge~$q_5$
because the right- and left- handed fermions have the same spectrum.
This is expected because the anomalous current equation does not receive radiative corrections~\cite{Adler:1969er}.

\paragraph{Higher Landau levels and Schwinger effect.}

Here we discuss the fermion production in the HLLs.
Contrary to the LLL, HLLs are gapped and hence we cannot create particles in a smooth way.
Nevertheless, the quantum tunneling allows a pair creation of particle and anti-particle, as in the Schwinger effect.
Before discussing the particle production in the HLLs,
we briefly summarize the basics of the Schwinger effect
by turning off the magnetic field.
Equipped with some intuition, we move on to the particle production in the HLLs.
In the following, we for simplicity take $Q > 0$, $\lambda = +$ unless otherwise stated.

Suppose that we turn on a uniform electric field during $0 < t < \tau$ pointing along the $z$-axis as in Eq.~\eqref{eq:electric_field}.
For a fixed transverse momentum $p_\perp$, with $p_\perp^2 = p_x^2 + p_y^2$,
the dispersion relation as a function of $p_z$ is given by $\omega = \pm \sqrt{p_\perp^2 + p_z^2}$, \textit{i.e.\,} it is gapped by the effective transverse mass given by $\abs{p_\perp}$.
In the presence of the electric field, the quantum mechanical pair production of particles and anti-particles is favored while there is no classical path for this process due to this transverse mass.
As a result, the pair-production rate is exponentially suppressed by $e^{-\pi p_\perp^2 / g Q E}$.
By taking this tunneling suppression into account, one gets the well-known result~\cite{Heisenberg:1935qt,Schwinger:1951nm,Cohen:2008wz}:
\begin{align}
	\dot n_\text{R}  
	&\simeq \frac{1}{\tau} \int \frac{\dd^3 p}{(2 \pi)^3}
	\theta (- p_z) \theta (p_z + g Q E \tau) e^{- \frac{\pi p_\perp^2}{g Q E}} 
	= \frac{g Q E}{2 \pi}
	\int \frac{\dd^2 p_\perp}{(2\pi)^2} e^{- \frac{\pi p_\perp^2}{g Q E}}\\
	&=
	\frac{g^2 Q^2 E^2}{8 \pi^2}
	= \dot{\bar n}_\text{R} = \dot n_\text{L} = \dot{\bar n}_\text{L} \,,
	\label{eq:schwinger}
\end{align}
where $n_H$ ($\bar n_H$) is the number density of the particle (anti-particle) for the $H$-handed fermions.
These are related to the corresponding charges through $q_H = n_H - \bar n_H$.

Now we move back to the HLLs.
Contrary to the simplest case we have discussed,
the \textit{transverse mass} is no longer continuous,
rather it is discretized as the Landau levels.
This observation indicates
that the exponential suppression becomes $e^{- 2 \pi B n / E}$ for each HLL of $n$, and that the integration with respect to $p_\perp$ should be replaced with the summation over $n$.
Moreover, if the electric field is much larger than the magnetic field,
the electric field probes much higher levels and thus the spectrum can be well approximated with the continuous one.
Thus, we expect that the ordinary Schwinger effect is recovered in the limit of $B \ll E$. 
In the rest of this section, we explicitly derive the production rate and confirm this expectation.

We again assume that the $z$-component of the vector potential obeys Eq.~\eqref{eq:electric_field}.
For $t \leq 0$ and $\tau \leq t$, the positive/negative frequency modes are clearly separated.
Let us start with the explicit form of the wave function of the positive/negative frequency modes for $t < 0$.
As can be seen from Eqs.~\eqref{eq:ll_R} and \eqref{eq:ll_L},
the HLLs have two independent solutions $\omega = \pm\sqrt{p_z^2 + 2 g Q B n}$.
We define $g_\text{R}^{(\pm)}$ so that it becomes the positive/negative frequency modes for $t \leq 0$, namely
$g_\text{R/L}^{(\pm)} \to N e^{\mp i \omega t}$ for $t \leq 0$
with $N$ being the normalization factor.
By plugging this into Eq.~\eqref{eq:aux}, we can express the wave function of the positive frequency mode as
\begin{align}
	u_{n,p_y,p_z}^\text{(R)} &= \left[
		\left( \omega + p_z \right) \chi_+ 
		- i\sqrt{gQB}\left( 
			\frac{\partial}{\partial x_-} + x_- 
		\right) \chi_-
	\right] g_\text{R}^{(+)} h_n \nonumber \\
	&= \frac{e^{- i \omega t}}{\sqrt{2 \omega (\omega + p_z)}}
	\left[
		\left( \omega + p_z \right) h_n \chi_+ 
		- i\sqrt{2n gQB} h_{n-1} \chi_-
	\right]\,.
\end{align}
In the second equality, we have used $h_n' = - x h_n + \sqrt{2n} h_{n - 1}$.
We normalize the wave function so that 
$\int \dd x u^{\text{(R)}^\dag}_{n,p_y,p_z} u_{\bar n,p_y,p_z}^\text{(R)} = \delta_{n \bar n}$.
Similarly, the wave function for the negative frequency mode is
\begin{align}
	v_{n, p_y, p_z}^\text{(R)}
	= \frac{e^{i \omega t}}{\sqrt{2 \omega ( \omega - p_z )}}
	\left[
		\left( - \omega + p_z \right) h_n \chi_+
		- i \sqrt{2 n g Q B} h_{n-1} \chi_-
	\right] \,,
\end{align}
where the normalization is
$\int \dd x \, v^{\text{(R)}^\dag}_{n,p_y,p_z} v_{\bar n,p_y,p_z}^\text{(R)} = \delta_{n \bar n}$.
In addition, $u$ and $v$ are orthogonal.
Thanks to these properties, one may extract the annihilation operator from $\psi_\text{R}$ by multiplying $u^\text{(R)}$ from the left,
\textit{i.e.}, $\int \dd x u^{\text{(R)}\dag}_{n,p_y,p_z} \psi_\text{R}(x;p_y,p_z)$.
This is also true for the creation operator for the anti-particle via multiplying $v^\text{(R)}$ from the left.

Motivated by this observation, let us define a wave function for the positive/negative frequency mode during $0 < t < \tau$.
Since the $z$-component of the vector potential depends on time,
the positive/negative frequencies in $g_\text{R}^{(\pm)}$ get mixed,
which indicates the particle production.
By inserting $G_\text{R}^{(+)} \propto e^{- i \int^t \dd t'\, \omega (t')}$ into Eq.~\eqref{eq:aux} instead of $g_\text{R}$, one may obtain the wave function for the positive frequency mode at a given time $t$:
\begin{align}
	U_{n,p_y,p_z}^\text{(R)}
	= \frac{e^{-i \int^t \dd t' \, \omega (t')}}
	{\sqrt{2 \omega (t) [ \omega (t) + \Pi_z (t) ]}}
	\left\{
		\left[ \omega (t) + \Pi_z (t)  \right] h_n \chi_+
		- i \sqrt{2n g Q B} h_{n-1} \chi_-
	\right\}\,,
\end{align}
where $\omega (t) = \sqrt{\Pi_z^2 (t) + 2 n g Q B}$ and $\Pi_z (t) = p_z + g Q E t$.
Also, the wave function for the negative frequency mode can be obtained from $G_\text{R}^{(-)} \propto e^{+ i \int^t \dd t'\, \omega (t')}$
as follows:
\begin{align}
	V_{n,p_y,p_z}^\text{(R)}
	= \frac{ e^{ + i \int^t \dd t'\, \omega (t')} }
	{\sqrt{2 \omega (t) [ \omega (t) -  \Pi_z (t) ]}}
	\left\{
		\left[ - \omega (t) + \Pi_z (t) \right] h_n \chi_+
		- i \sqrt{2 n g Q B} h_{n - 1} \chi_-
	\right\} \,.
\end{align}
Their normalization and orthogonality are the same as those for $t \leq 0$.

By means of those functions, one may extract the coefficient of the positive/negative frequency modes from $g_\text{R}^{(\pm)}$,
which has the positive/negative frequency initially $t \leq 0$.
That is to say, the Bogolyubov coefficients are obtained:
\begin{align}
	\alpha_{n, p_y, p_z}^\text{(R)} (t) 
	&= 
	\int \dd x\, U^{\text{(R)}\dag}_{n,p_y,p_z} 
	\left\{
		\left[ i \partial_t + \Pi_z (t) \right] h_n \chi_+
		- i \sqrt{2 n g Q B} h_{n-1} \chi_- 
	\right\} g_\text{R}^{(+)}
	\nonumber \\
	&=
	\sqrt{2 n g Q B} G_\text{R}^{(-)} \left[ i \partial_t + \omega (t) \right] g_\text{R}^{(+)}, \\[.5em]
	\beta_{n, p_y, p_z}^\text{(R)} (t)
	& = 
	\int \dd x\, V^{\text{(R)}\dag}_{n,  p_y, p_z} 
	\left\{
		\left[ i \partial_t + \Pi_z (t) \right] h_n \chi_+
		- i \sqrt{2 n g Q B} h_{n-1} \chi_-
	\right\} g_\text{R}^{(+)}
	\nonumber \\
	& =
	\sqrt{2 n g Q B} G_\text{R}^{(+)} \left[ -  i \partial_t + \omega  (t) \right] g_\text{R}^{(+)}\,.
\end{align}
The Bogolyubov transformation is defined by
\begin{align}
	B_{n,p_y,p_z}^\text{(R)} 
	&= \alpha^\text{(R)}_{n, p_y, p_z} b^\text{(R)}_{n,p_y,p_z} - \beta^{\text{(R)}\ast}_{n,p_y,p_z} d^{\text{(R)}\dag}_{n, - p_y, - p_z}\,, \\
	D_{n, - p_y, - p_z}^{\text{(R)}\dag}
	&= \beta^\text{(R)}_{n, p_y, p_z} b^\text{(R)}_{n,p_y,p_z} +\alpha^{\text{(R)}\ast}_{n,p_y,p_z} d^{\text{(R)}\dag}_{n, - p_y, - p_z}\,,
\end{align}
where $|\alpha|^2 + |\beta|^2 = 1$.
Recall that we have defined the vacuum via $b \ket{0} = d \ket{0} = 0$,
and taken this vacuum to be the initial state.
This means that the initial condition for the coefficients is given by
$\alpha = 1$ and $\beta = 0$.
One may write down the equation of motion for $\alpha$ and $\beta$ explicitly, which could be useful for numerical studies:
\begin{align}
	\dot \alpha^\text{(R)}_{n, p_y, p_z} 
	&= 
	- \beta^\text{(R)}_{n, p_y, p_z}
	\frac{g Q E}{2 \omega^2} e^{2 i \int^t \dd \tau \omega}
	\times \sqrt{2 n g Q B}\,, \\
	\dot \beta^\text{(R)}_{n, p_y, p_z} 
	&= 
	\alpha^\text{(R)}_{n, p_y, p_z}
	\frac{g Q E}{2 \omega^2} e^{- 2 i \int^t \dd \tau \omega}
	\times \sqrt{2 n g Q B}\,.
\end{align}
If one replaces $\sqrt{2 n g Q B}$ with $\sqrt{p_\perp^2}$ (which has the interpretation of an effective mass),
the equation for the ordinary Schwinger effect is recovered.
Hence, the asymptotic behavior of $|\beta|^2$ at $t = \tau$,
which is related to the number of produced particles,
is essentially the same as that of the Schwinger effect~\cite{Cohen:2008wz}
except for the exponential suppression factor:
\begin{align}
	|\beta^\text{(R)}_{n,p_y,p_z} (\tau)|^2 
	\simeq \theta ( - p_z) \theta (p_z + g Q E \tau) e^{- \frac{2 \pi n B}{E}}\,.
	\label{eq:beta}
\end{align}

Now we can estimate the pair-production rate of a particle and anti-particle at the $n$-th Landau level:
\begin{align}
	\dot n_\text{R}^{(n)} 
	&= \frac{1}{\tau \text{vol}\, (\mathbb{R}^3)}
	\int \dd x \dd p_y \dd p_z 
	\vev{B_{n,p_y,p_z}^{\text{(R)}\dag} B_{n, - p_y,- p_z}^\text{(R)}} 
	\left[ \frac{\omega + \Pi_z}{2\omega} h_n^2 (x_-) + 
	\frac{\omega - \Pi_z}{2 \omega} h_{n-1}^2 (x_-) \right]
	\nonumber \\
	&= \frac{1}{\tau} \, \frac{g Q B}{2 \pi}
	\int \frac{\dd p_z}{2 \pi} \theta ( - p_z) \theta (p_z + g Q E \tau) e^{- \frac{2 \pi n B}{E}} \nonumber\\
	&= \frac{g^2 Q^2}{4 \pi^2} EB e^{-\frac{2 \pi n B}{E}}  \label{eq:nHLL_n}\\
	&= \dot{\bar n}_\text{R}^{(n)} 
	= \dot n_\text{L}^{(n)} = \dot{\bar n}_\text{L}^{(n)} \,.
\end{align}
In the second equality, we have used Eq.~\eqref{eq:beta}.
Since $\langle B^\dag B \rangle$ and $\langle D^\dag D \rangle$ are determined by the same coefficient $|\beta |^2$,
the production rates for particles and anti-particles are exactly the same.
This is expected because the process is a pair-production.
One can explicitly check that the result does not depend on the sign of $Q$, $\lambda$, or the chirality.
This is because the dispersion relation of HLLs is insensitive to these. 
Summing over the Landau levels $n \geq 1$,
we eventually get
\begin{align}
	\dot n_H
	&= \sum_{n=1}^\infty \dot n_H^{(n)} 
	= \frac{g^2 Q^2}{4 \pi^2} EB \frac{1}{e^{2 \pi B / E} - 1} 
	= \dot{\bar n}_H \,,
\end{align}
for $H = \text{R}, \text{L}$.
It is obvious that one may recover the result of the Schwinger effect given in Eq.~\eqref{eq:schwinger} in the limit of $2\pi B \ll E$:
\begin{align}
	\dot n_H = \dot{\bar n}_H \to
	\frac{g^2 Q^2}{8 \pi^3} E^2\,.
	\label{eq:dnHLL}
\end{align}
\paragraph{~~}
Throughout this section we have assumed that the fermion production is fast compared to the expansion of the Universe. We now have all the ingredients to confirm this a posteriori. Let us focus in the following on the regime $\xi \gtrsim 3$, in which the simple analytical formulas for the electric and magnetic field, Eqs.~\eqref{eq:E2} and \eqref{eq:B2} apply. In this case the fermion production rates (see Eq.~\eqref{eq:dnLLL} and \eqref{eq:dnHLL}) read
\begin{align}
  \dot n_\psi^\text{LLL} & = 2 \times \frac{g^2 Q^2}{4 \pi^2} E B \,, \label{eq:nLLL} \\
  \dot n_\psi^\text{HLL} & = 4 \times \frac{g^2 Q^2}{8 \pi^3} \left(E^2 - \pi EB + \frac{\pi^2}{3} B^2 + \cdots \right)\,. \label{eq:nHLL}
\end{align} 
Choosing as reference values $Q = 1$ and the SM GUT-scale gauge coupling $g = 1/\sqrt{2}$, we find both rates to be much larger than $H^4$, justifying the flat spacetime approximation of this section.

\begin{figure}
 \centering
 \includegraphics[width = 0.5 \textwidth]{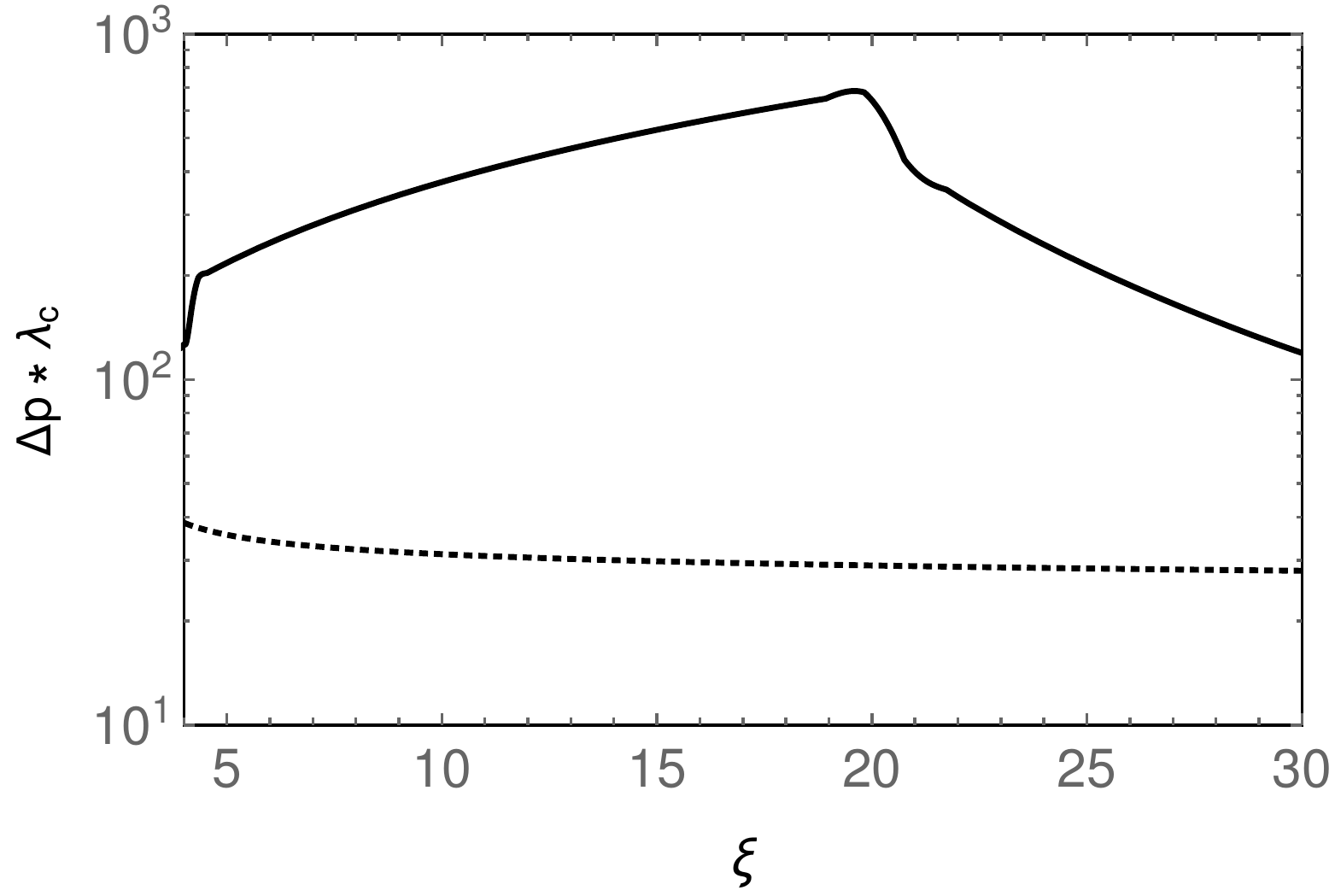}
 \caption{For the values of $E$ and $B$ obtained in Sec.~\ref{sec:br}, Pauli blocking is inefficient. The solid line corresponds to our estimate for $\Delta p \, \lambda_c$ for the upper bounds of the $E$ and $B$ fields as depicted in the left panel of Fig.~\ref{fig:EBbounds}, the dotted line corresponds to the estimates depicted in the right panel of Fig.~\ref{fig:EBbounds}.}
 \label{fig:PauliBlocking}
\end{figure}

Moreover, throughout this section we have neglected the possibility of Pauli blocking in the final HLL fermion states. To estimate the importance of this effect, consider the characteristic time scale for the production of one fermion within a volume $\lambda_c^3$, where $\lambda_c$ denotes the Compton wavelength of the fermion:
\begin{equation}
 t_\text{prod} = \dot n_\psi^{-1} \lambda_c^{-3} \,.
\end{equation}
During this time interval, the previous fermion generated in this phase space box will have been accelerated due to the force exerted by the constant {electric} field, $F = \dd p/\dd t = g Q E$. If this acceleration is large compared to the initial energy $\lambda_c^{-1}$, 
\begin{equation}
 \Delta p = g Q E t_\text{prod} \gg \lambda_c^{-1} \,, \label{eq:Pauli}
\end{equation}
the effect of Pauli blocking can be safely neglected. 
For the higher Landau levels, $\lambda_c$ is set by the transverse energy determining the level splitting, $p_\perp^2 = 2 n g |Q| B$, see Eq.~\eqref{eq:ll_L}. Evaluating Eq.~\eqref{eq:Pauli} requires the knowledge of the relative size of the $E$ and $B$ fields after taking into account the backreaction effects from the fermion production. This will be the main topic of the next section. Anticipating these results, we show that $\Delta p \lambda_c \gg 1$ holds in the entire parameter space of interest in Fig.~\ref{fig:PauliBlocking}, and hence Pauli Blocking can be safely neglected.

\section{Backreaction}
\label{sec:br}

In the previous section we computed the fermion production in a background of constant electric and magnetic fields, as sourced by the tachyonic instability in Eq.~\eqref{eq:wave_eq}. If produced abundantly enough, the produced particles (both fermions and gauge fields) may affect the production of the helical gauge field, as indicated by Eq.~\eqref{eq:eom_gauge} and also Eq.~\eqref{eq:dot_rho_gauge}. In this section, we derive the induced fermion current which allows us to estimate the gauge field production.

Before going into details, we would like to recall our basic assumptions. 
As explained at the beginning of Sec.~\ref{sec:fermion}, we employ an approximation for the gauge field,
focusing on its horizon-scale configuration. 
One may regard such a gauge field as a classical field 
which points in a random direction for each Hubble patch, 
analogous to the stochastic formalism in Ref.~\cite{Starobinsky:1994bd}.
We pick one Hubble patch
and adopt typical values of the electric/magnetic fields denoted as $\bm{E}$ and $\bm{B}$.\footnote{
	Note that, after averaging over all the Hubble patches, we get $\vev{E} = \vev{B} = 0$.
	The typical values at each Hubble patch are estimated from 
	$\sqrt{\overline{\vev{\bm{E}^2}}}$ and $\sqrt{\overline{\vev{\bm{B}^2}}}$
	with the overline denoting the spatial average. 
}
Under these assumptions, we expect
$\vev{\bm{E} \cdot \bm{B}} \simeq \bm{E} \cdot \bm{B}$, 
$\vev{\bm{E} \cdot \bm{E}} \simeq \bm{E}^2$, 
$\vev{\bm{B} \cdot \bm{B}} \simeq \bm{B}^2$, and
$\vev{\bm{E} \cdot \bm{J}_\psi} \simeq \bm{E} \cdot \vev{\bm{J}_\psi}$.
In particular, we will focus on the charged current $\vev{\bm{J}_\psi}$ 
induced by the presence of the classical background of $\bm{E}$ and $\bm{B}$.

\subsection{Consistency conditions}
\label{sec:consistenct}

Unfortunately, the backreaction makes the equations of motion non-linear and difficult to solve.
Hence we would like to give criteria that enable us to estimate the maximal amount of the gauge field without solving the equations of motion explicitly.
Here, we summarize the consistency conditions to have gauge fields ranging over superhorizon scales
which must be fulfilled regardless of the details of the fermion production.

\paragraph{Energy conservation.}
Let us look at the energy equation for the scalar field given in Eq.~\eqref{eq:dot_rho_scalar}:
\begin{align}
	\dot \rho_\phi + 3 H \vev{\dot \phi^2} 
	&= - 2 \xi H
	\hat{\bm{E}} \cdot \hat{\bm{B}}\,.
\end{align}
Here we have picked up typical values of the electric/magnetic fields of one Hubble patch,
and hence we can drop $\vev{\bullet}$ (and also the spatial average).
The term in the right-hand-side represents the energy conversion from the scalar field to the gauge field.
It is obvious that we cannot extract energy larger than that carried by $\rho_\phi$.
Since we have required that the change of $\dot \phi$ is slower than the cosmic expansion throughout our analysis (in other words we impose the existence of an equilibrium configuration with negligible $\ddot \phi$),
the gauge field production must not consume all the energy within a Hubble time.
This consideration puts the following  bound:
\begin{align}
\label{eq:bound1}
	\xi \hat{\bm{E}} \cdot \hat{\bm{B}} \lesssim \rho_\phi \simeq V\,.
\end{align}
In the second step, we have assumed $V \gg \dot\phi^2$.

\paragraph{Non-trivial attractor.}
Let us first neglect the fermion production and examine the solution \eqref{eq:gauge_desitter} from another viewpoint.
Eq.~\eqref{eq:gauge_desitter} indicates constant electric/magnetic fields over horizon scales.
We would like to understand its meaning through Eq.~\eqref{eq:dot_rho_gauge}:
\begin{align}
	\dot \rho_A \simeq - 4 H \rho_A + 2 \xi H \hat{\bm{E}} \cdot \hat{\bm{B}}\,.
\end{align}
To have such constant solutions, the production must compensate the decrease due to the cosmic expansion, \textit{i.e.}, 
\begin{align}
	0 &= \dot \rho_A \\
	&= 2 H \left( \xi \hat{\bm{E}}\cdot \hat{\bm{B}} - \hat{\bm{E}}^2 - \hat{\bm{B}}^2 \right)\,.
	\label{eq:bound2_nobr}
\end{align}
For $\xi \gg 1$, one finds two branches; $\hat{\bm{E}} \simeq \xi \hat{\bm{B}}$ and $\hat{\bm{B}} \simeq \xi \hat{\bm{E}}$,
which reflects an electromagnetic duality in the absence of matter.
In the end, we introduce the electric matter, $\psi$, and couple it with the gauge field perturbatively, which implies $\hat{\bm{E}} \simeq \xi \hat{\bm{B}}$. 
This is consistent with Eqs.~\eqref{eq:E2} and \eqref{eq:B2} as expected.

Now, we turn on the coupling with $\psi$.
We have an energy transfer from the gauge field to the fermion:
\begin{align}
	\dot \rho_A = - 4 H \rho_A + 2 \xi H \hat{\bm{E}} \cdot \hat{\bm{B}}
	- \hat{\bm{E}} \cdot 
	g Q \vev{\bm{J}_\psi} 
	\,.
\end{align}
The last term represents the energy reduction by the fermion production.
At the same time,
the gauge field is reproduced immediately from the scalar field. If the fermion production significantly drains the energy of the gauge field configuration, the background electric and magnetic fields decrease, which then leads to the reduction of the fermion production. Owing to this negative feedback, we expect that there exists a non-trivial attractor of constant gauge field even in the presence of $\psi$, where these processes have reached a dynamical equilibrium.
The condition to have such a constant gauge field is given by
\begin{align}
	0 &= \dot \rho_A \\
	&= - 4 H \rho_A + 2 \xi H \hat{\bm{E}} \cdot \hat{\bm{B}}
	- \hat{\bm{E}} \cdot 
	g Q \vev{\bm{J}_\psi} 
	\,.
	\label{eq:bound2}
\end{align}
\paragraph{~~}
To sum up, Eqs.~\eqref{eq:bound1} and \eqref{eq:bound2} must be satisfied in order to have approximately constant helical gauge fields over horizon-scales. The remaining question is how the induced current $\vev{\bm{J}_\psi}$ behaves as a function of $\hat{\bm{E}}$ and $\hat{\bm{B}}$.

\subsection{Induced current and backreaction}
\label{sec:induced}

It is instructive to first consider the induced current in general before discussing our particular setup.
Suppose that we have charged particles whose phase-space distribution is $f_\psi (p)$ and impose an electric field $\hat{\bm{E}}$.
The induced current in such a system is estimated by
\begin{align}
	g Q \vev{\bm{J}_\psi} 
	&\simeq N_\text{dof}\, g Q \int \frac{\dd^3 p}{(2 \pi)^3} \frac{\bm{\Pi}}{\omega} f_\psi (p), \nonumber\\
	& = 
	N_\text{dof}\, \left( g Q \right)^2 \hat{\bm{E}}\tau \int \frac{\dd^3 p}{(2 \pi)^3} \frac{f_\psi (p)}{\omega}\,,
\end{align}
where $\bm{\Pi} = \bm{p} + g Q \hat{\bm{E}} \tau$,
$\omega = \sqrt{\bm{p}^2 + g^2 Q^2 \hat{\bm{E}}^2 \tau^2}$,
and $N_\text{dof}$ counts the degrees of freedom for $\psi$.
Note that $\tau$ represents a typical time scale of acceleration until it is disrupted by large angle scatterings.
In the second line, we have assumed that the phase-space distribution is isotropic:
$f_\psi (p)$.

Let us estimate the behavior of the induced current.
Suppose that the phase-space distribution is dominated by a typical momentum of $\bar p$.
If the typical momentum, $\bar p$, is larger than the one acquired by the acceleration, $g Q \hat{\bm{E}} \tau$,
the induced current is proportional to the scattering time scale, $\tau$.
On the other hand, if it is not, the induced current is independent of $\tau$.
Hence, one may estimate the induced current as
\begin{align}
	g Q \vev{\bm{J}_\psi} 
	\sim 
	\begin{cases}
		\cfrac{g^2 Q^2 \hat{\bm{E}}\tau}{\bar p}  n_\psi 
		&\text{for}~~~ g \abs{Q} \hat{E} \tau \ll \bar p\,, \\
		g \abs{Q} n_\psi \bm{e}_{\hat E} 
		&\text{for}~~~ g \abs{Q} \hat{E} \tau \gg \bar p\,,
	\end{cases}
\end{align}
where $\bm{e}_{\hat E} \equiv \hat{\bm{E}} /\hat E$ denotes the unit vector in the direction of the electric field.

As an illustration, we confirm that this equation reproduces the electric conductivity in the thermal plasma.
We assume that the system is thermalized 
and would like to see how the system responses to a weak electric field.
In this setup, the typical momentum is just the temperature $T$
and the time scale of large angle scatterings would be $\tau \sim (\alpha^2 T)^{-1}$.
If the electric field is so weak that $g \abs{Q} \hat E \tau \ll T$, the induced current can be estimated as 
$\sim \hat{\bm{E}}\, T / \alpha$, which reproduces the well-known result of the electric conductivity, \textit{i.e.}, $\sigma \sim T / \alpha $~\cite{Baym:1997gq,Arnold:2000dr}.
From this demonstration, one can see that the typical momentum $\bar p$ and the time scale of scatterings play an essential role in determining the behavior of the induced current.

\paragraph{Neglecting scatterings among particles.}

Here we estimate the induced current by assuming that the scatterings among particles are so slow that $g \abs{Q} \hat E \tau \gg \bar p$.
A concrete value of $\bar p$ in our setup will be specified soon and the validity of this approximation will be justified in the next subsection.
Under this approximation, one may compute the induced current just by plugging the solutions [see \textit{e.g.}, Eqs.~\eqref{eq:sol_LLL} and \eqref{eq:beta}] into the definition of the current \eqref{eq:current}.

Again, let us take the electric and magnetic fields along the $z$-axis without loss of generality,
$\bm{E} = (0,0,E)$ and $\bm{B} = (0,0,\lambda B)$ with $\lambda = \pm$;
and assume that the gauge field evolves as Eq.~\eqref{eq:electric_field}.
By using the asymptotic solution of \eqref{eq:sol_LLL},
one may estimate the induced current of the LLL as
\begin{align}
\label{eq:induced_LLL}
	\frac{1}{\tau}
	g Q \left. \vev{J^z_\psi} \right|_\text{LLL}
	\simeq \frac{\left( g \abs{Q} \right)^3}{2 \pi^2} E B \,.
\end{align}
On the other hand, the induced current of the HLLs reads
\begin{align}
\label{eq:induced_HLL}
	\frac{1}{\tau} g Q \left. \vev{J^z_\psi} \right|_\text{HLLs}
	& \simeq
	\frac{1}{\tau} \frac{g^2 Q^2 B}{2 \pi}
	\sum_{n = 1}^\infty
	\int \frac{\dd p_z}{2 \pi}
	\frac{\Pi_z}{\omega} \times
	4 \abs{\beta_{n,p_z,p_y}}^2 \nonumber \\
	&\simeq
	\frac{\left( g \abs{Q} \right)^3}{\pi^2} EB
	\frac{1}{e^{2 \pi B / E} - 1} \,,
\end{align}
where $\Pi_z = p_z + g Q E \tau$
and $\omega = \sqrt{\Pi_z^2 + 2 n g \abs{Q} B}$.
Contrary to the LLL, the HLLs have transverse momentum, which sets a characteristic scale of $\sqrt{n g \abs{Q} B}$.
In the second step, we have assumed $g \abs{Q} E \tau \gg \sqrt{g \abs{Q} B}$.
Note that the induced current is dominated by the level $n$ which saturates $g \abs{Q} E  \tau \sim \sqrt{2 n g \abs{Q} B}$.
Summing Eqs.~\eqref{eq:induced_LLL} and \eqref{eq:induced_HLL}, we get~\cite{Abramchuk:2016afc}
\begin{align}
	\partial_\eta \left(
	g Q \vev{J^z_\psi} \right)
	\simeq 
	\frac{\left( g \abs{Q} \right)^3}{2 \pi^2}
	\coth \left( \frac{\pi B}{E} \right) 
	E B \,.
\end{align}

Let us turn on the cosmic expansion.
As mentioned at the beginning of Sec.~\ref{sec:fermion},
all particle production processes are much faster than the cosmic expansion for the parameters of our interest.
Hence, the cosmic expansion can be treated adiabatically by just replacing $E$ and $B$ with $a^2 \hat E$ and $a^2 \hat B$.
Assuming constant physical electric/magnetic fields,
we can perform the time integral, which reads
\begin{align}
\label{eq:induced_noscat}
	\frac{1}{a^3}
	g Q \vev{J^z_\psi}
	&\simeq 
	\frac{\left( g \abs{Q} \right)^3}{6 \pi^2}
		\coth \left( \frac{\pi \hat B}{\hat E} \right) 
	\hat E  \hat B \frac{1}{H} \\
	&\to \frac{\left( g \abs{Q} \right)^3}{6 \pi^3}
	\frac{\hat E^2}{H}
	~~~\text{for}~~~\hat E \gg \hat B \,.
\end{align}
In the second line,
we check that the result is consistent with the one known in the literature~\cite{Kobayashi:2014zza,Hayashinaka:2016qqn} for $B \to 0$.

\begin{figure}
\subfigure{
  \includegraphics[width = 0.45 \textwidth]{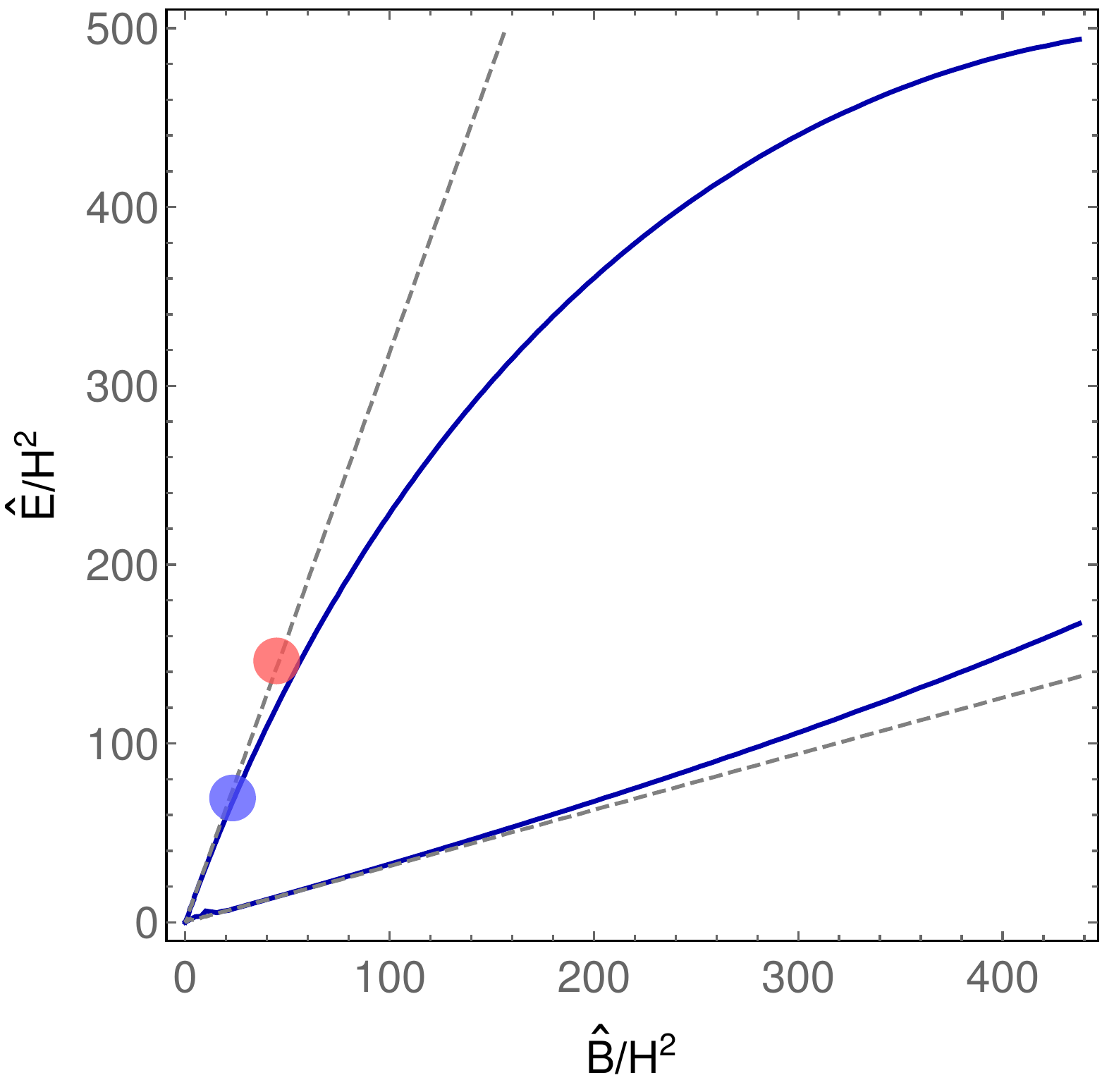}}
 \hfill
\subfigure{
  \includegraphics[width = 0.45  \textwidth]{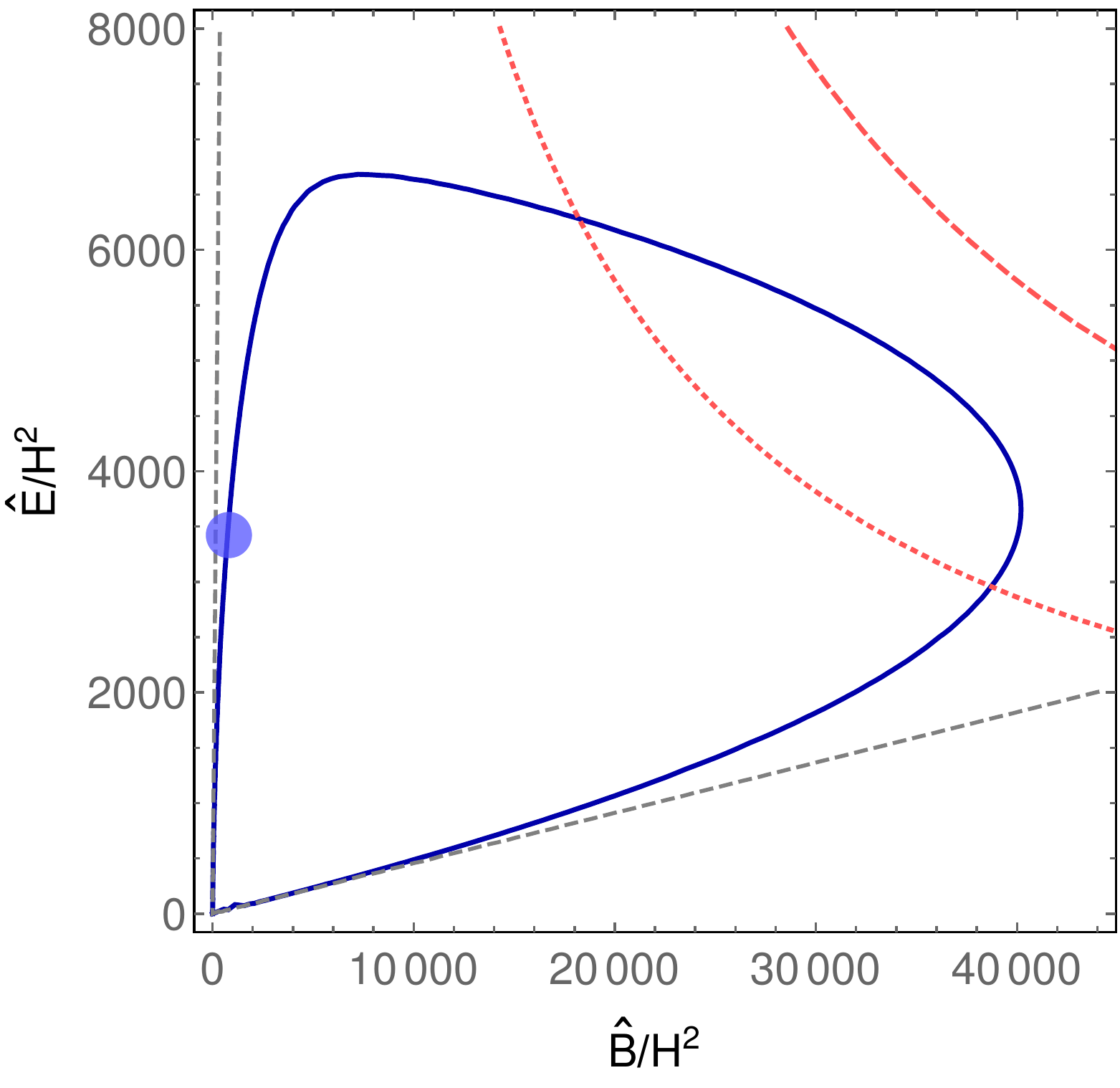}}
  \caption{Consistency conditions on the magnitude of the $\hat E$ and $\hat B$ fields for $\xi = 3.5$ (\textbf{left panel}) and $\xi = 22$ (\textbf{right panel}). The solid blue line indicates the consistency condition to have the stationary solution~\eqref{eq:bound2}. The red circle denotes the analytical solution without backreaction, in the right panel this solution is far outside the plotted range. The blue circle indicates the estimate obtained by taking $\xi_\text{eff}$ to be constant. The dotted (dashed) red contour indicate the upper bound from energy conservation in slow-roll inflation [Eq.~\eqref{eq:bound1}] for $r = 0.05$ ($r = 0.1$), relevant only for large $\xi$. Here we have set $Q = 1$ and $g = 1/\sqrt{2}$. } 
  \label{fig:consistency}
\end{figure}

\paragraph{Upper bounds on gauge fields.} Now we are in a position to discuss how the backreaction modifies the helical gauge field production by using the explicit expression for the induced current [Eq.~\eqref{eq:induced_noscat}]. 
The condition for the non-trivial attractor [Eq.~\eqref{eq:bound2}] defines a curve in the $(\hat E, \hat B)$ plane,
\begin{align}
\label{eq:curve}
	0 = - 2 H \left( \hat E^2 + \hat B^2 \right) + 2 \xi_\text{eff} H \hat{E} \hat{B} \,,
\end{align}
where
\begin{align}
\label{eq:xi_eff}
	\xi_\text{eff} = \xi - \frac{\left( g \abs{Q} \right)^3}{12 \pi^2} 
	\coth \left( \frac{\pi \hat B}{\hat E} \right) \frac{\hat E}{H^2}\,.
\end{align}
One may roughly estimate the maximum values of electric/magnetic fields on this curve:
\begin{align}
	\hat B_\text{max} \sim \frac{3 \pi^2}{\left( g \abs{Q} \right)^3} \xi^2 H^2 \,,
	\quad
	\hat E_\text{max} \sim \frac{12 \pi^2}{\left( g \abs{Q} \right)^3} \xi H^2\,.
	\label{eq:EBmax}
\end{align}
The curve is depicted as a blue solid line in Fig.~\ref{fig:consistency}.
For comparison, we also show the analytic solution without the backreaction as the red circle,
and the condition for a non-trivial attractor without the backreaction as the gray dashed line.

As suggested by the introduction of $\xi_\text{eff}$,
the equation of motion for the gauge field is obtained by simply replacing $\xi$ with $\xi_\text{eff}$.
In a crude estimation, we can estimate $\hat E$ and $\hat B$ as follows.
Taking $\xi_\text{eff}$ to be a time-independent constant [in line with the assumption of the existence of an attractor with constant $\hat E$ and $\hat B$, see Eq.~\eqref{eq:bound2}],
one may estimate the solution of $\hat E$ and $\hat B$ by just replacing $\xi \mapsto \xi_\text{eff}$ in Eqs.~\eqref{eq:E2} and \eqref{eq:B2}.
Then, using this $\hat E$ and $\hat B$, one may compute $\xi_\text{eff}$ according to Eq.~\eqref{eq:xi_eff}. Finally, requiring this $\xi_\text{eff}$ to be the same as the input $\xi_\text{eff}$ (which in turn depends on $\hat E$ and $\hat B$), we can find a self-consistent solution.
We indicate this estimation with a blue circle in Fig.~\ref{fig:consistency}.
Finally, the energy conservation condition~\eqref{eq:bound1} adds an upper bound on the electric/magnetic fields, shown as red curves in Fig.~\ref{fig:consistency}.

\begin{figure}
\centering
\subfigure{
  \includegraphics[width = 0.48  \textwidth]{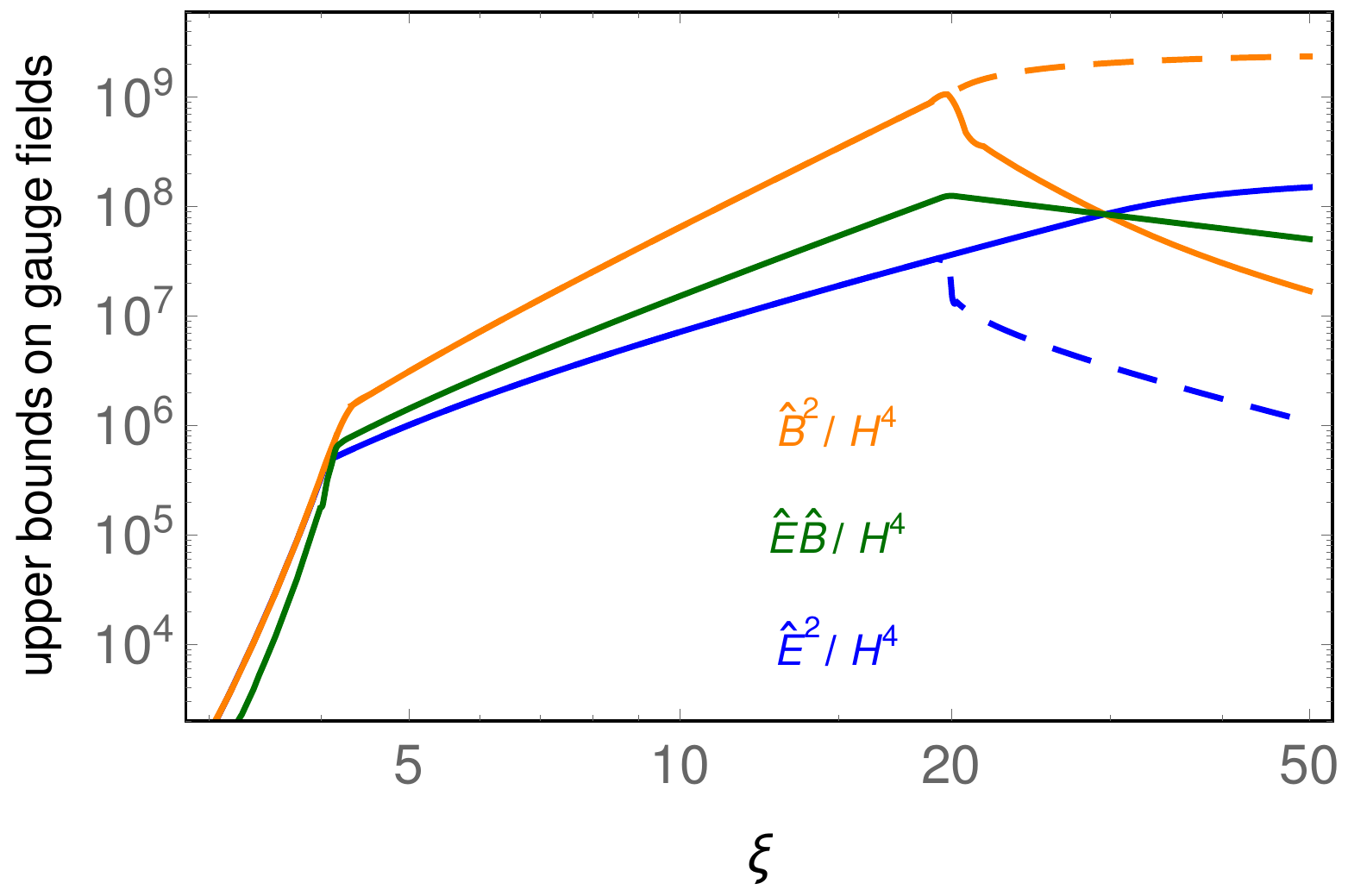}} \hfill
  \subfigure{
  \includegraphics[width = 0.48  \textwidth]{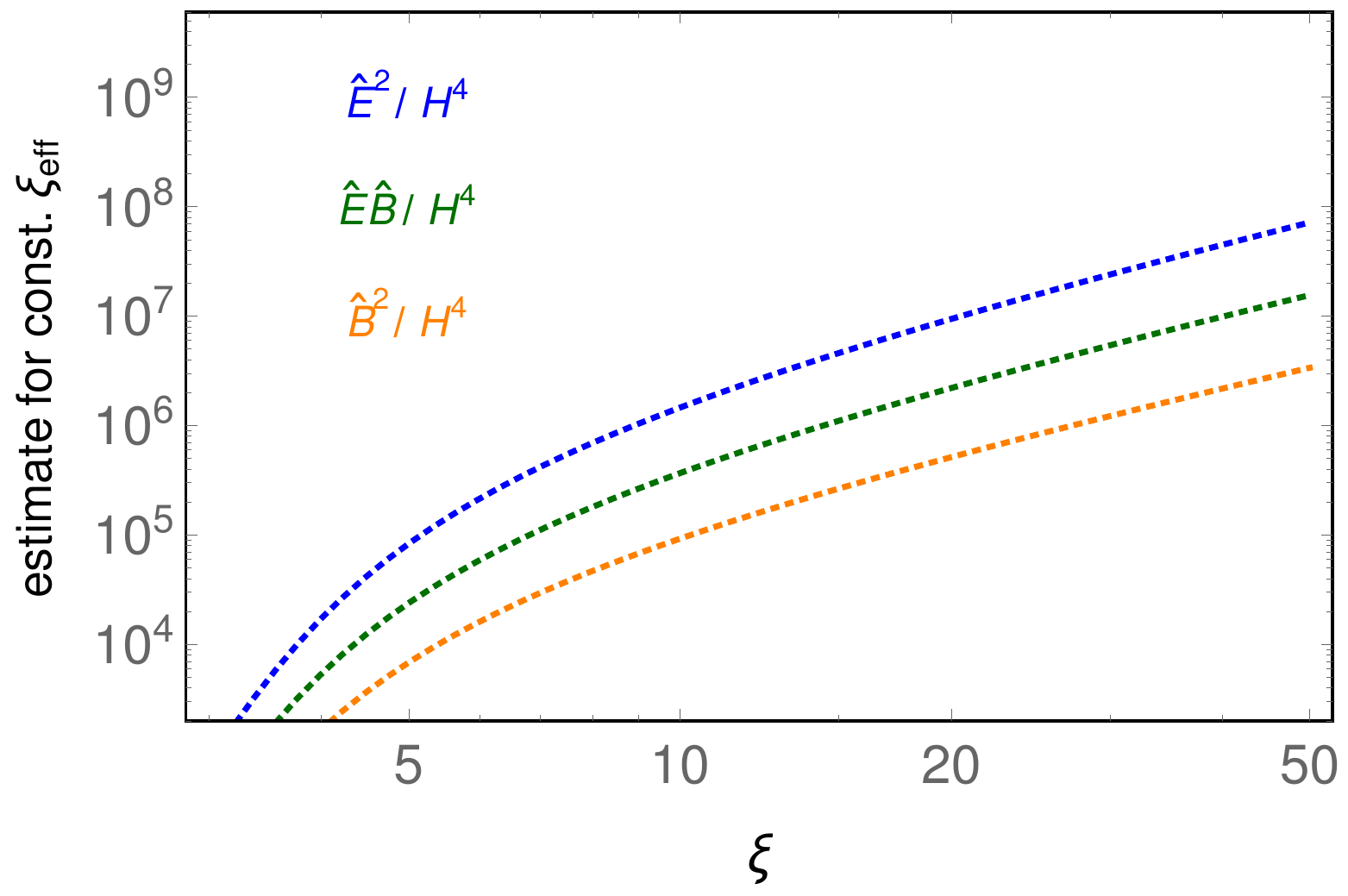}}
 \caption{Magnitude of the gauge fields including backreaction. \textbf{Left panel}: 
 Maximally allowed values for $\hat E$, $\hat B$ and $ {\hat E \hat B}$ requiring the condition to have a stationary solution [Eq.~\eqref{eq:curve}] and energy conservation in slow-roll inflation [Eq.~\eqref{eq:bound1}] for $r = 0.1$.  In the following we will focus left branch in Fig.~\ref{fig:consistency}, indicated by the solid lines. \textbf{Right panel}: Estimate of $\hat E$, $\hat B$ and $ {\hat E \hat B}$ assuming an attractor solution with constant $\xi_\text{eff}$, see Eq.~\eqref{eq:xi_eff}. Here we have set $Q = 1$ and $g = 1/\sqrt{2}$.} 
  \label{fig:EBbounds}
\end{figure}

In summary, we obtain upper limits on the electric/magnetic fields without solving the equation of motion explicitly, cf.\ left panel of Fig.~\ref{fig:EBbounds}. Taking $Q = 1$ and $g = 1/\sqrt{2}$, we numerically determine the maximal values of $E$ and $B$ (independently) allowed by Eq.~\eqref{eq:curve}. For $\xi \lesssim 4$, we  recover Eqs.~\eqref{eq:E2} to \eqref{eq:EB},  implying an exponential growth of the gauge fields as a function of $\xi$. For $\xi \gtrsim 4$, the backreaction becomes important, limiting the growth of the gauge fields. For $4 \lesssim \xi \lesssim 20$, we find that the maximally allowed $B$-field is well described by Eq.~\eqref{eq:EBmax}, whereas the maximally allowed $E$-field is slightly overestimated by this expression. For $\xi \gtrsim 20$, Eq.~\eqref{eq:bound1} becomes relevant, splitting the non-zero solutions of Eq.~\eqref{eq:curve} into two disconnected branches. Both the analytical solution in the absence of backreaction as well as our estimate of $\hat E$ and $\hat B$ for constant $\xi_\text{eff}$ hint towards values of gauge fields on the leftmost of these two branches, \textit{i.e.}, preferring larger values of $\hat E$ and smaller values of $\hat B$. In the following we will thus for definiteness focus on this branch. We have however checked explicitly that the results of this section, in particular the conclusions about thermalization below, do not depend on this choice. For reference, the right panel of Fig.~\ref{fig:EBbounds} shows the estimate obtained by taking $\xi_\text{eff}$ to be constant in Eq.~\eqref{eq:curve}.

In the absence of fermions, values of $\xi \gtrsim 10$ are not reached in slow-roll inflation with an axionic inflation - gauge field coupling, since the backreaction of the gauge fields on the inflaton acts a friction term, limiting the inflaton velocity, see \textit{e.g.}, \cite{Barnaby:2011qe}. As demonstrated above, the backreaction of the fermions however significantly reduces the efficiency of the gauge field production. This results in a power-law instead of an exponential dependence of the generated gauge fields on $\xi$. Consequently, we expect much larger values of $\xi$ to be reached. A first estimate for the maximal value of $\xi$ reached can be obtained as follows: 
In single field slow-roll inflation, where inflation ends at $\varepsilon = \dot \phi^2/(2 H^2 M_P^2) = 1$, the change in $\xi$ is bounded by the tensor-to-scalar ratio~$r$:
\begin{equation}
 \frac{\xi_\text{e}}{\xi_\text{CMB}} = \left( \frac{\varepsilon_\text{e}}{\varepsilon_\text{CMB}} \right)^{1/2} = \left( \frac{16}{r} \right)^{1/2} \,,
\end{equation}
where the indices `e' and `CMB' denote the end of inflation and the time when the CMB modes exited the horizon, respectively. Imposing the upper bound from the non-observation of non-gaussianities in the CMB, $\xi_\text{CMB} < 2.5$~\cite{Barnaby:2011qe}, this yields $\xi_\text{max} \lesssim 32$ in the case of $r = 0.1$. Hence at least for high-scale inflation models, we will be mainly interested in the regime where the gauge fields are bounded by Eq.~\eqref{eq:EBmax}. This is in particular the case for the example we will discuss in Sec.~\ref{sec:inflation}.

\paragraph{Thermalization in the fermion sector.} 
In the analysis above we neglected scattering among the produced fermions. The purpose of this section is to confirm the validity of this approximation.
To study this question, we will in the following assume $\xi \gtrsim 3$ and when giving numerical results set $Q$ and $g$ to the reference values $Q = 1$, $g = 1/\sqrt{2}$. The analysis presented here is based on gauge fields saturating the upper bounds depicted by the solid lines in Fig.~\ref{fig:EBbounds}, however we have checked that the conclusions remain unchanged both when employing the other branch of the upper bounds (dashed line line the left panel of Fig.~\ref{fig:EBbounds}) and when employing the self-consistent estimate shown in the right panel of Fig.~\ref{fig:EBbounds}.

\begin{figure}
\subfigure{
  \includegraphics[width = 0.55 \textwidth]{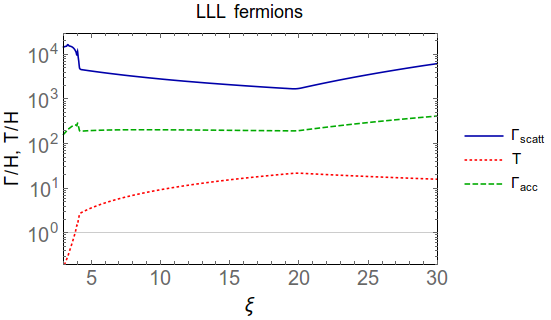}}
 \hfill
\subfigure{
  \includegraphics[width = 0.43  \textwidth]{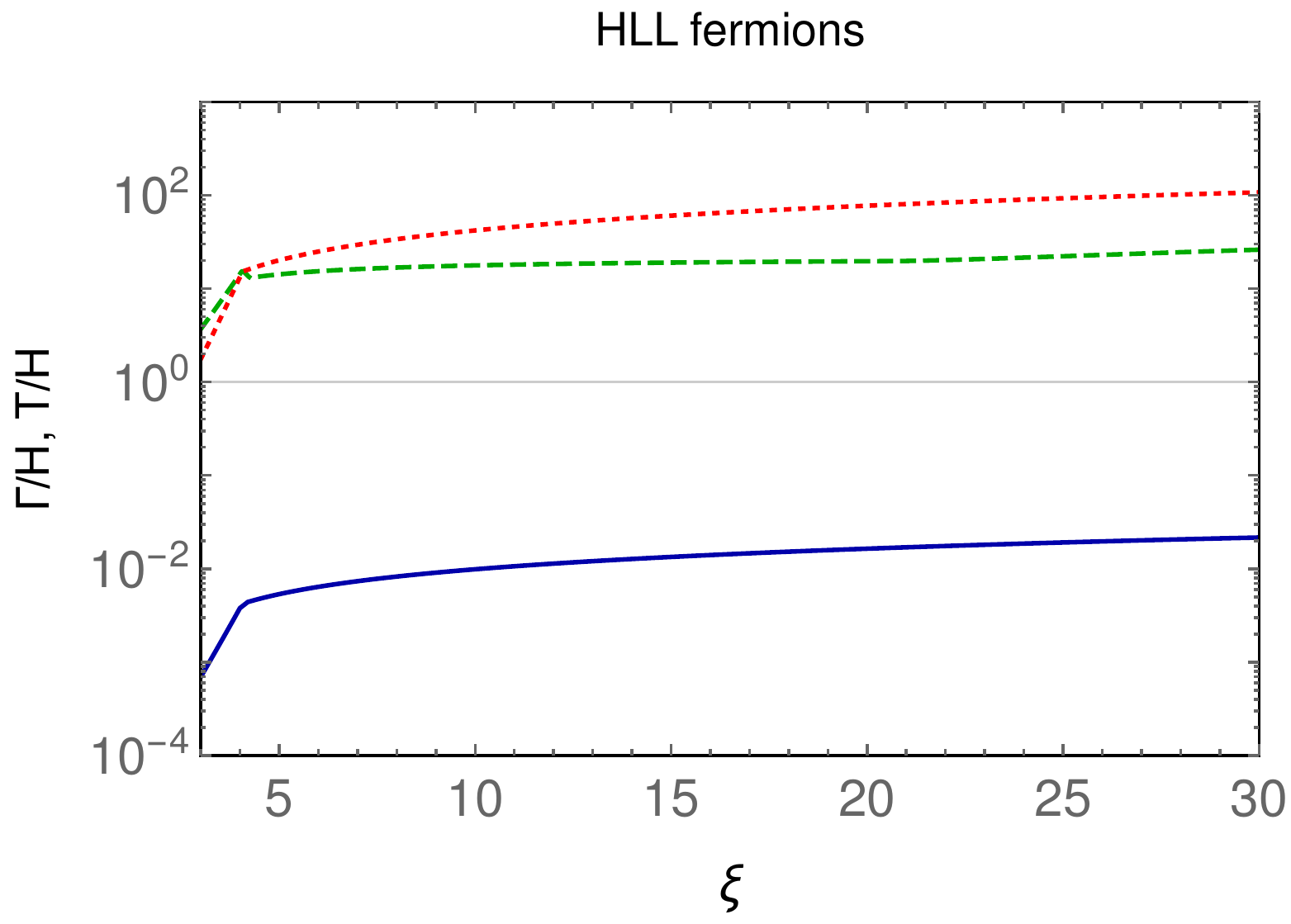}}
  \caption{Scattering rates and (would-be) temperature for the LLL fermions (\textbf{left}) and the HLL fermions (\textbf{right}). }
  \label{fig:rates}
\end{figure}

Let us start with the fermions in the lowest Landau level. We can estimate their scattering rate as
\begin{equation}
 \Gamma^\text{LLL}_\text{sc} = \tau_\text{sc}^{-1} = n_\psi \sigma_\text{sc} \quad \text{with  }  \sigma_\text{sc} = \frac{ 4 \pi \alpha^2}{3 s} \,.
\end{equation}
Here $s$ denotes the center of mass energy and is determined by the acceleration in the electric field, $s = 2(g Q E \tau_\text{sc})^2$ and $n_\psi = \dot n_\psi T$ is determined by Eq.~\eqref{eq:nLLL} with $T$ indicating the duration of continuous fermion production.
Solving for $\tau_\text{sc}$, we obtain
\begin{equation}
 \tau_\text{sc} = \frac{\alpha^2}{3 \pi} \frac{\hat B}{\hat E} \,  T \simeq 1.7 \times 10^{-4} \; \frac{\hat B}{\hat E} \;  H^{-1}\,.
\end{equation}
where in the last step we have set  $T \simeq H^{-1}$. From Fig.~\ref{fig:EBbounds} we expect $\hat B/\hat E \lesssim {\cal O}(5)$ for $\xi < 50$, implying that scattering rate is much faster than the Hubble rate.
Hence the LLL fermions thermalize and we can estimate the temperature of the resulting fermion gas as
\begin{equation}
 (T_\psi^\text{LLL})^4 = \frac{30}{\pi^2 g_*} n_\psi \sqrt{s/2} \,,
\end{equation}
where for the number of relativistic degrees of freedom we take the SM value $g_* = 427/4$.
Both the scattering rate and temperature are depicted in Fig.~\ref{fig:rates} as a function of $\xi$. For completeness we also include the acceleration rate which indicates the time-scale the particle would double its typical (thermal) energy due to acceleration in the electric field in the absence of scattering.

The situation is a bit more complicated for the fermions in the higher Landau levels. The energy density of the produced HLL fermions (before acceleration in the electric field) is given by $\sum_n p_\perp n^{(n)}_\psi$. Inserting Eq.~\eqref{eq:nHLL_n}, we find that the main contribution to the fermion energy density in the HLL (at the time of production) arises from the Landau level with $2 \pi n B/E \simeq 1$. For the range of $\xi$ of interest, this implies that the most import level is the $n=1$ level. Estimating the scattering rate as above, but now taking into account the transverse energy, $s = 2 (p_\perp^2 + (g Q E \tau_\text{sc})^2)$, we find the scattering rate to be highly suppressed compared to the Hubble expansion rate, $\Gamma_\text{sc}/H \sim 10^{-5}$. However, this picture may change when taking into account multiple soft gauge boson scatterings. In a realistic scenario, we expect the presence of both Abelian and non-Abelian gauge fields. We will hence in the following estimate the thermalization by employing the Landau-Pomeranchuk-Migdal scattering rate~\cite{Landau:1953um,Migdal:1956tc} for non-Abelian gauge groups~\cite{Gyulassy:1993hr,Arnold:2001ba,Arnold:2002ja,Kurkela:2011ti},\footnote{
	See also \cite{Harigaya:2013vwa,Mukaida:2015ria,Ellis:2015jpg} in the context of reheating after inflation.
}
\begin{equation}
 \Gamma_\text{sc}^\text{HLL} = \tau_\text{sc}^{-1} = \begin{cases}
                                            \alpha^2 T_\text{wb} & \text{for } T_\text{wb} > \omega \\
                                            \alpha^2 T_\text{wb} \left(T_\text{wb}/\omega\right)^{1/2}
                                            & \text{for } T_{wb} < \omega
                                           \end{cases} \,,
\end{equation}
where $T_\text{wb}^4 = 30/(\pi^2 g_*) n_\psi \omega$ denotes the 'would-be' temperature if the fermions did thermalize and $\omega = \sqrt{s/2}$.  In the parameter range of interest, we have $T_\text{wb}/\omega < 1$. The resulting scattering rate, together with the temperature $T_\text{wb}$ is depicted in the right panel of Fig.~\ref{fig:rates}.  In addition, we also show the acceleration rate $\Gamma_\text{acc}$, which indicates the time scale for the acceleration due to the $E$-field to overcome the initial energy $p_\perp$. For the entire $\xi$-range of interest, the scattering rate is negligible compared to both the Hubble expansion and the acceleration rate, indicating that contrary to the LLL fermions, the kinetic energy of the HLL fermions is dominated by acceleration in the electric field and not by random thermal motion.

\begin{figure}
\subfigure{
  \includegraphics[width = 0.48 \textwidth]{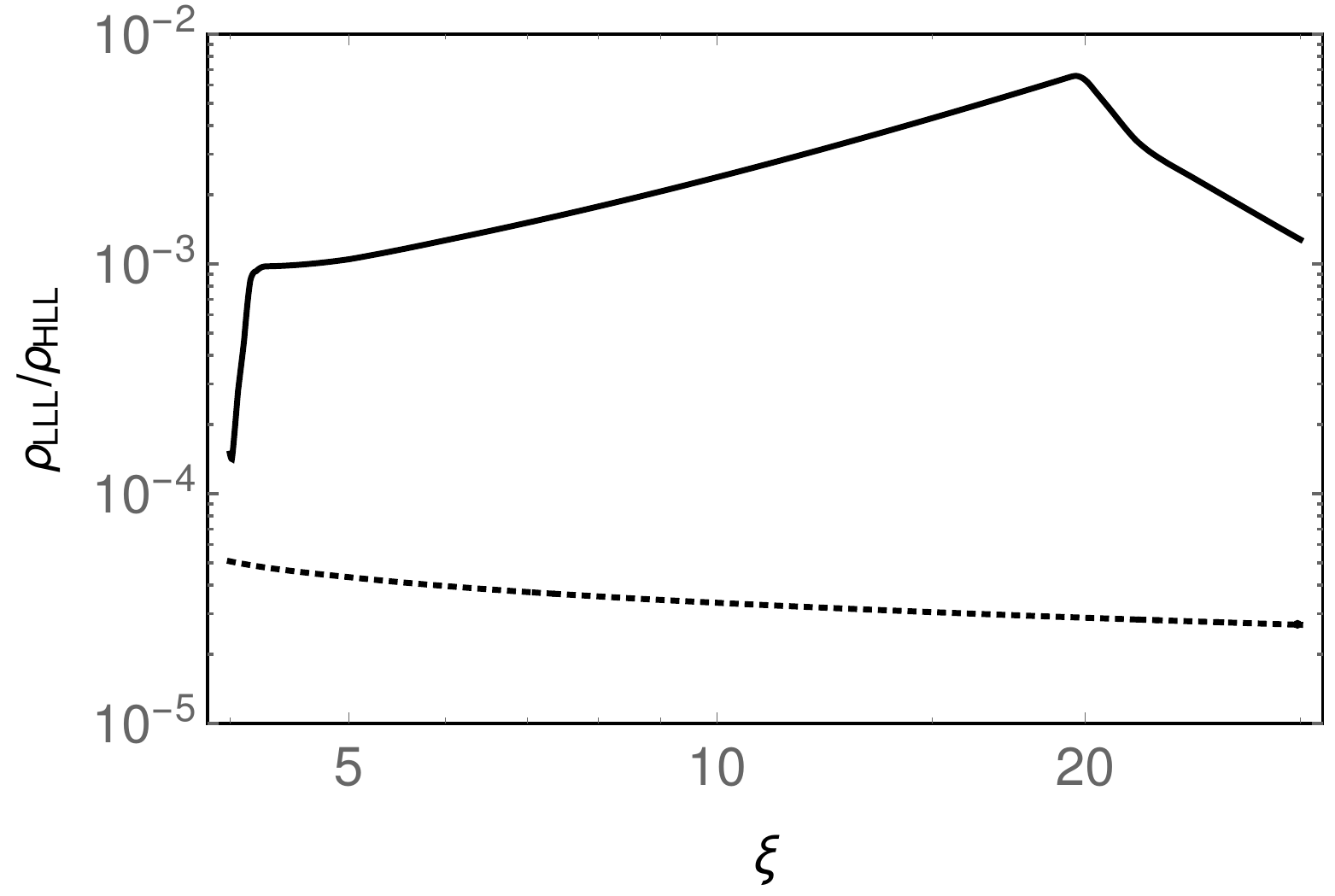}}
 \hfill
\subfigure{
  \includegraphics[width = 0.48  \textwidth]{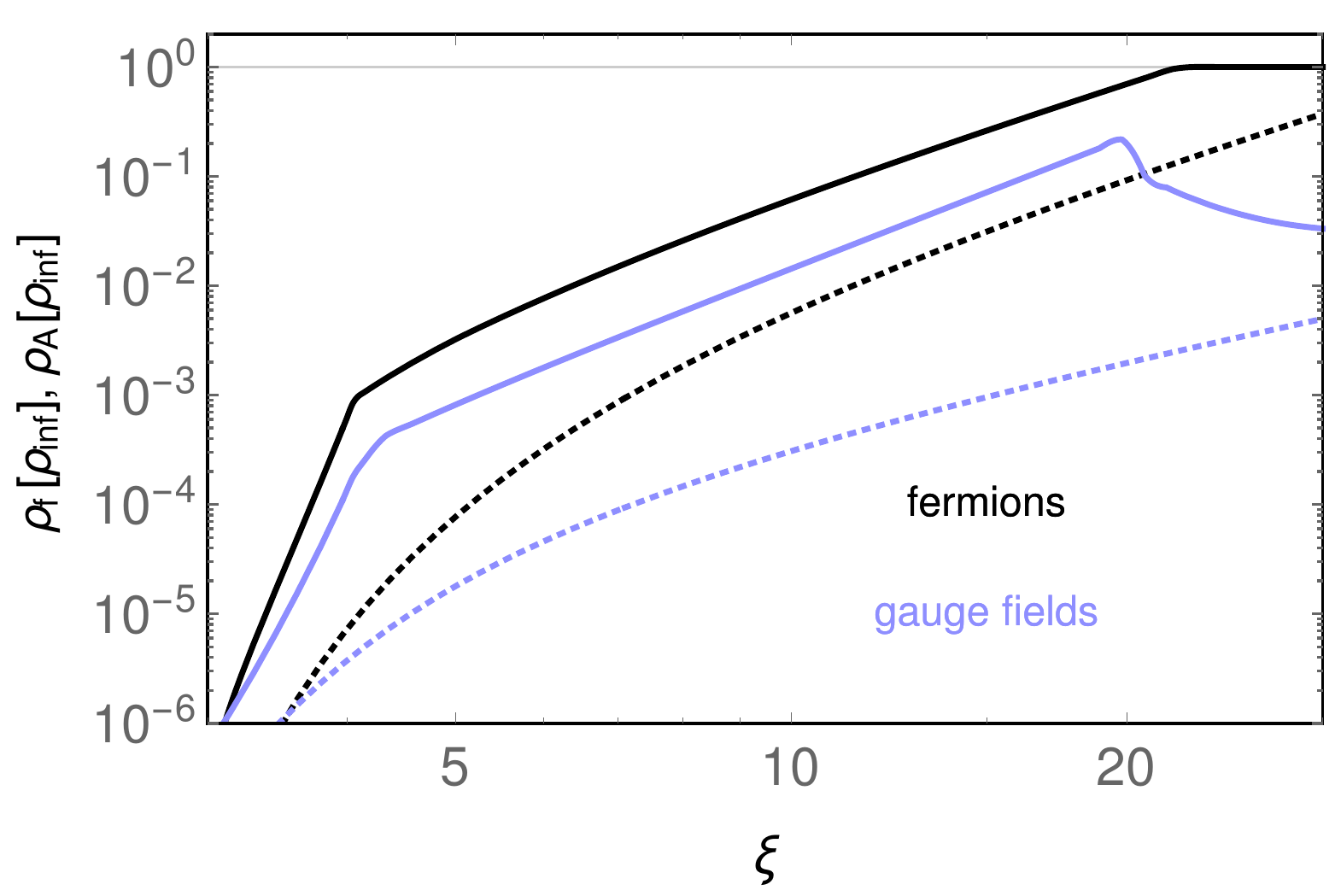}}
  \caption{Relative energie densities. \textbf{Left panel}: fraction of the fermion energy density in the lowest Landau level compared to the $n=1$ level. \textbf{Right panel}: upper bounds on energy density stored in fermions and gauge fields. In both panels, the solid curves correspond to gauge fields saturating the upper bonds (see left panel of Fig.~\ref{fig:EBbounds}, whereas the dotted curves correspond to the estimate of an attractor solution with constant $\hat E$ and $\hat B$ (see right panel of Fig.~\ref{fig:EBbounds}). }
  \label{fig:rhoLLLvsHLL}
\end{figure}

With these results, we can now quantify the relative energy density in the LLL compared to the HLL ($n=1$) for values of the gauge fields saturating the bound in Fig.~\ref{fig:EBbounds}. We show this in the left panel of Fig.~\ref{fig:rhoLLLvsHLL}. Clearly the HLL population  is responsible for nearly the entire fermion energy. Note that due to energy conservation, here we have imposed that the maximal acceleration of fermions is limited by the energy scale of inflation. The relative number densities in the LLL and HLLs are given by Eq.~\eqref{eq:nLLL} and \eqref{eq:nHLL}. Our estimate of the gauge field magnitudes (see right panel of Fig.~\ref{fig:EBbounds}) indicates $E B \ll E^2$, implying that the total number density is dominated by the HLL population. In addition, the efficient pair annihilation in the LLL will reduce this number density compared to the result of Eq.~\eqref{eq:nLLL}. In summary, the induced current is dominated by the HLL fermions (which exhibit negligible scattering rates), thus a posteriori justifying our assumption of omitting the fermion scattering above.

Finally, the right panel of Fig.~\ref{fig:rhoLLLvsHLL} shows upper bounds on the energy density of the gauge field and the fermion as a function of $\xi$.
The energy density of the fermion is always larger than that of the gauge field. 
This implies that the gauge field converts
most of its energy to the fermions.

The effects of fermion production was recently studied in a similar context in Ref.~\cite{Ferreira:2017lnd}: if the gauge fields obtain a thermal mass,  their amplitude grows as a power-law of $\xi$ instead of exponentially with $\xi$, similarly to what we find here. Note however that the transverse mode of U(1) gauge theory, which exhibits the tachyonic instability, never acquires the magnetic mass, and hence a naive application of their analysis to our case is not clear. In addition, here we find (within the approximations of our study) that the dominant contributions of the fermion population (the HLL fermions) do not thermalize efficiently since their large kinetic energy suppresses the scattering cross section. Nevertheless, the induced current limits the growth of the gauge fields to the expressions given in Eq.~\eqref{eq:EBmax}.

\section{Phenomenological implications}
\label{sec:pheno}

\subsection{Inflation \label{sec:inflation}}

The pseudo-scalar coupling $\phi F \tilde F$ as in Eq.~\eqref{eq:frame1} is the key ingredient of phenomenologically rich inflation model discussed \textit{e.g.},  in Ref.~\cite{Barnaby:2010vf}. In the absence of fermions, the tachyonic instability in the gauge field equation of motion leads to an exponential production of low-momentum gauge fields~\cite{Turner:1987vd,Garretson:1992vt, Anber:2006xt}. These gauge fields form an additional, classical source for scalar and tensor perturbations, dramatically modifying the predictions obtained from the usual vacuum contributions only. This leads to a large range of possible observable consequences: non-gaussianities in the scalar power spectrum in the CMB~\cite{Barnaby:2010vf,Barnaby:2011qe,Barnaby:2011vw}, a distortion of the CMB black body spectrum~\cite{Meerburg:2012id}, primordial black black hole (PBH) production~\cite{Linde:2012bt, Garcia-Bellido:2016dkw, Domcke:2017fix} and an enhanced chiral gravitational wave signal in the frequency band of LIGO and LISA~\cite{Cook:2011hg,Barnaby:2011qe,Barnaby:2011vw,Anber:2012du,Domcke:2016bkh,Bartolo:2016ami}. This provides a unique window to probe a coupling of the inflaton to other (\textit{e.g.,} Standard Model) particles. In this section we show that the presence of massless fermions, related to the gauge fields through the anomaly equation~\eqref{eq:AnomalyIntro}, significantly changes some of these predictions.

The slow-roll equation of motion for the homogeneous inflaton field is\footnote{Here we are using the convention $\phi < 0$, $\dot \phi > 0$ and hence $\langle \hat 
{\bm E} \hat{\bm B} \rangle > 0$. }
\begin{equation}
 \ddot \phi + 3 H \dot \phi + V'(\phi) = -  \frac{\alpha}{\pi f_a} \langle \hat 
{\bm E} \cdot \hat{\bm B} \rangle \,.
\label{eq:eom_inflaton}
\end{equation}
Here $ \langle \hat {\bm E} \cdot \hat{\bm B} \rangle $ is directly related to the chiral current and the Chern-Simons charge,
\begin{equation}
  a^4(\eta) \langle \hat {\bm E} \cdot \hat{\bm B} \rangle  = - \frac{1}{4} \langle F_{\mu \nu} \tilde F^{\mu \nu} \rangle = - \frac{\pi}{2 \alpha Q^2} \langle \partial_\mu J^\mu_5 \rangle =  \frac{\pi}{2 \alpha} \langle \partial_\mu K^\mu_\text{CS} \rangle  =  \frac{\pi}{2 \alpha}  \partial_0 q_\text{CS} \,.
\end{equation}
Consequently, in the presence of fermions, it suffices to implement the modified dependence of $ \langle \hat {\bm E} \cdot \hat{\bm B} \rangle $ on $\xi$, Eq.~\eqref{eq:eom_inflaton} is insensitive to the distribution of the chiral charge between the gauge field and fermion sector. Decoupling the equations of motion of the inflaton and the gauge fields in this way relies on the assumption that the variation of $\xi$ is small, which is well justified up to the very last few e-folds of inflation (slow-roll approximation).
\begin{figure}
\subfigure{
  \includegraphics[width = 0.48 \textwidth]{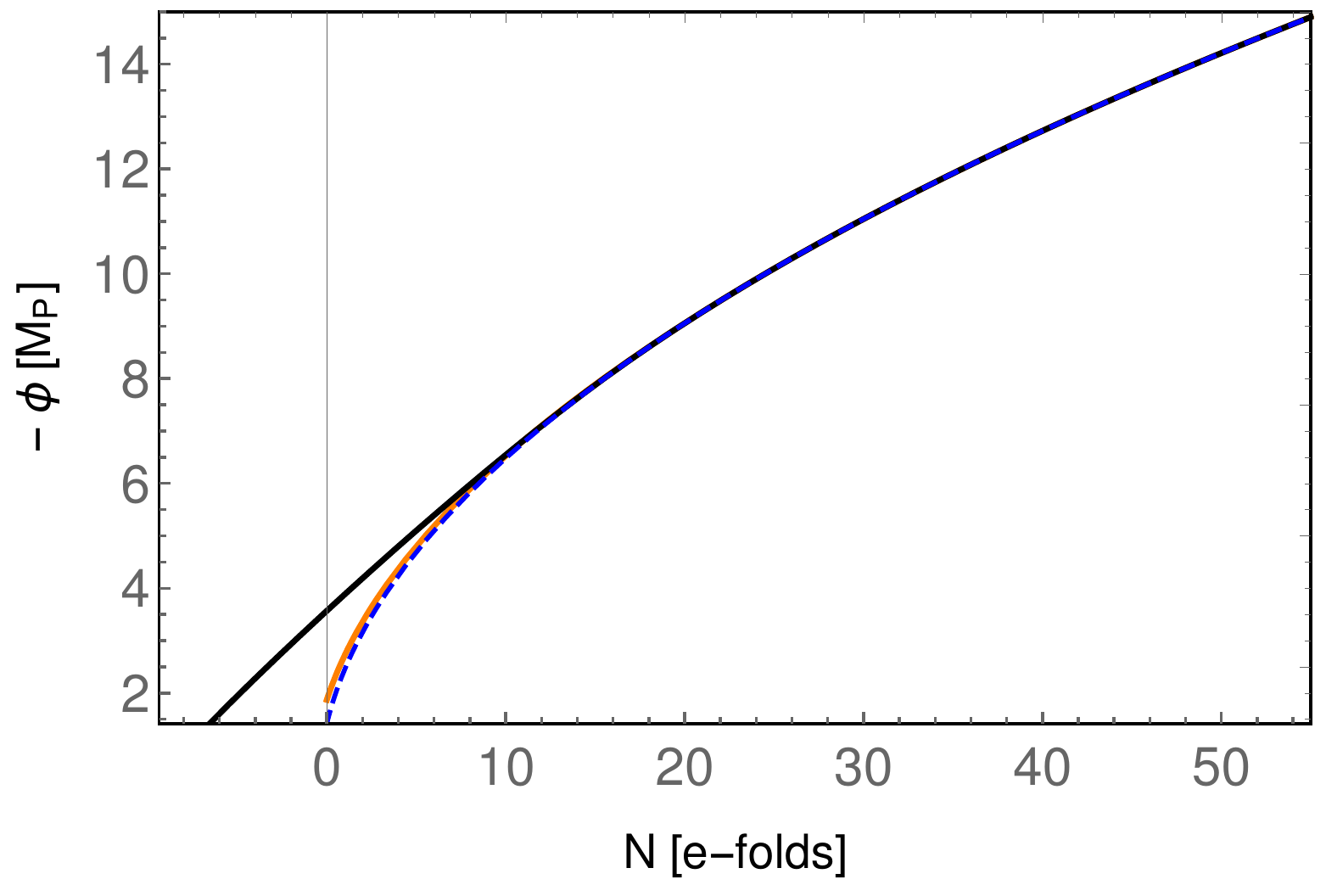}}
 \hfill
\subfigure{
  \includegraphics[width = 0.48  \textwidth]{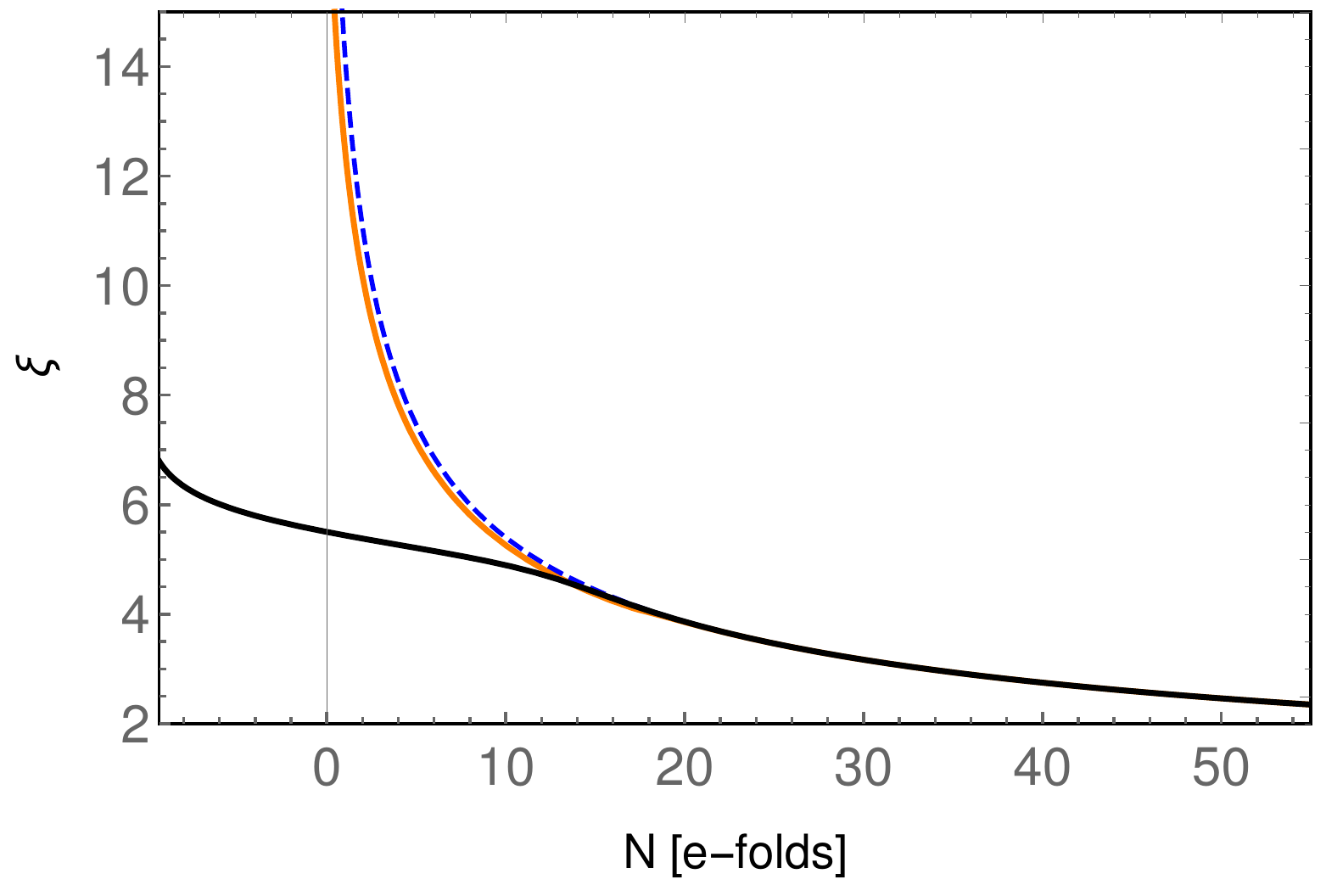}}
  \caption{Evolution of the homogeneous inflaton in the absence of any couplings (dashed blue), including gauge fields (black) and including gauge fields and fermions (orange).  The parameters chosen for this figure are $m = 6.5 \times 10^{-6} \, M_P$, $Q = 1$, $\alpha / (\pi f_a) = 35$ and ${\cal N} = 8 $. }
  \label{fig:inflaton}
\end{figure}

 Fig.~\ref{fig:inflaton} shows the evolution of the homogeneous inflaton background in a scalar potential\footnote{Despite being disfavoured by the latest Planck data, we use this reference model for simplicity. Agreement with the Planck data (\textit{i.e., } a reduced tensor-to-scalar ratio) can easily be obtained by adding a small negative quartic term, which would not impact the main results presented here.} $V(\phi) = \frac{1}{2} m^2 \phi^2$ for three cases, taking into account (i) the inflaton only (\textit{i.e.}, $\alpha = 0$), (ii) the inflaton and gauge fields and (iii) the inflaton, gauge fields and fermions. In the latter two cases we have chosen the maximal value of the inflaton-gauge field coupling in accordance with current bounds on the non-gaussianity of the scalar power spectrum ($\alpha / (\pi f_a) = 35$, corresponding to $\xi_\text{CMB} = 2.5$). In the absence of fermions, the gauge field production is exponentially sensitive to the inflaton velocity (encoded in $\xi$). The right-hand side of Eq.~\eqref{eq:eom_inflaton} thus acts as a very efficient friction term once $\xi$ crosses some critical value. This leads to the flattening of the growth of $\xi$ and it also delays the end of inflation by about 6 e-folds. In the presence of fermions, the backreaction reduces the efficiency of the gauge field production. The resulting upper bound on $ \langle \hat {\bm E} \cdot \hat{\bm B} \rangle $ is determined by Eq.~\eqref{eq:curve}, see also Fig.~\ref{fig:EBbounds}. (Note that for the values of $\xi$ arising in Fig.~\ref{fig:inflaton}, the bound~\eqref{eq:bound1} is irrelevant.) For $\xi \gtrsim 4$, the exponential dependence on $\xi$ reduces to a power-law dependence. Correspondingly the resulting `friction' in Eq.~\eqref{eq:eom_inflaton} is reduced and the backreaction of the gauge field on the inflaton becomes negligible over the entire course of inflation. Interestingly, this implies that in the presence of fermions, this setup is less sensitive to the theoretical uncertainties of the strong backreaction regime, which (in the absence of fermions) may become important at $\xi \gtrsim 4.7$~\cite{Peloso:2016gqs}.

Taking $\dot \xi$ to be small, the equation of motion for the inflaton fluctuations $\delta \phi$ can be estimated as~\cite{Linde:2012bt}
\begin{equation}
 \delta \ddot{\phi} + 3 \beta H \delta \dot{\phi} - \frac{\nabla^2}{a^2} \delta \phi + V_{,\phi \phi}(\phi) \delta \phi = -  \frac{\alpha}{\pi f_a} \left(\hat{\bm E} \cdot \hat{\bm B} -  \langle \hat 
{\bm E} \cdot \hat{\bm B} \rangle \right) \,,
\end{equation}
with $\beta = 1 + 2 \xi \alpha \vev{  \hat{\bm E} \cdot \hat{\bm B} } /(3 \pi H \dot \phi f_a)$. The total scalar power spectrum then reads
\begin{equation}
 \Delta_s^2 = \frac{H^2}{\dot \phi^2} \langle \delta \phi^2 \rangle  \simeq 
 \left( \frac{H^2}{2 \pi \dot \phi} \right)^2 + 
 \left( \frac{\alpha \langle  \hat{\bm E} \cdot \hat{\bm B} \rangle }{3 \pi \beta H \dot \phi f_a \sqrt{{\cal N}}} \right)^2 \,,
\end{equation}
where the first term is the standard vacuum contribution and the second term in sourced by the Chern-Simons charge. In the limit of large gauge fields,
\begin{equation}
 \beta \approx \frac{2 \xi \alpha \vev{  \hat{\bm E} \cdot \hat{\bm B} }}{3 \pi H \dot \phi f_a} \,,
 \label{eq:betaApprox}
\end{equation}
and we obtain the simple  expression
\begin{equation}
 \Delta_s^2 \simeq \frac{1}{{\cal N} (2 \pi \xi)^2} \,.
 \label{eq:Ds_limit}
\end{equation}
Here we have introduced the parameter ${\cal N}$ to denote the number of Abelian gauge groups (or equivalently the number generators of weakly coupled non-Abelian gauge groups)~\cite{Anber:2012du,Domcke:2016bkh}. As can be seen from Eq.~\eqref{eq:Ds_limit}, this suppresses the amplitude of the scalar power spectrum, avoiding a dangerous overproduction of primordial black holes~\cite{Linde:2012bt}. Alternatively, this parameter may be seen as a parametrization of the theoretical uncertainties in the calculation of the scalar power spectrum due the breakdown of perturbation theory in the strong backreaction regime~\cite{Ferreira:2015omg,Peloso:2016gqs}.

\begin{figure}
 \centering
 \includegraphics[width = 0.6 \textwidth]{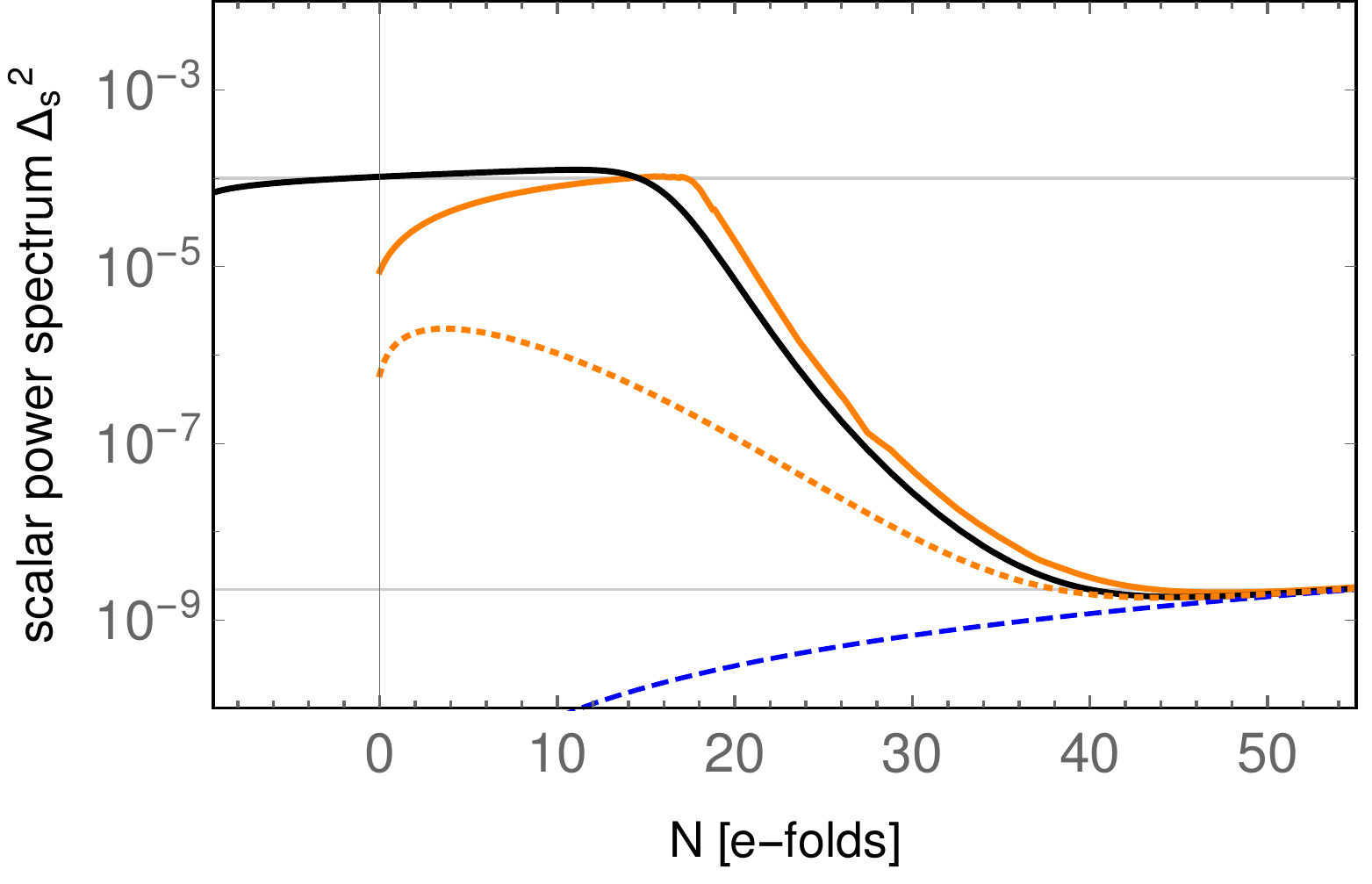}
 \caption{Scalar power spectrum in the absence of any couplings (dashed blue), including gauge fields (black) and including gauge fields and fermions (orange). The solid orange curve corresponds to gauge fields saturating the upper bound obtained from Eq.~\eqref{eq:curve}, the dotted orange curve corresponds to the estimate for the gauge fields based on an attractor solution with constant gauge fields, cf.~\eqref{eq:xi_eff}.
 Color-coding and parameters as in Fig.~\ref{fig:inflaton}.}
 \label{fig:spectrum}
\end{figure}

Fig.~\ref{fig:spectrum} shows the scalar power spectra with/without gauge fields and with/without fermions. The correct normalization at the CMB scales ($N = 55$) is ensured by our choice of $m$ in the scalar potential (lower horizontal gray line). The inflaton - gauge field coupling leads to a dramatic enhancement of the scalar power spectrum at small scales. In the absence of fermions, Eq.~\eqref{eq:Ds_limit} leads to nearly scale-invariant spectrum. In the presence of fermions, the value of $\xi$ grows rapidly towards the end of inflation since the friction term in the equation of motion for the inflaton is less efficient. Consequently, Eq.~\eqref{eq:Ds_limit} leads to a rapid drop of the scalar power spectrum as $N \rightarrow 0$. The upper horizontal gray line in Fig.~\ref{fig:spectrum} indicates (a rough estimate of) the critical threshold of PBH formation. Note that a large non-Gaussianity of this model slightly reduces the threshold value for creating PBHs, and this effect is already taken into account in Fig.~\ref{fig:spectrum}. See \textit{e.g.}, Ref.~\cite{Domcke:2017fix} for a more detailed analysis. In the absence of fermions, strong bounds on the presence of relatively light PBHs $(\sim 10^{15}~\text{g})$ exclude the possibility of arranging for any significant PBH component of dark matter. In the presence of fermions and assuming the gauge fields saturate the upper bound given by Eq.~\eqref{eq:curve}, 
the scalar power spectrum is suppressed, barely touching the PBH bound without the need to employ more than ${\cal N} = 1$ U(1) gauge fields. Moreover, we note that a more peaked spectrum arises, opening up the possibility to evade the strong bounds on light PBHs while obtaining a population of heavier black holes with could contribute significantly to dark matter. This effect becomes stronger if one goes to scalar potential which come with a stronger acceleration of the inflaton $\xi \propto (1 + N)^{-p/2}$ with $p > 1$~\cite{Domcke:2016bkh}. If the gauge fields do not saturate this upper bound but are instead given by the self-consistent estimate for constant $\hat E$ and $\hat B$, cf.~Eq.~\eqref{eq:xi_eff}, we do not reach the regime indicated in Eq.~\eqref{eq:betaApprox} and the scalar power spectrum is significantly suppressed. Obviously, also this result is extremely relevant for PBH production.

The source term for the tensor power spectrum is the transverse traceless part of the energy momentum tensor in the linearized Einstein equation. This is not immediately related to one of our conserved charges, and hence the distribution of the energy density between the gauge sector and fermion sector becomes important. We leave a full computation of the tensor power spectrum to future work, and restrict ourselves here to a simple and very conservative estimate: assuming that the fermions do not source any gravitational waves, and taking $\hat E$ and $\hat B$ to be approximately constant for $\xi \geq 4$, we can estimate the amplitude of gravitational waves by evaluating the corresponding expression in the absence of fermions at $\xi = 4$~\cite{Barnaby:2011qe,Barnaby:2011vw}:
\begin{equation}
 \Omega_\text{GW}  |_{\xi \geq 4} >  \Omega_\text{GW} (\xi = 4)|_\text{no fermions} = \left. \frac{\Omega_r}{12}  \left( \frac{H}{\pi M_P } \right)^2 \left( 1 + 4.3 \cdot 10^{-7} \, {\cal N} \frac{H^2}{M_P^2} \xi^{-6} e^{4 \pi \xi} \right)\right|_{\xi = 4} \,,
\end{equation}
where $\Omega_r = 8.6 \cdot 10^{-5}$ denotes the radiation energy density today.
For the model we are considering here, this yields $\Omega_\text{GW} h^2 > 10^{-13}$, which is about two orders of magnitude above the vacuum contribution and within the sensitivity range of the planned space-based interferometer LISA~\cite{Audley:2017drz}. A more detailed study of this signal, including also possible contributions from the chiral fermion sector via the gravitational anomaly~\cite{Anber:2016yqr},\footnote{
	Roughly speaking, this is the opposite process as discussed in gravi-leptogenesis~\cite{Alexander:2004us,Adshead:2017znw}.
} would thus be very interesting.

\subsection{Leptogenesis}

As is well known, the SM fermion exhibits chiral anomaly which renders the global $B + L$ symmetry anomalous. This opens up the possibility that the primordial helical gauge field/chiral asymmetry generated during inflation could be eventually converted into the  baryon asymmetry of the Universe,
if it survives the wash-out induced by the electroweak Sphaleron processes.
In the following, we briefly discuss leptogenesis as one of the interesting phenomenological applications originating from the pseudo-scalar coupling.
{Determining the final net baryon asymmetry of the Universe requires solving kinetic equations for all SM particles, including in particular the wash-out by the Sphaleron processes and the effect of  Yukawa interactions. We postpone this challenging task to future work, instead illustrating by means of our toy model the qualitatively new effects compared to related studies (see \textit{e.g.}, \cite{Giovannini:1997eg,Boyarsky:2012ex,Kamada:2016cnb,Kamada:2016eeb,Jimenez:2017cdr,Kamada:2018tcs}), focusing in particular to the impact of the chiral charge $q_5$ generated during inflation.  }

\begin{figure}
 \centering
 \includegraphics[width = 0.6 \textwidth]{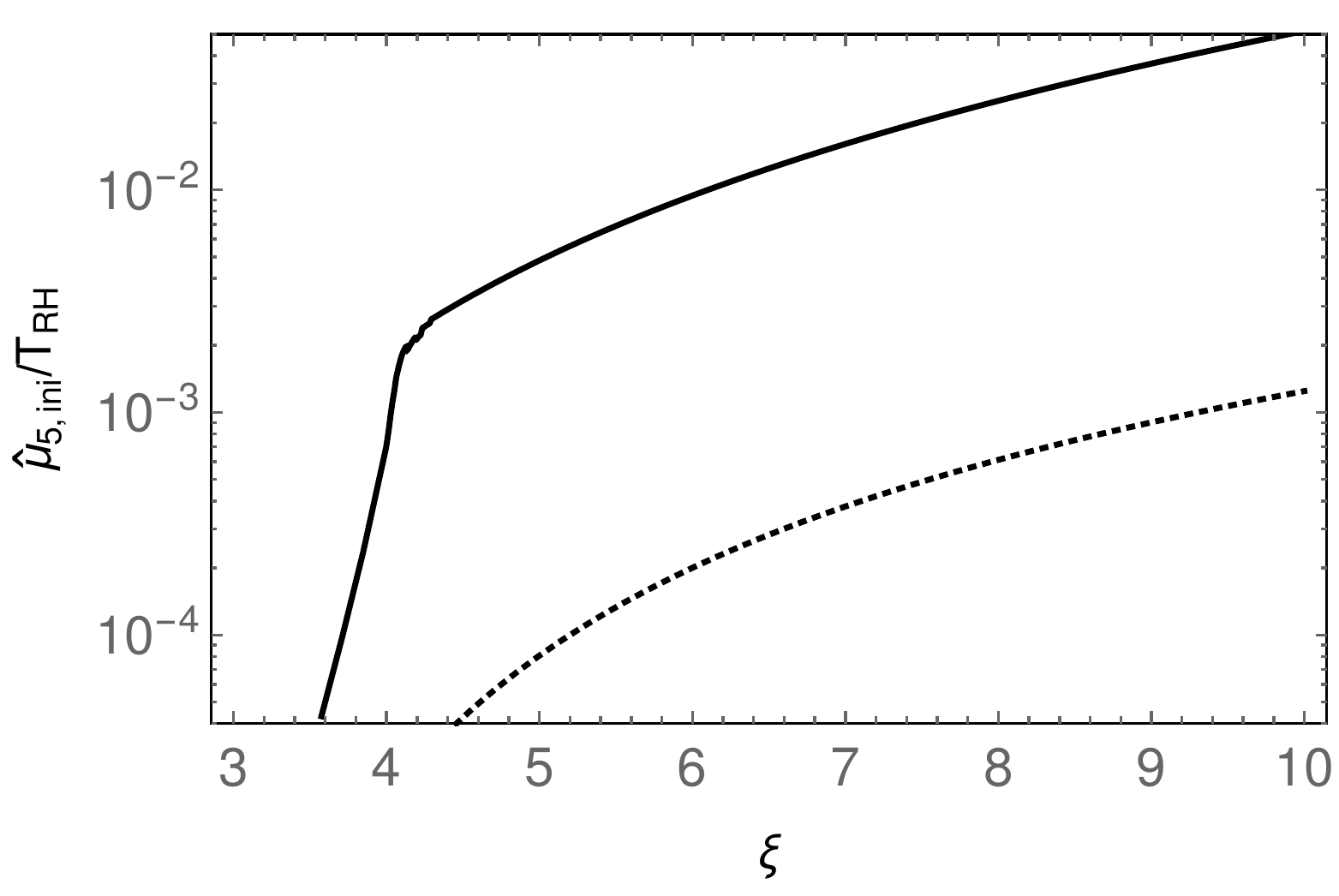}
 \caption{The chemical potential over the temperature right after inflation as a function of $\xi$. The solid line shows the upper bound obtained from Eqs.~\eqref{eq:curve} and \eqref{eq:muoverT}. The dotted line corresponds to the estimate based on the assumption of an attractor solution with constant gauge fields, cf.~\eqref{eq:xi_eff}. }
 \label{fig:chemical_pot}
\end{figure}

As a crude approximation, we assume instant reheating and thermalization after inflation.
The plasma is characterized by the temperature $\hat T$ and chemical potential $\hat \mu_5$.
Note that both are diluted by the cosmic expansion as $\hat T = T / a$ and $\hat \mu_5 = \mu_5 / a$.
Throughout this section, we take $Q = 1$ to avoid unnecessary complications.
The temperature right after inflation is $\hat T_\text{ini} \sim \sqrt{\Mpl H_\text{inf}}$.
The chiral chemical potential right after inflation can be estimated by using the relation $\hat q_5 \sim \hat \mu_5 \hat T^2$ which holds in kinetic equilibrium for $\hat \mu_5 < \hat T$:
\begin{align}
	\hat \mu_{5, \text{ini}} \sim \left. \frac{\hat q_5}{\hat T^2} \right|_\text{ini}
	\sim 
	\frac{\alpha}{\pi} \frac{\vev{\hat{\bm{E}} \cdot \hat{\bm{B}}}_\text{ini}}{H_\text{inf}^2 \Mpl}.
	\label{eq:muoverT}
\end{align}
If one inserts the analytic solution given in Eq.~\eqref{eq:EB0}, the estimate of Eq.~\eqref{eq:Q5_dS} is recovered.
However, notice that this solution is not valid for $\xi \gtrsim 4$, because of the backreaction as discussed in Sec.~\ref{sec:br}.
Fig.~\ref{fig:chemical_pot} shows the upper bound of $\hat \mu_5 / \hat T$ as a function of $\xi$ and also the rough self-consistent estimate depicted as a blue circle in Fig.~\ref{fig:consistency}.

The question is how the system would evolve with this specific initial condition.
We further assume that the system is in the regime of magnetohydrodynamics (MHD), which suffices in the most cases.\footnote{
	Note that the MHD description holds if $\mu_5 / T < \alpha$.  See Ref.~\cite{Figueroa:2017hun} for a nice summary of the range of the validity.
}
Then one may estimate the induced current by
\begin{align}
	g \vev{\bm{J}_\psi} \simeq a \hat \sigma \bm{E} + \frac{\alpha}{\pi} 2 a \hat \mu_5 \bm{B},
	\label{eq:MHDcurrent}
\end{align}
with the electric conductivity being~\cite{Baym:1997gq,Arnold:2000dr}
\begin{align}
	\hat \sigma \sim \frac{\hat T}{\alpha}.
\end{align}
The first term  in Eq.~\eqref{eq:MHDcurrent} is just  Ohm's law and the second one comes from the chiral magnetic effect~\cite{Fukushima:2008xe}.
Here we have neglected the fluid velocity for simplicity.\footnote{
	However note that the velocity field grows due to the Lorentz force, which leads to  turbulence and may affect the dynamics~\cite{Schober:2017cdw,Kamada:2018tcs,Schober:2018ojn}. We leave the detailed discussion for our future work.
	\label{ft:velocity}
} 
Let us recall that the comoving $\bm{E}$ and $\bm{B}$ remain constant against the cosmic expansion while the physical ones, $\hat{\bm{E}}$ and $\hat{\bm{B}}$, decay via $\hat{\bm{E}} = \bm{E} / a^2$ and $\hat{\bm{B}} = \bm{B} / a^2$.
The equations governing MHD read
\begin{align}
	\label{eq:mhd_b}
	\frac{\partial}{\partial \eta} \bm{B} &\simeq
	\frac{1}{a \hat \sigma}
	\left(
		\bm{\nabla}^2 \bm{B} + \frac{2 \alpha}{\pi} a \hat \mu_5 \bm{\nabla} \times \bm{B}
	\right), \\
	0 & \simeq \bm{\nabla} \times \bm{B}  - a \hat \sigma \bm{E} - \frac{2 \alpha}{\pi} a\hat \mu_5 \bm{B}.
\end{align}
The notable difference with respect to  Eq.~\eqref{eq:helical_vac} is that 
the opposite helicity of the gauge field exhibits the instability due to the primordial $\hat \mu_5$.
Thus, this process erases the primordial helicity of the gauge field generated during inflation as well as the primordial chiral asymmetry.
We can see this by expanding the gauge field in the same polarization  modes defined below Eq.~\eqref{eq:Adecom} and rewrite Eq.~\eqref{eq:mhd_b}:
\begin{align}
\label{eq:mhd_inst}
	\frac{\partial}{\partial \eta} \bm{B}_{\pm} (\eta, \bm{k})
	\simeq - \frac{1}{a \hat \sigma} \left( k^2 \mp \frac{2 \alpha}{\pi} a \hat \mu_5 \right) \bm{B}_\pm (\eta, \bm{k}).
\end{align}
In the absence of $\hat \mu_5$, the magnetic field just decays by the diffusion $k^2/a \hat\sigma$ which becomes slow for a large conductivity, \textit{i.e.,} high temperature. The presence of $\hat \mu_5$ may activate a more violent process as we discuss below. 
Recalling that $\hat \mu_5 \gtrless 0$ for $\dot \phi \gtrless 0$ right after inflation [see Eq.~\eqref{eq:Q5_dS} for instance], 
one can see that $B_\lambda$ exhibits the instability with $\lambda = \pm$ for $\dot \phi \gtrless 0$,
which has an opposite polarization to the solution during inflation \eqref{eq:gauge_desitter}.
We can see this directly by keeping the CS term sourced by $\dot \phi \neq 0$ while assuming the MHD approximation.
Then, we readily find an additional term with the opposite sign, $- (\alpha a \dot \phi / \pi f_a) \bm{\nabla}\times \bm{B}$, in the right-hand-side of Eq.~\eqref{eq:mhd_b}.

Note that this behavior is expected from the viewpoint of the conservation laws discussed in Sec.~\ref{sec:overall}.
The background of $\dot \phi \neq 0$ can be regarded as an external chemical potential.
In the presence of $\dot \phi \neq 0$, the system would like to approach $2 \hat \mu_5 \to \dot \phi / f_a$
by creating helical gauge fields and correspondingly chiral fermions
to fulfill the anomaly equation.
However, after inflation, the external chemical potential vanishes.  The system then tries to move back to $\hat \mu_5 \to 0$ by
erasing the chemical potential and helical gauge fields since there is no external driving force.
Since our initial condition at inflation is $q_5 = q_\text{CS} = 0$, the anomaly equation allows the complete erasure of the helical gauge fields and of the chiral charge.

As a final remark of this section, let us estimate the time scale of this erasure process by $\hat \mu_5$.
From Eq.~\eqref{eq:mhd_inst}, one easily finds that the typical time scale of the erasure is $\eta_\text{ers} = \pi^2 \sigma / \alpha^2 \mu_5^2$, or equivalently the typical temperature at which the erasure process becomes efficient is
\begin{align}
	T_\text{ers} &\sim \Mpl^{1/2}H_\text{ers}^{1/2} = \left(\frac{\pi^2 \hat \sigma_\text{ini}}{\alpha^2 \hat \mu_{5,\text{ini}}^2} \frac{ H_\text{inf}^{1/2} }{\Mpl^{1/2}} \right)^{-1}
	\sim \frac{\alpha^5}{\pi^4} 
	\frac{\vev{\hat{\bm{E}}\cdot \hat{\bm{B}}}_\text{ini}^2}{H_\text{inf}^5 \Mpl^{2}} \\
	& \sim 10^6 \GeV \times 
	\left( \frac{\alpha}{0.04}  \right)^5
	\left( \frac{H_\text{inf}}{10^{14} \GeV} \right)^3
	\left( \frac{ \vev{\hat{\bm{E}}\cdot \hat{\bm{B}}}_\text{ini} / H_\text{inf}^4}{10^5} \right)^2 \label{eq:T_ers}\\
	&\sim 10^7 \GeV \times \left( \frac{\alpha}{0.04} \right)^3 \left( \frac{\hat \mu_{5,\text{ini}} / \hat T_\text{ini}}{10^{-3}} \right)^2.
\end{align}
See Figs.~\ref{fig:chemical_pot} for the upper bound and estimate of $\hat \mu_\text{ini}/\hat T_\text{ini}$. 
One can see that the erasure process is quite efficient in our toy model, driving the asymmetry to zero already in the early Universe.
Since we have started from vanishing chiral/CS charges in the infinite past, this erasure process can continue until both charges become zero, which is a consequence of the anomaly equation~\eqref{eq:anomalous}.
This erasure process is generic for all the models which create helical gauge fields without modifying the anomalous current equation in the fermion-gauge system.
Note again that the velocity field may affect the dynamics as mentioned in the footnote~\ref{ft:velocity}, and hence this time scale is regarded as an indication where the erasure process becomes relevant.

Based on these observations, let us briefly speculate what might happen in a more realistic scenario. Suppose that we have a CS coupling of U$(1)_\text{Y}$ with the inflaton $\phi$.
In the case of the SM, the SU$(2)_\text{W}$ Sphaleron explicitly breaks Eq.~\eqref{eq:anomalous} in the left-handed sector and the Yukawa interactions mediate this breaking to the right-handed sector. The electroweak Sphaleron becomes efficient already at temperatures around $10^{12}$~GeV, before the erasure process discussed above becomes relevant. Consequently, the balance between helical hyper gauge fields and chiral fermions is disrupted. This may enable some amount of the hypercharge magnetic field to survive until the EW phase transition\footnote{
	See also Refs.~\cite{Boyarsky:2011uy,Boyarsky:2012ex}.
}, where it can regenererate a chiral asymmetry~\cite{Fujita:2016igl,Kamada:2016eeb,Kamada:2016cnb}. {Moreover, if all the SM processes are efficient (at least below $T \sim 10^{5}\GeV$ when the electron Yukawa becomes also efficient), they can re-shuffle all the chemical potentials. 
In this case, the final baryon asymmetry may become independent of the initial conditions for the chemical potentials.} In summary, the specific initial conditions imposed by the anomaly equation as well as the erasure process discussed above may impact scenarios which relate the baryon asymmetry of the Universe to primordial magnetic fields~\cite{Giovannini:1997eg,Boyarsky:2012ex,Kamada:2018tcs}. The final verdict depends on the details of the interplay of all the processes involved and we leave a detailed study to future work.

\section{Conclusions and outlook}
\label{sec:conclusions}

Particle production in the early Universe plays a crucial role in a wide range of processes, \textit{e.g.},  during inflation, during (p)reheating and for baryogenesis. In this paper, we clarify the duality between helical gauge field and chiral fermion production in the presence of a chiral anomaly. We demonstrate the equivalence of the two actions, Eq.~\eqref{eq:frame1} and Eq.~\eqref{eq:frame2}, at the level of the equations of motion and illustrate the intimate connection between gauge field and fermion production by means of conservation equations and Noether charges in Sec.~\ref{sec:setup}. This qualitative understanding of the system is then confirmed by explicit computations of the fermion and gauge field production, including backreaction effects, in Secs.~\ref{sec:prod} and \ref{sec:br}.

The helical gauge fields are produced through a tachyonic instability arising from the Chern-Simons term $\phi F \tilde F$ in the presence of a rolling scalar field, $\dot \phi \neq 0$. For the most part of this paper, we take this scalar field to be the inflaton, but our formalism applies also to more general situations. The fermions are produced by two distinct mechanisms. On the one hand, similar to Schwinger production (but taking into account a non-vanishing magnetic field), pairs of fermions and anti-fermions are created with vanishing net chiral charge. Quantizing the energy states of the fermions in Landau levels with label $n$, this corresponds to populating the higher landau levels ($n \geq 1$). On the other hand, in the presence of helical gauge fields, the population of the lowest Landau level ($n = 0$), leads to fermion production with a non-vanishing net chiral charge. We find that fermions produced through the first production channel typically dominate the fermion energy density.

Energy conservation dictates that the production of chiral fermions must reduce the energy in the helical gauge fields. The fermions are accelerated in the gauge field background, leading to an induced current which counteracts the gauge field background. Additionally, the Schwinger-type pair production drains the gauge field energy. We derive consistency equations to take this backreaction into account, based on energy conservation [see Eq.~\eqref{eq:bound1}] and on postulating a dynamical equilibrium state with constant gauge field energy density~\eqref{eq:curve}. Along the way, we obtain an analytical estimate for the induced current, see Eq.~\eqref{eq:induced_noscat}. Assuming the existence of a dynamical equilibrium between particle production and backreaction, 
we derive upper bounds on the gauge field amplitudes, leading immediately to two important consequences: Firstly, the gauge field amplitudes now exhibit a power-law dependence on $|\dot \phi|$ instead of the exponential dependence known in the absence of this backreaction, leading to a completely different picture in the regime of increasing $|\dot \phi|$. Secondly, due to the dominance of the pair production over the chiral fermion production, we find that thermalization is typically inefficient. Besides obtaining upper bounds for the gauge field, we also provide an estimate for their magnitude.
Depending on the value of $\dot \phi$, this estimate lies about 1 - 2 orders of magnitude below the upper bound, rendering the effects of the fermion backreaction even more pronounced. 

As an illustrative example, we study the impact of these effects on axion inflation. As is common for single field slow-roll inflation models, the velocity of the inflaton (an axion-like particle) increases over the course of inflation. This initially triggers the gauge field and fermion production, but also leads to a strong backreaction towards the end of inflation. Correspondingly, for gauge fields saturating the bounds imposed by the fermion backreaction, the scalar power spectrum is enhanced at intermediate scales, but is suppressed at the largest and smallest scales. This structure, arising naturally in this setup, could be interesting in the context of producing primordial black holes in a mass range where they could contribute to dark matter. One the other hand, if the gauge fields do not saturate the upper bounds, the scalar power spectrum is further suppressed,  eliminating the possibility of producing primordial black holes in this setup.
We expect similarly important effects on the gravitational wave spectrum. However, the analytical estimates performed here are not sufficient to address this question, since contrary to the scalar power spectrum, the tensor power spectrum is not directly related to a chiral charge. We hence leave this interesting analysis for future work.

Other applications in the post-inflationary Universe require tracking the evolution of our Noether charges far into the radiation dominated regime. As an example, we considered possible implications for leptogenesis. In the toy model considered in the main part of this paper, the balance between chiral fermions and helical gauge fields, imposed by the anomaly equation, leads to the full erasure of both the chiral and the Chern-Simons charge. {The physical process responsible for this is the chiral plasma instability which reduces both the chiral charge and helical gauge fields.} Moreover, the thermal plasma may reduce these charges  analogously to the Sphaleron in non-Abelian gauge theory\cite{Figueroa:2017hun}.
In a more realistic setup, involving the Standard Model thermal plasma, the balance between the chiral and Chern-Simons charge is disrupted by the electroweak Sphalerons and Yukawa interactions, which might allow for a final net asymmetry surviving till today. Nevertheless, the symmetric initial conditions for the chiral fermions and helical gauge fields, dictated by the anomaly equation, will in general have an impact on the final value of the predicted baryon asymmetry.  Further possible applications in the post-inflationary universe include the (p)reheating process as well as implementations of the relaxion  using the $\phi F \tilde F$ coupling, see \textit{e.g.}, \cite{Hook:2016mqo, Fonseca:2018xzp}. We hope that our work will stimulate further research in this direction.

Finally, we emphasize that while the analytical estimates performed here lead to  upper bounds on the particle production, we do not know if these bounds are actually saturated. We illustrated the effect of this uncertainty by showing our results both for gauge fields saturating these bounds and for and an estimated solution which turns out to lie significantly lower. Moreover, both our upper bound and the estimated solution are based on assuming the existence of non-trivial attractor. While this is plausible and well motivated, it would be important to check the dynamical approach of the system to this attractor explicitly.
Resolving this question requires solving the non-linear equations of motion for the gauge-fields involving the backreaction, possibly by means of a lattice study. The application of inflation moreover requires performing such a study in de-Sitter space, a challenging task, which is far beyond the scope of this paper.
Also, 
note that,
throughout this paper, we have for simplicity limited ourselves to the CS coupling with an Abelian gauge field and studied the decay of the gauge field via the production of fermions. An extension to a non-Abelian case immediately comes to mind. There, another violent process called Nielsen-Olesen instability~\cite{Nielsen:1978rm} may accelerate the decay of the gauge field~\cite{Tanji:2011di}.
We leave the discussion on this effect to future work.

\section*{Acknowledgements}
We thank Ricardo Z.\ Ferreira, Daniel Figueroa, Kohei Kamada, Alessio Notari and Kai Schmitz for valuable discussions at various stages of this project.

\appendix
\section{Notations and conventions}
\label{sec:nandc}

\subsection{Metric}
We adopt 
\begin{align}
	\left( \eta_{ab} \right)
	= \text{diag}\,\left( +,-,-,- \right)
	\label{eq:sign}
\end{align}
for the metric in Minkowski spacetime. For the totally anti-symmetric tensor, we take the following convention
\begin{align}
	\epsilon^{0123} = - \epsilon_{0123} = +1.
\end{align}
Throughout this paper, we consider the Friedmann-Lema\^itre-Robertson-Walker metric with zero curvature:
\begin{align}
	\dd s^2 = \dd t^2 - a^2 (t) 
		\dd \bm{x}^2
	= a^2 (\eta) 
	\left( \dd \eta^2 - \dd \bm{x}^2 \right).
	\label{eq:flrw}
\end{align}
We usually adopt the conformal time $\eta$ unless otherwise stated. This metric implies the following vierbein:
\begin{align}
	e^a_\mu = a \delta^a_\mu,
	~~~
	e^\mu_a = \frac{1}{a} \delta^\mu_a,
	\label{eq:vierbein}
\end{align}
where $\mu$ runs over $\eta, x, y, z$ 
while $a$ runs over the flat coordinate.

\subsection{Clifford algebras}
The Clifford algebras in Minkowski spacetime is given by
\begin{align}
	\left\{
		\gamma^a, \gamma^b
	\right\} 
	= 2 \eta^{ab}.
\end{align}
Together with the sign convention \eqref{eq:sign}, one can see that $\gamma^0$ is hermitian while $\gamma^i$ is anti-hermitian. 
We use the chiral representation for the gamma matrices
\begin{align}
	\gamma^0 = 
	\left(\begin{array}{cc}0 & 1 \\1 & 0\end{array}\right),~~~
	\bm{\gamma} =
	\left(\begin{array}{cc}0 & \bm{\sigma} \\ -\bm{\sigma} & 0\end{array}\right),~~~
	\gamma_5 =
	\left(\begin{array}{cc} -1 & 0 \\0 & 1\end{array}\right).
\end{align}
The left-/right-handed fermions are obtained from acting with the projection operator  
$\mathcal P_\text{L/R} = (1 \mp \gamma_5)/2$.

The Clifford algebras in the curved spacetime is
\begin{align}
	\left\{ \hat \gamma^\mu, \hat \gamma^\nu \right\}
	= 2 g^{\mu\nu}.
\end{align}
The gamma matrices with a hat represent those defined in the curved spacetime. They are related to the flat spacetime ones via
\begin{align}
	\hat \gamma^\mu = e^\mu_a \gamma^a
	= \frac{1}{a} \gamma^\mu.
\end{align}
In the second equality, we have plugged in Eq.~\eqref{eq:vierbein}. In the curved spacetime, the covariant derivative acting on fermions includes the spin connection $\omega_\mu^{ab}$:
\begin{align}
	\hat{\slashed{\mathcal D}} \hat\psi 
	& = \hat \gamma^\mu 
	\left( 
		\partial_\mu + i g Q \hat A_\mu + \frac{1}{4} \omega_\mu{}^{ab}
		\gamma_{ab}
	\right) \hat \psi \\
	& =
	\left[ \hat \gamma^\mu \left( \partial_\mu + i g Q \hat A_\mu \right)
	+ \frac{3}{2} a H \hat \gamma^0 \right] \hat \psi,
\end{align}
where $\gamma_{ab} = [\gamma_a, \gamma_b]/2$.
In the second line, we have inserted Eq.~\eqref{eq:vierbein} to compute the spin connection explicitly.

\small
\bibliographystyle{utphys}
\bibliography{refs}{}

\providecommand{\href}[2]{#2}\begingroup\raggedright\begin{thebibliography}{10}

\bibitem{Adshead:2012kp}
P.~Adshead and M.~Wyman, ``{Chromo-Natural Inflation: Natural inflation on a
  steep potential with classical non-Abelian gauge fields},''
  \href{http://dx.doi.org/10.1103/PhysRevLett.108.261302}{{\em Phys. Rev.
  Lett.} {\bfseries 108} (2012) 261302},
\href{http://arxiv.org/abs/1202.2366}{{\ttfamily arXiv:1202.2366 [hep-th]}}.

\bibitem{Turner:1987vd}
M.~S. Turner and L.~M. Widrow, ``{Gravitational Production of Scalar Particles
  in Inflationary Universe Models},''
\href{http://dx.doi.org/10.1103/PhysRevD.37.3428}{{\em Phys. Rev.} {\bfseries
  D37} (1988) 3428}.

\bibitem{Garretson:1992vt}
W.~D. Garretson, G.~B. Field, and S.~M. Carroll, ``{Primordial magnetic fields
  from pseudoGoldstone bosons},''
  \href{http://dx.doi.org/10.1103/PhysRevD.46.5346}{{\em Phys. Rev.} {\bfseries
  D46} (1992) 5346--5351},
\href{http://arxiv.org/abs/hep-ph/9209238}{{\ttfamily arXiv:hep-ph/9209238
  [hep-ph]}}.

\bibitem{Anber:2006xt}
M.~M. Anber and L.~Sorbo, ``{N-flationary magnetic fields},''
  \href{http://dx.doi.org/10.1088/1475-7516/2006/10/018}{{\em JCAP} {\bfseries
  0610} (2006) 018},
\href{http://arxiv.org/abs/astro-ph/0606534}{{\ttfamily arXiv:astro-ph/0606534
  [astro-ph]}}.

\bibitem{Cook:2011hg}
J.~L. Cook and L.~Sorbo, ``{Particle production during inflation and
  gravitational waves detectable by ground-based interferometers},''
  \href{http://dx.doi.org/10.1103/PhysRevD.86.069901,
  10.1103/PhysRevD.85.023534}{{\em Phys. Rev.} {\bfseries D85} (2012) 023534},
  \href{http://arxiv.org/abs/1109.0022}{{\ttfamily arXiv:1109.0022
  [astro-ph.CO]}}.
[Erratum: Phys. Rev.D86,069901(2012)].

\bibitem{Barnaby:2011qe}
N.~Barnaby, E.~Pajer, and M.~Peloso, ``{Gauge Field Production in Axion
  Inflation: Consequences for Monodromy, non-Gaussianity in the CMB, and
  Gravitational Waves at Interferometers},''
  \href{http://dx.doi.org/10.1103/PhysRevD.85.023525}{{\em Phys. Rev.}
  {\bfseries D85} (2012) 023525},
\href{http://arxiv.org/abs/1110.3327}{{\ttfamily arXiv:1110.3327
  [astro-ph.CO]}}.

\bibitem{Barnaby:2011vw}
N.~Barnaby, R.~Namba, and M.~Peloso, ``{Phenomenology of a Pseudo-Scalar
  Inflaton: Naturally Large Nongaussianity},''
  \href{http://dx.doi.org/10.1088/1475-7516/2011/04/009}{{\em JCAP} {\bfseries
  1104} (2011) 009},
\href{http://arxiv.org/abs/1102.4333}{{\ttfamily arXiv:1102.4333
  [astro-ph.CO]}}.

\bibitem{Anber:2012du}
M.~M. Anber and L.~Sorbo, ``{Non-Gaussianities and chiral gravitational waves
  in natural steep inflation},''
  \href{http://dx.doi.org/10.1103/PhysRevD.85.123537}{{\em Phys. Rev.}
  {\bfseries D85} (2012) 123537},
\href{http://arxiv.org/abs/1203.5849}{{\ttfamily arXiv:1203.5849
  [astro-ph.CO]}}.

\bibitem{Linde:2012bt}
A.~Linde, S.~Mooij, and E.~Pajer, ``{Gauge field production in supergravity
  inflation: Local non-Gaussianity and primordial black holes},''
  \href{http://dx.doi.org/10.1103/PhysRevD.87.103506}{{\em Phys. Rev.}
  {\bfseries D87} no.~10, (2013) 103506},
\href{http://arxiv.org/abs/1212.1693}{{\ttfamily arXiv:1212.1693 [hep-th]}}.

\bibitem{Adshead:2015pva}
P.~Adshead, J.~T. Giblin, T.~R. Scully, and E.~I. Sfakianakis,
  ``{Gauge-preheating and the end of axion inflation},''
  \href{http://dx.doi.org/10.1088/1475-7516/2015/12/034}{{\em JCAP} {\bfseries
  1512} no.~12, (2015) 034},
\href{http://arxiv.org/abs/1502.06506}{{\ttfamily arXiv:1502.06506
  [astro-ph.CO]}}.

\bibitem{Anber:2009ua}
M.~M. Anber and L.~Sorbo, ``{Naturally inflating on steep potentials through
  electromagnetic dissipation},''
  \href{http://dx.doi.org/10.1103/PhysRevD.81.043534}{{\em Phys. Rev.}
  {\bfseries D81} (2010) 043534},
\href{http://arxiv.org/abs/0908.4089}{{\ttfamily arXiv:0908.4089 [hep-th]}}.

\bibitem{Hook:2016mqo}
A.~Hook and G.~Marques-Tavares, ``{Relaxation from particle production},''
  \href{http://dx.doi.org/10.1007/JHEP12(2016)101}{{\em JHEP} {\bfseries 12}
  (2016) 101},
\href{http://arxiv.org/abs/1607.01786}{{\ttfamily arXiv:1607.01786 [hep-ph]}}.

\bibitem{Tangarife:2017vnd}
W.~Tangarife, K.~Tobioka, L.~Ubaldi, and T.~Volansky, ``{Relaxed Inflation},''
\href{http://arxiv.org/abs/1706.00438}{{\ttfamily arXiv:1706.00438 [hep-ph]}}.

\bibitem{Tangarife:2017rgl}
W.~Tangarife, K.~Tobioka, L.~Ubaldi, and T.~Volansky, ``{Dynamics of Relaxed
  Inflation},'' \href{http://dx.doi.org/10.1007/JHEP02(2018)084}{{\em JHEP}
  {\bfseries 02} (2018) 084},
\href{http://arxiv.org/abs/1706.03072}{{\ttfamily arXiv:1706.03072 [hep-ph]}}.

\bibitem{Fonseca:2018xzp}
N.~Fonseca, E.~Morgante, and G.~Servant, ``{Higgs relaxation after
  inflation},''
\href{http://arxiv.org/abs/1805.04543}{{\ttfamily arXiv:1805.04543 [hep-ph]}}.

\bibitem{Giovannini:1997eg}
M.~Giovannini and M.~E. Shaposhnikov, ``{Primordial hypermagnetic fields and
  triangle anomaly},'' \href{http://dx.doi.org/10.1103/PhysRevD.57.2186}{{\em
  Phys. Rev.} {\bfseries D57} (1998) 2186--2206},
\href{http://arxiv.org/abs/hep-ph/9710234}{{\ttfamily arXiv:hep-ph/9710234
  [hep-ph]}}.

\bibitem{Anber:2015yca}
M.~M. Anber and E.~Sabancilar, ``{Hypermagnetic Fields and Baryon Asymmetry
  from Pseudoscalar Inflation},''
  \href{http://dx.doi.org/10.1103/PhysRevD.92.101501}{{\em Phys. Rev.}
  {\bfseries D92} no.~10, (2015) 101501},
\href{http://arxiv.org/abs/1507.00744}{{\ttfamily arXiv:1507.00744 [hep-th]}}.

\bibitem{Fujita:2016igl}
T.~Fujita and K.~Kamada, ``{Large-scale magnetic fields can explain the baryon
  asymmetry of the Universe},''
  \href{http://dx.doi.org/10.1103/PhysRevD.93.083520}{{\em Phys. Rev.}
  {\bfseries D93} no.~8, (2016) 083520},
\href{http://arxiv.org/abs/1602.02109}{{\ttfamily arXiv:1602.02109 [hep-ph]}}.

\bibitem{Kamada:2016eeb}
K.~Kamada and A.~J. Long, ``{Baryogenesis from decaying magnetic helicity},''
  \href{http://dx.doi.org/10.1103/PhysRevD.94.063501}{{\em Phys. Rev.}
  {\bfseries D94} no.~6, (2016) 063501},
\href{http://arxiv.org/abs/1606.08891}{{\ttfamily arXiv:1606.08891
  [astro-ph.CO]}}.

\bibitem{Cado:2016kdp}
Y.~Cado and E.~Sabancilar, ``{Asymmetric Dark Matter and Baryogenesis from
  Pseudoscalar Inflation},''
  \href{http://dx.doi.org/10.1088/1475-7516/2017/04/047}{{\em JCAP} {\bfseries
  1704} no.~04, (2017) 047},
\href{http://arxiv.org/abs/1611.02293}{{\ttfamily arXiv:1611.02293 [hep-ph]}}.

\bibitem{Jimenez:2017cdr}
D.~Jim\'enez, K.~Kamada, K.~Schmitz, and X.-J. Xu, ``{Baryon asymmetry and
  gravitational waves from pseudoscalar inflation},''
  \href{http://dx.doi.org/10.1088/1475-7516/2017/12/011}{{\em JCAP} {\bfseries
  1712} no.~12, (2017) 011},
\href{http://arxiv.org/abs/1707.07943}{{\ttfamily arXiv:1707.07943 [hep-ph]}}.

\bibitem{Dolgov:1994zq}
A.~Dolgov and K.~Freese, ``{Calculation of particle production by Nambu
  Goldstone bosons with application to inflation reheating and baryogenesis},''
  \href{http://dx.doi.org/10.1103/PhysRevD.51.2693}{{\em Phys. Rev.} {\bfseries
  D51} (1995) 2693--2702},
\href{http://arxiv.org/abs/hep-ph/9410346}{{\ttfamily arXiv:hep-ph/9410346
  [hep-ph]}}.

\bibitem{Kusenko:2014uta}
A.~Kusenko, K.~Schmitz, and T.~T. Yanagida, ``{Leptogenesis via Axion
  Oscillations after Inflation},''
  \href{http://dx.doi.org/10.1103/PhysRevLett.115.011302}{{\em Phys. Rev.
  Lett.} {\bfseries 115} no.~1, (2015) 011302},
\href{http://arxiv.org/abs/1412.2043}{{\ttfamily arXiv:1412.2043 [hep-ph]}}.

\bibitem{Adshead:2015jza}
P.~Adshead and E.~I. Sfakianakis, ``{Leptogenesis from left-handed neutrino
  production during axion inflation},''
  \href{http://dx.doi.org/10.1103/PhysRevLett.116.091301}{{\em Phys. Rev.
  Lett.} {\bfseries 116} no.~9, (2016) 091301},
\href{http://arxiv.org/abs/1508.00881}{{\ttfamily arXiv:1508.00881 [hep-ph]}}.

\bibitem{Adshead:2015kza}
P.~Adshead and E.~I. Sfakianakis, ``{Fermion production during and after axion
  inflation},'' \href{http://dx.doi.org/10.1088/1475-7516/2015/11/021}{{\em
  JCAP} {\bfseries 1511} (2015) 021},
\href{http://arxiv.org/abs/1508.00891}{{\ttfamily arXiv:1508.00891 [hep-ph]}}.

\bibitem{DeSimone:2016ofp}
A.~De~Simone and T.~Kobayashi, ``{Cosmological Aspects of Spontaneous
  Baryogenesis},'' \href{http://dx.doi.org/10.1088/1475-7516/2016/08/052}{{\em
  JCAP} {\bfseries 1608} no.~08, (2016) 052},
\href{http://arxiv.org/abs/1605.00670}{{\ttfamily arXiv:1605.00670 [hep-ph]}}.

\bibitem{Anber:2016yqr}
M.~M. Anber and E.~Sabancilar, ``{Chiral Gravitational Waves from Chiral
  Fermions},'' \href{http://dx.doi.org/10.1103/PhysRevD.96.023501}{{\em Phys.
  Rev.} {\bfseries D96} no.~2, (2017) 023501},
\href{http://arxiv.org/abs/1607.03916}{{\ttfamily arXiv:1607.03916 [hep-th]}}.

\bibitem{Adshead:2018oaa}
P.~Adshead, L.~Pearce, M.~Peloso, M.~A. Roberts, and L.~Sorbo, ``{Phenomenology
  of fermion production during axion inflation},''
\href{http://arxiv.org/abs/1803.04501}{{\ttfamily arXiv:1803.04501
  [astro-ph.CO]}}.

\bibitem{Adler:1969gk}
S.~L. Adler, ``{Axial vector vertex in spinor electrodynamics},''
  \href{http://dx.doi.org/10.1103/PhysRev.177.2426}{{\em Phys. Rev.} {\bfseries
  177} (1969) 2426--2438}.
[,241(1969)].

\bibitem{Bell:1969ts}
J.~S. Bell and R.~Jackiw, ``{A PCAC puzzle: pi0 --> gamma gamma in the sigma
  model},''
\href{http://dx.doi.org/10.1007/BF02823296}{{\em Nuovo Cim.} {\bfseries A60}
  (1969) 47--61}.

\bibitem{Peskin:1995ev}
M.~E. Peskin and D.~V. Schroeder, {\em {An Introduction to quantum field
  theory}}.
\newblock Addison-Wesley, Reading, USA, 1995.
\newblock
\url{http://www.slac.stanford.edu/~mpeskin/QFT.html}.
\newblock

\bibitem{Nielsen:1983rb}
H.~B. Nielsen and M.~Ninomiya, ``{ADLER-BELL-JACKIW ANOMALY AND WEYL FERMIONS
  IN CRYSTAL},''
\href{http://dx.doi.org/10.1016/0370-2693(83)91529-0}{{\em Phys. Lett.}
  {\bfseries 130B} (1983) 389--396}.

\bibitem{Heisenberg:1935qt}
W.~Heisenberg and H.~Euler, ``{Folgerungen aus der Diracschen Theorie des
  Positrons},'' \href{http://dx.doi.org/10.1007/BF01343663}{{\em Z. Phys.}
  {\bfseries 98} (1936) 714--732},
\href{http://arxiv.org/abs/physics/0605038}{{\ttfamily arXiv:physics/0605038
  [physics]}}.

\bibitem{Schwinger:1951nm}
J.~S. Schwinger, ``{On gauge invariance and vacuum polarization},''
  \href{http://dx.doi.org/10.1103/PhysRev.82.664}{{\em Phys. Rev.} {\bfseries
  82} (1951) 664--679}.
[,116(1951)].

\bibitem{Abramchuk:2016afc}
R.~A. Abramchuk and M.~A. Zubkov, ``{Schwinger pair creation in Dirac
  semimetals in the presence of external magnetic and electric fields},''
  \href{http://dx.doi.org/10.1103/PhysRevD.94.116012}{{\em Phys. Rev.}
  {\bfseries D94} no.~11, (2016) 116012},
\href{http://arxiv.org/abs/1605.02379}{{\ttfamily arXiv:1605.02379
  [cond-mat.mes-hall]}}.

\bibitem{Bavarsad:2017oyv}
E.~Bavarsad, S.~P. Kim, C.~Stahl, and S.-S. Xue, ``{Effect of a magnetic field
  on Schwinger mechanism in de Sitter spacetime},''
  \href{http://dx.doi.org/10.1103/PhysRevD.97.025017}{{\em Phys. Rev.}
  {\bfseries D97} no.~2, (2018) 025017},
\href{http://arxiv.org/abs/1707.03975}{{\ttfamily arXiv:1707.03975 [hep-th]}}.

\bibitem{Kobayashi:2014zza}
T.~Kobayashi and N.~Afshordi, ``{Schwinger Effect in 4D de Sitter Space and
  Constraints on Magnetogenesis in the Early Universe},''
  \href{http://dx.doi.org/10.1007/JHEP10(2014)166}{{\em JHEP} {\bfseries 10}
  (2014) 166},
\href{http://arxiv.org/abs/1408.4141}{{\ttfamily arXiv:1408.4141 [hep-th]}}.

\bibitem{Hayashinaka:2016qqn}
T.~Hayashinaka, T.~Fujita, and J.~Yokoyama, ``{Fermionic Schwinger effect and
  induced current in de Sitter space},''
  \href{http://dx.doi.org/10.1088/1475-7516/2016/07/010}{{\em JCAP} {\bfseries
  1607} no.~07, (2016) 010},
\href{http://arxiv.org/abs/1603.04165}{{\ttfamily arXiv:1603.04165 [hep-th]}}.

\bibitem{Fukushima:2008xe}
K.~Fukushima, D.~E. Kharzeev, and H.~J. Warringa, ``{The Chiral Magnetic
  Effect},'' \href{http://dx.doi.org/10.1103/PhysRevD.78.074033}{{\em Phys.
  Rev.} {\bfseries D78} (2008) 074033},
\href{http://arxiv.org/abs/0808.3382}{{\ttfamily arXiv:0808.3382 [hep-ph]}}.

\bibitem{Akamatsu:2013pjd}
Y.~Akamatsu and N.~Yamamoto, ``{Chiral Plasma Instabilities},''
  \href{http://dx.doi.org/10.1103/PhysRevLett.111.052002}{{\em Phys. Rev.
  Lett.} {\bfseries 111} (2013) 052002},
\href{http://arxiv.org/abs/1302.2125}{{\ttfamily arXiv:1302.2125 [nucl-th]}}.

\bibitem{Parker:2009uva}
L.~E. Parker and D.~Toms,
  \href{http://dx.doi.org/10.1017/CBO9780511813924}{{\em {Quantum Field Theory
  in Curved Spacetime}}}.
\newblock Cambridge Monographs on Mathematical Physics. Cambridge University
  Press, 2009.
\newblock
\url{http://www.cambridge.org/de/knowledge/isbn/item2327457}.
\newblock

\bibitem{Fujikawa:1979ay}
K.~Fujikawa, ``{Path Integral Measure for Gauge Invariant Fermion Theories},''
\href{http://dx.doi.org/10.1103/PhysRevLett.42.1195}{{\em Phys. Rev. Lett.}
  {\bfseries 42} (1979) 1195--1198}.

\bibitem{Fujikawa:1980eg}
K.~Fujikawa, ``{Path Integral for Gauge Theories with Fermions},''
  \href{http://dx.doi.org/10.1103/PhysRevD.21.2848,
  10.1103/PhysRevD.22.1499}{{\em Phys. Rev.} {\bfseries D21} (1980) 2848}.
[Erratum: Phys. Rev.D22,1499(1980)].

\bibitem{Turner:1987bw}
M.~S. Turner and L.~M. Widrow, ``{Inflation Produced, Large Scale Magnetic
  Fields},''
\href{http://dx.doi.org/10.1103/PhysRevD.37.2743}{{\em Phys. Rev.} {\bfseries
  D37} (1988) 2743}.

\bibitem{Kasper:2014uaa}
V.~Kasper, F.~Hebenstreit, and J.~Berges, ``{Fermion production from real-time
  lattice gauge theory in the classical-statistical regime},''
  \href{http://dx.doi.org/10.1103/PhysRevD.90.025016}{{\em Phys. Rev.}
  {\bfseries D90} no.~2, (2014) 025016},
\href{http://arxiv.org/abs/1403.4849}{{\ttfamily arXiv:1403.4849 [hep-ph]}}.

\bibitem{Fukushima:2015tza}
K.~Fukushima, ``{Simulating net particle production and chiral magnetic current
  in a $CP$-odd domain},''
  \href{http://dx.doi.org/10.1103/PhysRevD.92.054009}{{\em Phys. Rev.}
  {\bfseries D92} no.~5, (2015) 054009},
\href{http://arxiv.org/abs/1501.01940}{{\ttfamily arXiv:1501.01940 [hep-ph]}}.

\bibitem{Mueller:2016aao}
N.~Mueller, F.~Hebenstreit, and J.~Berges, ``{Anomaly-induced dynamical
  refringence in strong-field QED},''
  \href{http://dx.doi.org/10.1103/PhysRevLett.117.061601}{{\em Phys. Rev.
  Lett.} {\bfseries 117} no.~6, (2016) 061601},
\href{http://arxiv.org/abs/1605.01413}{{\ttfamily arXiv:1605.01413 [hep-ph]}}.

\bibitem{Adler:1969er}
S.~L. Adler and W.~A. Bardeen, ``{Absence of higher order corrections in the
  anomalous axial vector divergence equation},''
  \href{http://dx.doi.org/10.1103/PhysRev.182.1517}{{\em Phys. Rev.} {\bfseries
  182} (1969) 1517--1536}.
[,268(1969)].

\bibitem{Cohen:2008wz}
T.~D. Cohen and D.~A. McGady, ``{The Schwinger mechanism revisited},''
  \href{http://dx.doi.org/10.1103/PhysRevD.78.036008}{{\em Phys. Rev.}
  {\bfseries D78} (2008) 036008},
\href{http://arxiv.org/abs/0807.1117}{{\ttfamily arXiv:0807.1117 [hep-ph]}}.

\bibitem{Starobinsky:1994bd}
A.~A. Starobinsky and J.~Yokoyama, ``{Equilibrium state of a selfinteracting
  scalar field in the De Sitter background},''
  \href{http://dx.doi.org/10.1103/PhysRevD.50.6357}{{\em Phys. Rev.} {\bfseries
  D50} (1994) 6357--6368},
\href{http://arxiv.org/abs/astro-ph/9407016}{{\ttfamily arXiv:astro-ph/9407016
  [astro-ph]}}.

\bibitem{Baym:1997gq}
G.~Baym and H.~Heiselberg, ``{The Electrical conductivity in the early
  universe},'' \href{http://dx.doi.org/10.1103/PhysRevD.56.5254}{{\em Phys.
  Rev.} {\bfseries D56} (1997) 5254--5259},
\href{http://arxiv.org/abs/astro-ph/9704214}{{\ttfamily arXiv:astro-ph/9704214
  [astro-ph]}}.

\bibitem{Arnold:2000dr}
P.~B. Arnold, G.~D. Moore, and L.~G. Yaffe, ``{Transport coefficients in high
  temperature gauge theories. 1. Leading log results},''
  \href{http://dx.doi.org/10.1088/1126-6708/2000/11/001}{{\em JHEP} {\bfseries
  11} (2000) 001},
\href{http://arxiv.org/abs/hep-ph/0010177}{{\ttfamily arXiv:hep-ph/0010177
  [hep-ph]}}.

\bibitem{Landau:1953um}
L.~D. Landau and I.~Pomeranchuk, ``{Limits of applicability of the theory of
  bremsstrahlung electrons and pair production at high-energies},''
{\em Dokl. Akad. Nauk Ser. Fiz.} {\bfseries 92} (1953) 535--536.

\bibitem{Migdal:1956tc}
A.~B. Migdal, ``{Bremsstrahlung and pair production in condensed media at
  high-energies},''
\href{http://dx.doi.org/10.1103/PhysRev.103.1811}{{\em Phys. Rev.} {\bfseries
  103} (1956) 1811--1820}.

\bibitem{Gyulassy:1993hr}
M.~Gyulassy and X.-n. Wang, ``{Multiple collisions and induced gluon
  Bremsstrahlung in QCD},''
  \href{http://dx.doi.org/10.1016/0550-3213(94)90079-5}{{\em Nucl. Phys.}
  {\bfseries B420} (1994) 583--614},
\href{http://arxiv.org/abs/nucl-th/9306003}{{\ttfamily arXiv:nucl-th/9306003
  [nucl-th]}}.

\bibitem{Arnold:2001ba}
P.~B. Arnold, G.~D. Moore, and L.~G. Yaffe, ``{Photon emission from
  ultrarelativistic plasmas},''
  \href{http://dx.doi.org/10.1088/1126-6708/2001/11/057}{{\em JHEP} {\bfseries
  11} (2001) 057},
\href{http://arxiv.org/abs/hep-ph/0109064}{{\ttfamily arXiv:hep-ph/0109064
  [hep-ph]}}.

\bibitem{Arnold:2002ja}
P.~B. Arnold, G.~D. Moore, and L.~G. Yaffe, ``{Photon and gluon emission in
  relativistic plasmas},''
  \href{http://dx.doi.org/10.1088/1126-6708/2002/06/030}{{\em JHEP} {\bfseries
  06} (2002) 030},
\href{http://arxiv.org/abs/hep-ph/0204343}{{\ttfamily arXiv:hep-ph/0204343
  [hep-ph]}}.

\bibitem{Kurkela:2011ti}
A.~Kurkela and G.~D. Moore, ``{Thermalization in Weakly Coupled Nonabelian
  Plasmas},'' \href{http://dx.doi.org/10.1007/JHEP12(2011)044}{{\em JHEP}
  {\bfseries 12} (2011) 044},
\href{http://arxiv.org/abs/1107.5050}{{\ttfamily arXiv:1107.5050 [hep-ph]}}.

\bibitem{Harigaya:2013vwa}
K.~Harigaya and K.~Mukaida, ``{Thermalization after/during Reheating},''
  \href{http://dx.doi.org/10.1007/JHEP05(2014)006}{{\em JHEP} {\bfseries 05}
  (2014) 006},
\href{http://arxiv.org/abs/1312.3097}{{\ttfamily arXiv:1312.3097 [hep-ph]}}.

\bibitem{Mukaida:2015ria}
K.~Mukaida and M.~Yamada, ``{Thermalization Process after Inflation and
  Effective Potential of Scalar Field},''
  \href{http://dx.doi.org/10.1088/1475-7516/2016/02/003}{{\em JCAP} {\bfseries
  1602} no.~02, (2016) 003},
\href{http://arxiv.org/abs/1506.07661}{{\ttfamily arXiv:1506.07661 [hep-ph]}}.

\bibitem{Ellis:2015jpg}
J.~Ellis, M.~A.~G. Garcia, D.~V. Nanopoulos, K.~A. Olive, and M.~Peloso,
  ``{Post-Inflationary Gravitino Production Revisited},''
  \href{http://dx.doi.org/10.1088/1475-7516/2016/03/008}{{\em JCAP} {\bfseries
  1603} no.~03, (2016) 008},
\href{http://arxiv.org/abs/1512.05701}{{\ttfamily arXiv:1512.05701
  [astro-ph.CO]}}.

\bibitem{Ferreira:2017lnd}
R.~Z. Ferreira and A.~Notari, ``{Thermalized Axion Inflation},''
  \href{http://dx.doi.org/10.1088/1475-7516/2017/09/007}{{\em JCAP} {\bfseries
  1709} no.~09, (2017) 007},
\href{http://arxiv.org/abs/1706.00373}{{\ttfamily arXiv:1706.00373
  [astro-ph.CO]}}.

\bibitem{Barnaby:2010vf}
N.~Barnaby and M.~Peloso, ``{Large Nongaussianity in Axion Inflation},''
  \href{http://dx.doi.org/10.1103/PhysRevLett.106.181301}{{\em Phys. Rev.
  Lett.} {\bfseries 106} (2011) 181301},
\href{http://arxiv.org/abs/1011.1500}{{\ttfamily arXiv:1011.1500 [hep-ph]}}.

\bibitem{Meerburg:2012id}
P.~D. Meerburg and E.~Pajer, ``{Observational Constraints on Gauge Field
  Production in Axion Inflation},''
  \href{http://dx.doi.org/10.1088/1475-7516/2013/02/017}{{\em JCAP} {\bfseries
  1302} (2013) 017},
\href{http://arxiv.org/abs/1203.6076}{{\ttfamily arXiv:1203.6076
  [astro-ph.CO]}}.

\bibitem{Garcia-Bellido:2016dkw}
J.~Garcia-Bellido, M.~Peloso, and C.~Unal, ``{Gravitational waves at
  interferometer scales and primordial black holes in axion inflation},''
  \href{http://dx.doi.org/10.1088/1475-7516/2016/12/031}{{\em JCAP} {\bfseries
  1612} no.~12, (2016) 031},
\href{http://arxiv.org/abs/1610.03763}{{\ttfamily arXiv:1610.03763
  [astro-ph.CO]}}.

\bibitem{Domcke:2017fix}
V.~Domcke, F.~Muia, M.~Pieroni, and L.~T. Witkowski, ``{PBH dark matter from
  axion inflation},''
  \href{http://dx.doi.org/10.1088/1475-7516/2017/07/048}{{\em JCAP} {\bfseries
  1707} (2017) 048},
\href{http://arxiv.org/abs/1704.03464}{{\ttfamily arXiv:1704.03464
  [astro-ph.CO]}}.

\bibitem{Domcke:2016bkh}
V.~Domcke, M.~Pieroni, and P.~Bin\'etruy, ``{Primordial gravitational waves for
  universality classes of pseudoscalar inflation},''
  \href{http://dx.doi.org/10.1088/1475-7516/2016/06/031}{{\em JCAP} {\bfseries
  1606} (2016) 031},
\href{http://arxiv.org/abs/1603.01287}{{\ttfamily arXiv:1603.01287
  [astro-ph.CO]}}.

\bibitem{Bartolo:2016ami}
N.~Bartolo {\em et~al.}, ``{Science with the space-based interferometer LISA.
  IV: Probing inflation with gravitational waves},''
  \href{http://dx.doi.org/10.1088/1475-7516/2016/12/026}{{\em JCAP} {\bfseries
  1612} no.~12, (2016) 026},
\href{http://arxiv.org/abs/1610.06481}{{\ttfamily arXiv:1610.06481
  [astro-ph.CO]}}.

\bibitem{Peloso:2016gqs}
M.~Peloso, L.~Sorbo, and C.~Unal, ``{Rolling axions during inflation:
  perturbativity and signatures},''
  \href{http://dx.doi.org/10.1088/1475-7516/2016/09/001}{{\em JCAP} {\bfseries
  1609} no.~09, (2016) 001},
\href{http://arxiv.org/abs/1606.00459}{{\ttfamily arXiv:1606.00459
  [astro-ph.CO]}}.

\bibitem{Ferreira:2015omg}
R.~Z. Ferreira, J.~Ganc, J.~Nore\~na, and M.~S. Sloth, ``{On the validity of
  the perturbative description of axions during inflation},''
  \href{http://dx.doi.org/10.1088/1475-7516/2016/10/E01,
  10.1088/1475-7516/2016/04/039}{{\em JCAP} {\bfseries 1604} no.~04, (2016)
  039}, \href{http://arxiv.org/abs/1512.06116}{{\ttfamily arXiv:1512.06116
  [astro-ph.CO]}}.
[Erratum: JCAP1610,no.10,E01(2016)].

\bibitem{Audley:2017drz}
{\bfseries LISA} Collaboration, H.~Audley {\em et~al.}, ``{Laser Interferometer
  Space Antenna},''
\href{http://arxiv.org/abs/1702.00786}{{\ttfamily arXiv:1702.00786
  [astro-ph.IM]}}.

\bibitem{Alexander:2004us}
S.~H.-S. Alexander, M.~E. Peskin, and M.~M. Sheikh-Jabbari, ``{Leptogenesis
  from gravity waves in models of inflation},''
  \href{http://dx.doi.org/10.1103/PhysRevLett.96.081301}{{\em Phys. Rev. Lett.}
  {\bfseries 96} (2006) 081301},
\href{http://arxiv.org/abs/hep-th/0403069}{{\ttfamily arXiv:hep-th/0403069
  [hep-th]}}.

\bibitem{Adshead:2017znw}
P.~Adshead, A.~J. Long, and E.~I. Sfakianakis, ``{Gravitational Leptogenesis,
  Reheating, and Models of Neutrino Mass},''
  \href{http://dx.doi.org/10.1103/PhysRevD.97.043511}{{\em Phys. Rev.}
  {\bfseries D97} no.~4, (2018) 043511},
\href{http://arxiv.org/abs/1711.04800}{{\ttfamily arXiv:1711.04800 [hep-ph]}}.

\bibitem{Boyarsky:2012ex}
A.~Boyarsky, O.~Ruchayskiy, and M.~Shaposhnikov, ``{Long-range magnetic fields
  in the ground state of the Standard Model plasma},''
  \href{http://dx.doi.org/10.1103/PhysRevLett.109.111602}{{\em Phys. Rev.
  Lett.} {\bfseries 109} (2012) 111602},
\href{http://arxiv.org/abs/1204.3604}{{\ttfamily arXiv:1204.3604 [hep-ph]}}.

\bibitem{Kamada:2016cnb}
K.~Kamada and A.~J. Long, ``{Evolution of the Baryon Asymmetry through the
  Electroweak Crossover in the Presence of a Helical Magnetic Field},''
  \href{http://dx.doi.org/10.1103/PhysRevD.94.123509}{{\em Phys. Rev.}
  {\bfseries D94} no.~12, (2016) 123509},
\href{http://arxiv.org/abs/1610.03074}{{\ttfamily arXiv:1610.03074 [hep-ph]}}.

\bibitem{Kamada:2018tcs}
K.~Kamada, ``{Return of grand unified theory baryogenesis: Source of helical
  hypermagnetic fields for the baryon asymmetry of the universe},''
  \href{http://dx.doi.org/10.1103/PhysRevD.97.103506}{{\em Phys. Rev.}
  {\bfseries D97} no.~10, (2018) 103506},
\href{http://arxiv.org/abs/1802.03055}{{\ttfamily arXiv:1802.03055 [hep-ph]}}.

\bibitem{Figueroa:2017hun}
D.~G. Figueroa and M.~Shaposhnikov, ``{Anomalous non-conservation of
  fermion/chiral number in Abelian gauge theories at finite temperature},''
  \href{http://dx.doi.org/10.1007/JHEP04(2018)026}{{\em JHEP} {\bfseries 04}
  (2018) 026},
\href{http://arxiv.org/abs/1707.09967}{{\ttfamily arXiv:1707.09967 [hep-ph]}}.

\bibitem{Schober:2017cdw}
J.~Schober, I.~Rogachevskii, A.~Brandenburg, A.~Boyarsky, J.~Fr{\"o}hlich,
  O.~Ruchayskiy, and N.~Kleeorin, ``{Laminar and turbulent dynamos in chiral
  magnetohydrodynamics. II. Simulations},''
  \href{http://dx.doi.org/10.3847/1538-4357/aaba75}{{\em Astrophys. J.}
  {\bfseries 858} no.~2, (2018) 124},
\href{http://arxiv.org/abs/1711.09733}{{\ttfamily arXiv:1711.09733
  [physics.flu-dyn]}}.

\bibitem{Schober:2018ojn}
J.~Schober, A.~Brandenburg, I.~Rogachevskii, and N.~Kleeorin, ``{Magnetic
  Prandtl number dependence of turbulence generated by chiral MHD dynamos},''
\href{http://arxiv.org/abs/1803.06350}{{\ttfamily arXiv:1803.06350
  [physics.flu-dyn]}}.

\bibitem{Boyarsky:2011uy}
A.~Boyarsky, J.~Frohlich, and O.~Ruchayskiy, ``{Self-consistent evolution of
  magnetic fields and chiral asymmetry in the early Universe},''
  \href{http://dx.doi.org/10.1103/PhysRevLett.108.031301}{{\em Phys. Rev.
  Lett.} {\bfseries 108} (2012) 031301},
\href{http://arxiv.org/abs/1109.3350}{{\ttfamily arXiv:1109.3350
  [astro-ph.CO]}}.

\bibitem{Nielsen:1978rm}
N.~K. Nielsen and P.~Olesen, ``{An Unstable Yang-Mills Field Mode},''
\href{http://dx.doi.org/10.1016/0550-3213(78)90377-2}{{\em Nucl. Phys.}
  {\bfseries B144} (1978) 376--396}.

\bibitem{Tanji:2011di}
N.~Tanji and K.~Itakura, ``{Schwinger mechanism enhanced by the Nielsen-Olesen
  instability},'' \href{http://dx.doi.org/10.1016/j.physletb.2012.05.043}{{\em
  Phys. Lett.} {\bfseries B713} (2012) 117--121},
\href{http://arxiv.org/abs/1111.6772}{{\ttfamily arXiv:1111.6772 [hep-ph]}}.

\end{thebibliography}\endgroup
  
\end{document}